\documentclass[a4paper, 10pt]{article}
\pdfoutput=1

\usepackage{jheppub}

\usepackage{amsthm}

\usepackage[numbers,sort&compress]{natbib}
\usepackage{dsfont}
\usepackage{slashed}
\usepackage{color}
\usepackage{empheq}

\usepackage[export]{adjustbox}

\usepackage{bibgerm}

\usepackage{soul}

\usepackage[english]{babel}

\usepackage[utf8x]{inputenc}

\usepackage{graphicx,epsfig}
\usepackage{pstricks}

\usepackage{tikz}

\usepackage{bashful}

\usepackage{subfigure}

\usepackage{setspace}

\usepackage{booktabs}

\usepackage{float}

\usepackage{nextpage}

\usepackage{pdfpages}

\usepackage{hypcap}

\usepackage[small,bf]{caption}

\usepackage{longtable}

\usepackage{fancyhdr}

\usepackage[makeroom]{cancel}

\usepackage{microtype}

\usepackage{xparse}
\usepackage{etoolbox}

\usepackage[subfigure]{tocloft}

\usepackage{listings}

\usepackage{ifmtarg}

\usepackage{multirow}

\newcommand{\nn}{\nonumber\\}

\newcommand{\C}{\mathcal{C}}
\newcommand{\M}{\mathcal{M}}
\newcommand{\A}{\mathcal{A}}
\newcommand{\B}{\mathcal{B}}
\newcommand{\F}{\mathcal{F}}
\newcommand{\p}{\partial}

\renewcommand{\Im}{\text{Im}}

\newcommand{\<}{\langle}
\renewcommand{\>}{\rangle}

\newcommand{\mpio}{M_{\pi^0}}

\newcommand{\beq}{\begin{equation}}
\newcommand{\eeq}{\end{equation}}

\newcommand{\remark}[1]{}

\makeatletter
\newcommand{\Cr}[2]{\@ifmtarg{#2}{\C_{#1}}{\C_{#1}\big[#2\big]}}
\makeatother

\makeatletter
  \def\my@tag@font{\normalsize}
  \def\maketag@@@#1{\hbox{\m@th\normalfont\my@tag@font#1}}
  \let\amsmath@eqref\eqref
  \renewcommand\eqref[1]{{\let\my@tag@font\relax\amsmath@eqref{#1}}}
\makeatother

\makeatletter
\renewcommand\paragraph{\@startsection{paragraph}{4}{\z@}%
  {-3.25ex\@plus -1ex \@minus -.2ex}%
  {1.5ex \@plus .2ex}%
  {\normalfont\normalsize\bfseries}}
\makeatother

\cftsetindents{section}{0em}{2em}
\cftsetindents{subsection}{2em}{3em}
\cftsetindents{subsubsection}{5em}{4em}

\allowdisplaybreaks[1]
\title{\boldmath Dispersion relations for hadronic light-by-light scattering in triangle kinematics}

\author[a]{Jan Lüdtke,}
\author[a]{Massimiliano Procura,}
\author[b,c]{Peter Stoffer}

\affiliation[a]{Faculty of Physics, University of Vienna, Boltzmanngasse 5, 1090 Vienna, Austria}
\affiliation[b]{Physik-Institut, Universit\"at Z\"urich, Winterthurerstrasse 190, 8057 Z\"urich, Switzerland}
\affiliation[c]{Paul Scherrer Institut, 5232 Villigen PSI, Switzerland}

\abstract{
We present a new strategy for the dispersive evaluation of the hadronic light-by-light contribution to the anomalous magnetic moment of the muon $a_\mu$. The new approach directly applies in the kinematic limit relevant for $a_\mu$: one of the photons is treated as an external electromagnetic field with vanishing momentum, so that the kinematics corresponds to a triangle. We derive expressions for the relevant single-particle intermediate states, as well as the tensor decompositions of the two-pion sub-processes that appear in addition to those needed in the established dispersive approach. The existing approach is based on a set of dispersion relations for the hadronic light-by-light tensor in four-point kinematics. At present it is not known how to consistently include in this framework resonant intermediate states of spin 2 or larger, due to the appearance of kinematic singularities that can be traced back to the redundancy of the tensor decomposition. We show that our new approach circumvents this problem and enables dispersion relations in the limit of triangle kinematics that are manifestly free from kinematic singularities, paving the way towards a data-driven evaluation of all relevant exclusive hadronic intermediate states.
}

\numberwithin{equation}{section}

\begin{document}

\preprint{
\mbox{}\hfill{} PSI-PR-23-5 \\
\mbox{}\hfill{} UWThPh 2023-5 \\
\mbox{}\hfill{} ZU-TH 10/23
}
	\maketitle



\section{Introduction}

Achieving a stringent comparison between an accurate Standard Model evaluation of the muon $g-2$~\cite{Aoyama:2020ynm,Aoyama:2012wk,Aoyama:2019ryr,Czarnecki:2002nt,Gnendiger:2013pva,Davier:2017zfy,Keshavarzi:2018mgv,Colangelo:2018mtw,Hoferichter:2019mqg,Davier:2019can,Keshavarzi:2019abf,Kurz:2014wya,Melnikov:2003xd,Masjuan:2017tvw,Colangelo:2017fiz,Hoferichter:2018kwz,Gerardin:2019vio,Bijnens:2019ghy,Colangelo:2019uex,Blum:2019ugy,Colangelo:2014qya} with robust theory uncertainties and its increasingly precise experimental measurements~\cite{Muong-2:2006rrc,Muong-2:2021ojo} is a key goal in particle physics.  Hadronic contributions play a central role in this context since they are responsible for the bulk of the theory uncertainty. According to the current consensus~\cite{Aoyama:2020ynm}, a substantial part of this uncertainty is due to the hadronic light-by-light contribution (HLbL)~\cite{Melnikov:2003xd,Pauk:2014rta,Danilkin:2016hnh,Jegerlehner:2017gek,Masjuan:2017tvw,Colangelo:2017fiz,Colangelo:2017qdm,Hoferichter:2018kwz,Hoferichter:2018dmo,Knecht:2018sci,Gerardin:2019vio,Bijnens:2019ghy,Colangelo:2019uex,Roig:2019reh,Eichmann:2019bqf,Blum:2019ugy,Zanke:2021wiq,Chao:2021tvp,Leutgeb:2021mpu,Stamen:2022uqh,Colangelo:2022jxc,Chao:2022xzg,Leutgeb:2022lqw}, which is the subject of the present study. In the framework of a data-driven determination of HLbL based on dispersion relations, achieving control over the contributions from intermediate states with masses between 1 and 2 GeV is crucial to reduce the theory error to the size of the projected precision of the final experimental results at Fermilab~\cite{Muong-2:2015xgu,Aoyama:2020ynm}. A model-independent evaluation of these effects is not available yet, also due to the fact that it is not known how to unambiguously include contributions from resonant intermediate states of spin two or larger within the standard dispersive representation of HLbL in general four-point kinematics~\cite{Colangelo:2014dfa,Colangelo:2015ama,Colangelo:2017fiz}. In this paper, we introduce a novel formalism that overcomes this issue. Our framework employs dispersion relations formulated directly in the limit of a soft external photon (triangle kinematics), which is free of the aforementioned ambiguities. Compared to the established approach, contributions from different intermediate states get reshuffled, unitarity relations become more involved, and the dispersive reconstructions of additional hadronic sub-processes, most importantly $\gamma^* \gamma^* \gamma \to \pi \pi$ and $\pi \pi \to \pi \pi \gamma$, are required. Here we explicitly derive analytic expressions for the single-particle intermediate state contributions to HLbL in triangle kinematics, including tensor resonances, as well as Lorentz decompositions for the two-pion sub-processes  $\gamma^* \gamma^* \gamma \to \pi \pi$ and $\pi \pi \to \pi \pi \gamma$ leading to scalar functions free of kinematic singularities. This opens a path towards the first complete data-driven evaluation of all exclusive hadronic contributions to HLbL that are relevant at an accuracy adequate for the comparison with the forthcoming measurements of the muon $g-2$.

The paper is organized as follows. After a brief review of the Lorentz decomposition of the HLbL tensor and the master formula to extract the HLbL contribution to the muon anomalous magnetic moment (Sect.~\ref{sec:TensorDecomposition}), we discuss the dispersion relations in triangle kinematics and highlight similarities, differences, and advantages with respect to the established dispersive approach to HLbL in Sect.~\ref{sec:DRs}. Unitarity relations, also in comparison with four-point kinematics, are the subject of Sect.~\ref{sec:URs}. Our results for the single-particle intermediate states are collected in Sect.~\ref{sec:SingleParticle}. Sect.~\ref{sec:TwoPionSubprocesses} is devoted to the tensor decompositions for the two-pion sub-processes required to solve two-pion unitarity and to the discussion of the relevant soft-photon limits. Conclusions are drawn in Sect.~\ref{sec:Conclusion}.


\section{The HLbL tensor}
\label{sec:TensorDecomposition}

In this section, we briefly review the Lorentz decomposition of the HLbL tensor and the master formula for the HLbL contribution to $a_\mu$, in the notation of Refs.~\cite{Colangelo:2015ama,Colangelo:2017fiz}.

\subsection{BTT decomposition of the HLbL tensor}

The HLbL tensor is defined as the hadronic Green's function of four electromagnetic currents in pure QCD:
\begin{align}
	\label{eq:HLbLTensorDefinition}
	\Pi^{\mu\nu\lambda\sigma}(q_1,q_2,q_3) = -i \int d^4x \, d^4y \, d^4z \, e^{-i(q_1 \cdot x + q_2 \cdot y + q_3 \cdot z)} \< 0 | T \{ j_\mathrm{em}^\mu(x) j_\mathrm{em}^\nu(y) j_\mathrm{em}^\lambda(z) j_\mathrm{em}^\sigma(0) \} | 0 \> \, ,
\end{align}
where the electromagnetic current includes the three lightest quarks:
\begin{align}
	j_\mathrm{em}^\mu := \bar q Q \gamma^\mu q \, , \quad q = ( u , d, s )^T \, , \quad Q = \mathrm{diag}\left(\frac{2}{3}, -\frac{1}{3}, -\frac{1}{3}\right) \, .
\end{align}
The hadronic contribution to the helicity amplitudes for (off-shell) photon--photon scattering is given by the contraction of the HLbL tensor with polarization vectors:
\begin{align}
	\label{eq:HLbLHelicityAmplitudesDefinition}
	H_{\lambda_1\lambda_2,\lambda_3\lambda_4} = \epsilon_\mu^{\lambda_1}(q_1) \epsilon_\nu^{\lambda_2}(q_2) {\epsilon_\lambda^{\lambda_3}}^*(-q_3) {\epsilon_\sigma^{\lambda_4}}^*(q_4) \Pi^{\mu\nu\lambda\sigma}(q_1,q_2,q_3) \, ,
\end{align}
where $q_4 = q_1 + q_2 + q_3$. We use rescaled helicity amplitudes that remain finite in the limit $q_i^2\to0$:
\begin{align}
	\label{eq:HLbLFiniteHelicityAmplitudes}
	H_{\lambda_1\lambda_2,\lambda_3\lambda_4} =: \kappa^1_{\lambda_1} \kappa^2_{\lambda_2} \kappa^3_{\lambda_3} \kappa^4_{\lambda_4} \bar H_{\lambda_1\lambda_2,\lambda_3\lambda_4} \, , \quad
	\kappa^i_{\pm} = 1, \quad \kappa^i_0 = \frac{q_i^2}{\xi_i} ,
\end{align}
where $\xi_i$ refers to the normalization of the longitudinal polarization vectors, see Ref.~\cite{Colangelo:2017fiz}.

The usual Mandelstam variables
\begin{align}
	s := (q_1+q_2)^2 \, , \quad t := (q_1+q_3)^2 \, , \quad u := (q_2 + q_3)^2
\end{align}
fulfill the linear relation
\begin{align}
	s + t + u = \sum_{i=1}^4 q_i^2 =: \Sigma \, .
\end{align}
Gauge invariance requires the HLbL tensor to satisfy the Ward--Takahashi identities
\begin{align}
	\label{eq:WardIdentitiesHLbLTensor}
	\{q_1^\mu, q_2^\nu, q_3^\lambda, q_4^\sigma\} \Pi_{\mu\nu\lambda\sigma}(q_1,q_2,q_3) = 0 \, .
\end{align}

Based on a recipe by Bardeen, Tung~\cite{Bardeen:1969aw}, and Tarrach~\cite{Tarrach:1975tu} (BTT), in Ref.~\cite{Colangelo:2015ama} a decomposition of the HLbL tensor was derived:
\begin{align}
	\label{eqn:HLbLTensorKinematicFreeStructures}
	\Pi^{\mu\nu\lambda\sigma} &= \sum_{i=1}^{54} T_i^{\mu\nu\lambda\sigma} \Pi_i \, ,
\end{align}
where the tensor structures are given by
\begin{align}
	\label{eq:HLbLBTTStructures}
	T_1^{\mu\nu\lambda\sigma} &= \epsilon^{\mu\nu\alpha\beta} \epsilon^{\lambda\sigma\gamma\delta} {q_1}_\alpha {q_2}_\beta {q_3}_\gamma {q_4}_\delta \, , \nn
	T_4^{\mu\nu\lambda\sigma} &= \Big(q_2^\mu q_1^\nu - q_1 \cdot q_2 g^{\mu \nu} \Big) \Big( q_4^\lambda q_3^\sigma - q_3 \cdot q_4 g^{\lambda \sigma} \Big) \, , \nn
	T_7^{\mu\nu\lambda\sigma} &= \Big(q_2^\mu q_1^\nu - q_1 \cdot q_2 g^{\mu \nu } \Big) \Big( q_1 \cdot q_4 \left(q_1^\lambda q_3^\sigma -q_1 \cdot q_3 g^{\lambda \sigma} \right) + q_4^\lambda q_1^\sigma q_1 \cdot q_3 - q_1^\lambda q_1^\sigma q_3 \cdot q_4 \Big) \, , \nn
	 T_{19}^{\mu\nu\lambda\sigma} &= \Big( q_2^\mu q_1^\nu - q_1 \cdot q_2 g^{\mu \nu } \Big) \Big(q_2 \cdot q_4 \left(q_1^\lambda q_3^\sigma - q_1 \cdot q_3 g^{\lambda\sigma} \right)+q_4^\lambda q_2^\sigma q_1 \cdot q_3 - q_1^\lambda q_2^\sigma q_3 \cdot q_4 \Big) \, , \nn
	T_{31}^{\mu\nu\lambda\sigma} &= \Big(q_2^\mu q_1^\nu - q_1\cdot q_2 g^{\mu\nu}\Big) \Big(q_2^\lambda q_1\cdot q_3 - q_1^\lambda q_2\cdot q_3\Big) \Big(q_2^\sigma q_1\cdot q_4 - q_1^\sigma q_2\cdot q_4\Big) \, , \nn
	T_{37}^{\mu\nu\lambda\sigma} &= \Big( q_3^\mu q_1\cdot q_4 - q_4^\mu q_1\cdot q_3\Big) \begin{aligned}[t]
		& \Big( q_3^\nu q_4^\lambda q_2^\sigma - q_4^\nu q_2^\lambda q_3^\sigma + g^{\lambda\sigma} \left(q_4^\nu q_2\cdot q_3 - q_3^\nu q_2\cdot q_4\right) \\
		& + g^{\nu\sigma} \left( q_2^\lambda q_3\cdot q_4 - q_4^\lambda q_2\cdot q_3 \right) + g^{\lambda\nu} \left( q_3^\sigma q_2\cdot q_4 - q_2^\sigma q_3\cdot q_4 \right) \Big) \, , \end{aligned} \nn
	T_{49}^{\mu\nu\lambda\sigma} &= q_3^\sigma  \begin{aligned}[t]
			& \Big( q_1\cdot q_3 q_2\cdot q_4 q_4^\mu g^{\lambda\nu} - q_2\cdot q_3 q_1\cdot q_4 q_4^\nu g^{\lambda\mu} + q_4^\mu q_4^\nu \left( q_1^\lambda q_2\cdot q_3 - q_2^\lambda q_1\cdot q_3 \right) \\
			& + q_1\cdot q_4 q_3^\mu q_4^\nu q_2^\lambda - q_2\cdot q_4 q_4^\mu q_3^\nu q_1^\lambda + q_1\cdot q_4 q_2\cdot q_4 \left(q_3^\nu g^{\lambda\mu} - q_3^\mu g^{\lambda\nu}\right) \Big) \end{aligned} \nn
		& - q_4^\lambda \begin{aligned}[t]
			& \Big( q_1\cdot q_4 q_2\cdot q_3 q_3^\mu g^{\nu\sigma} - q_2\cdot q_4 q_1\cdot q_3 q_3^\nu g^{\mu\sigma} + q_3^\mu q_3^\nu \left(q_1^\sigma q_2\cdot q_4 - q_2^\sigma q_1\cdot q_4\right) \\
			& + q_1\cdot q_3 q_4^\mu q_3^\nu q_2^\sigma - q_2\cdot q_3 q_3^\mu q_4^\nu q_1^\sigma + q_1\cdot q_3 q_2\cdot q_3 \left( q_4^\nu g^{\mu\sigma} - q_4^\mu g^{\nu\sigma} \right) \Big) \end{aligned} \\
		& + q_3\cdot q_4 \Big(\left(q_1^\lambda q_4^\mu - q_1\cdot q_4 g^{\lambda\mu}\right) \left(q_3^\nu q_2^\sigma - q_2\cdot q_3 g^{\nu\sigma}\right) - \left(q_2^\lambda q_4^\nu - q_2\cdot q_4 g^{\lambda\nu}\right) \left(q_3^\mu q_1^\sigma - q_1\cdot q_3 g^{\mu\sigma}\right)\Big) \nonumber
\end{align}
and all remaining ones are crossed versions of the above structures~\cite{Colangelo:2015ama}.
The BTT decomposition has the property that on the one hand all the Lorentz structures fulfill the Ward--Takahashi identities, i.e.,
\begin{align}
	\{q_1^\mu, q_2^\nu, q_3^\lambda, q_4^\sigma\} T^i_{\mu\nu\lambda\sigma}(q_1,q_2,q_3) = 0 \, , \quad \forall i \in \{1, \ldots, 54 \} \, ,
\end{align}
on the other hand the scalar coefficient functions $\Pi_i$ are free of kinematic singularities and zeros.

Since the number of helicity amplitudes for fully off-shell photon--photon scattering is 41, the set of 54 structures $\{ T_i^{\mu\nu\lambda\sigma}\}$ does not form a basis, but exhibits a 13-fold redundancy, as discussed in detail in Ref.~\cite{Colangelo:2015ama}. While 11 linear relations hold in general, two additional ones are present in four space-time dimensions~\cite{Leo:1975fb,Eichmann:2014ooa}. They can be derived most easily using the relation
\begin{align}
	\label{eq:Schouten}
	0 &= q_1^\alpha q_2^\beta q_3^\gamma \Big( g_{\mu\nu} \epsilon_{\lambda\alpha\beta\gamma} + g_{\mu\lambda} \epsilon_{\alpha\beta\gamma\nu} + g_{\mu\alpha} \epsilon_{\beta\gamma\nu\lambda} + g_{\mu\beta} \epsilon_{\gamma\nu\lambda\alpha} + g_{\mu\gamma} \epsilon_{\nu\lambda\alpha\beta} \Big) q_1^{\alpha'} q_2^{\beta'} q_3^{\gamma'} \epsilon_{\sigma\alpha'\beta'\gamma'} \, ,
\end{align}
which holds because the bracket vanishes in $D=4$ space-time dimensions due to the Schouten identity. After expanding the right-hand side of Eq.~\eqref{eq:Schouten} and expressing the products of Levi-Civita tensors in terms of metric tensors, the projection onto the BTT set gives a linear relation between the Lorentz structures, while a second independent relation is obtained from a crossed version of Eq.~\eqref{eq:Schouten}.

Away from $D=4$ space-time dimensions, a subset of 43 Lorentz structures forms a basis:
\begin{align}
	\label{eqn:HLbLTensorBasisDecomposition}
	\Pi^{\mu\nu\lambda\sigma} &= \sum_{i=1}^{43} \mathcal{B}_i^{\mu\nu\lambda\sigma} \tilde\Pi_i \, ,
\end{align}
where the basis-coefficient functions $\tilde\Pi_i$ are no longer free of kinematic singularities. However, the explicit structure of their kinematic singularities follows from the projection of the BTT decomposition onto this basis in $D$ dimensions.

\subsection[Master formula for the HLbL contribution to $a_\mu$]{\boldmath Master formula for the HLbL contribution to $a_\mu$}

\label{sec:MasterFormula}

Based on projection techniques in Dirac space, one can extract the HLbL contribution to $a_\mu = (g-2)_\mu/2$ from the following expression~\cite{Aldins:1970id}:
\begin{align}
	\label{eq:BasicMasterFormula}
	a_\mu^\mathrm{HLbL} &= - \frac{e^6}{48 m_\mu}  \int \frac{d^4q_1}{(2\pi)^4} \frac{d^4q_2}{(2\pi)^4} \frac{1}{q_1^2 q_2^2 (q_1+q_2)^2} \frac{1}{(p+q_1)^2 - m_\mu^2} \frac{1}{(p-q_2)^2 - m_\mu^2} \nn
		& \quad \times \mathrm{Tr}\left( (\slashed p + m_\mu) [\gamma^\rho,\gamma^\sigma] (\slashed p + m_\mu) \gamma^\mu (\slashed p + \slashed q_1 + m_\mu) \gamma^\lambda (\slashed p - \slashed q_2 + m_\mu) \gamma^\nu \right)  \nn
		& \quad \times  \sum_{i=1}^{54} \left( \frac{\p}{\p q_4^\rho} T^i_{\mu\nu\lambda\sigma}(q_1,q_2,q_4-q_1-q_2) \right)  \bigg|_{q_4=0} \Pi_i(q_1,q_2,-q_1-q_2) \, .
\end{align}
There are only 19 independent linear combinations of the structures $T_i^{\mu\nu\lambda\sigma}$ that contribute to $(g-2)_\mu$, hence we can make a basis change in the 54 structures
\begin{align}
	\label{eq:HLbLTensorPiHatDecomposition}
	\Pi^{\mu\nu\lambda\sigma} = \sum_{i=1}^{54} T_i^{\mu\nu\lambda\sigma} \Pi_i =  \sum_{i=1}^{54} \hat T_i^{\mu\nu\lambda\sigma} \hat \Pi_i \, ,
\end{align}
in such a way that in the limit $q_4\to0$ the derivative of 35 structures $\hat T_i^{\mu\nu\lambda\sigma}$ vanishes. For the non-vanishing derivatives with indices $\{ g_i \} = \{ 1, \ldots, 11,13,14,16,17,39,50,51,54\}$, we define
\begin{align}
	\label{eq:HLbLThatTensorStructures}
	\hat T_{g_i}^{\mu\nu\lambda\sigma;\rho}(q_1, q_2) := \left( \frac{\p}{\p q_{4\rho}} \hat T_{g_i}^{\mu\nu\lambda\sigma}(q_1,q_2,q_4-q_1-q_2) \right)  \bigg|_{q_4=0} \, .
\end{align}

The 13-fold redundancy in the set of HLbL tensor structures $\{T_i^{\mu\nu\lambda\sigma}\}$ implies ambiguities in the scalar functions $\Pi_i$ in general kinematics and results in kinematic singularities in the basis elements $\tilde\Pi_i$. In contrast, in the limit $q_4\to0$ the 19 scalar functions $\hat\Pi_i$ that contribute to $(g-2)_\mu$ are free from ambiguities and kinematic singularities. This follows from the BTT construction and the fact that the $19\times19$ matrix
\begin{align}
	A_{ij}(q_1^2,q_2^2,q_3^2) := \hat T_{g_i}^{\mu\nu\lambda\sigma;\rho}(q_1,q_2) \hat T^{g_j}_{\mu\nu\lambda\sigma;\rho}(q_1,q_2)
\end{align}
is invertible and allows one to obtain a set of 19 projectors $\mathcal{P}_i^{\mu\nu\lambda\sigma;\rho}$ that fulfill
\begin{align}
	\hat\Pi_i(q_1, q_2, -q_1-q_2) = \mathcal{P}_i^{\mu\nu\lambda\sigma;\rho} \left( \frac{\p}{\p q_4^\rho} \Pi_{\mu\nu\lambda\sigma} \right) \bigg|_{q_4 = 0} \, .
\end{align}
Due to gauge invariance, the projectors are not unique. A possible choice has been given in Ref.~\cite{Bijnens:2020xnl}.

The set of 19 scalar functions that contribute to $(g-2)_\mu$ is defined by the six representatives
\begin{align}
	\label{eq:PiHatFunctions}
	\hat\Pi_1 &= \Pi_1 + q_1 \cdot q_2 \Pi_{47} \, , \nn
	\hat\Pi_4 &= \Pi_4 - q_1 \cdot q_3 \left( \Pi_{19} - \Pi_{42} \right) - q_2 \cdot q_3 \left( \Pi_{20} - \Pi_{43} \right) + q_1 \cdot q_3  q_2 \cdot q_3 \Pi_{31} \, , \nn
	\hat \Pi_7 &= \Pi_7 - \Pi_{19} + q_2 \cdot q_3 \Pi_{31} \, , \nn
	\hat \Pi_{17} &= \Pi_{17} + \Pi_{42} + \Pi_{43} - \Pi_{47} \, , \nn
	\hat \Pi_{39} &= \Pi_{39} + \Pi_{40} + \Pi_{46} \, , \nn
	\hat \Pi_{54} &= \Pi_{42} - \Pi_{43} + \Pi_{54} \, ,
\end{align}
together with the crossed versions
\begin{align}
	\label{eq:CrossingRelationsPiHat}
	\hat \Pi_2 &= \Cr{23}{\hat \Pi_1} \, , \quad \hat \Pi_3 = \Cr{13}{\hat \Pi_1} \, , \quad 
	\hat \Pi_5 = \Cr{23}{\hat \Pi_4} \, , \quad \hat \Pi_6 = \Cr{13}{\hat \Pi_4} \, , \nn
	\hat \Pi_8 &= \Cr{12}{\hat \Pi_7} \, , \quad \hat \Pi_9 = \Cr{12}{\Cr{13}{\hat \Pi_7}} \, , \quad
	\hat \Pi_{10} = \Cr{23}{\hat \Pi_7} \, , \quad \hat \Pi_{13} = \Cr{13}{\hat \Pi_7} \, , \quad
	\hat \Pi_{14} = \Cr{12}{\Cr{23}{\hat \Pi_7}} \, , \nn
	\hat \Pi_{11} &= \Cr{13}{\hat \Pi_{17}} \, , \quad \hat \Pi_{16} = \Cr{23}{\hat \Pi_{17}} \, , \quad
	\hat \Pi_{50} = -\Cr{23}{\hat \Pi_{54}} \, , \quad \hat \Pi_{51} = \Cr{13}{\hat \Pi_{54}} \, ,
\end{align}
where the crossing operators $\Cr{ij}{}$ exchange the photons $i$ and $j$~\cite{Colangelo:2017fiz}. Crossing symmetry in addition implies the intrinsic symmetries~\cite{Colangelo:2017fiz}
\begin{align}
	\label{eq:InternalCrossingSymmetriesPiHat}
	\hat\Pi_1 &= \Cr{12}{\hat\Pi_1} , \quad \hat\Pi_4 = \Cr{12}{\hat\Pi_4} , \quad \hat\Pi_{17} = \Cr{12}{\hat\Pi_{17}} \, , \nn
	\hat \Pi_{39} &= \Cr{12}{\hat \Pi_{39}} = \Cr{13}{\hat \Pi_{39}} = \ldots , \quad \hat \Pi_{54} = - \Cr{12}{\hat \Pi_{54}} \, .
\end{align}

After applying a Wick rotation, using Gegenbauer polynomial techniques~\cite{Rosner:1967zz,Knecht:2001qf} to perform
five of the eight integrals, and employing the crossing symmetries under $q_1 \leftrightarrow -q_2$, one arrives at the master formula for the HLbL contribution to
$(g-2)_\mu$ containing a sum of only 12 terms~\cite{Colangelo:2015ama,Colangelo:2017fiz}:
\begin{align}
	\label{eq:MasterFormula3Dim}
	a_\mu^\mathrm{HLbL} &= \frac{2 \alpha^3}{3 \pi^2} \int_0^\infty dQ_1 \int_0^\infty dQ_2 \int_{-1}^1 d\tau \sqrt{1-\tau^2} Q_1^3 Q_2^3 \sum_{i=1}^{12} T_i(Q_1,Q_2,\tau) \bar \Pi_i(Q_1,Q_2,\tau) \, ,
\end{align}
where $Q_1 := |Q_1|$ and $Q_2 := |Q_2|$ denote the norm of the Euclidean four-vectors. The 12 scalar functions $\bar \Pi_i$ are a subset of the functions $\hat \Pi_i$ and need to be evaluated for the reduced $(g-2)_\mu$ kinematics
\begin{align}
	\label{eq:Gm2Kinematics}
	s &= q_3^2 = - Q_3^2 = - Q_1^2 - 2 Q_1 Q_2 \tau - Q_2^2 \, , \quad
	t = q_2^2 = -Q_2^2 \, , \quad
	u = q_1^2 = -Q_1^2 \, , \quad
	q_4^2 = 0 \, .
\end{align}


\section{Dispersion relations in triangle kinematics}
\label{sec:DRs}

\subsection{Summary of the existing approach}
\label{sec:ExistingApproach}
	
The framework worked out in Refs.~\cite{Colangelo:2015ama,Colangelo:2017fiz} consists of dispersion relations for the HLbL tensor in general four-point kinematics, which can be derived from the Mandelstam double-spectral representation. The photon virtualities are treated as fixed external variables, while dispersion relations are written in terms of the Mandelstam variables. In particular, in Ref.~\cite{Colangelo:2017fiz} a basis of scalar functions $\check\Pi_i$ was derived that is suitable for dispersion relations in the singly-on-shell limit $q_4^2 = 0$. For $t=q_2^2$, the scalar functions are free from kinematic singularities in the Mandelstam variables $s$ and $u$, enabling fixed-$t$ dispersion relations. The representation is also manifestly free from contributions of unphysical helicity amplitudes. After writing the dispersion relation, the limit $q_4\to0$ is taken to arrive at the kinematics relevant for $(g-2)_\mu$.

One of the major difficulties in this approach is the fact that the BTT tensor decomposition does not directly provide a tensor basis free from kinematic singularities, but rather a redundant set of structures. Although the singly-on-shell basis functions $\check\Pi_i$ derived in Ref.~\cite{Colangelo:2017fiz} are free from singularities in the Mandelstam variables, the redundancies in the tensor basis result in spurious kinematic singularities in the photon virtualities. The residues of these apparent singularities vanish due to a set of sum rules: these follow directly from the fact that the tensor decomposition involves structures of different mass dimension and they guarantee that the result of the dispersion relation for the entire HLbL tensor is independent of the choice of tensor basis. At the same time, they imply that the apparent kinematic singularities drop out for contributions that satisfy the sum rules. This is guaranteed to happen only for the entire HLbL contribution, i.e., the sum over all intermediate states in the unitarity relation. In contrast, individual intermediate states do not necessarily satisfy the sum rules. These sum-rule violations make the contributions of individual states depend on the chosen basis~\cite{Aoyama:2020ynm,Danilkin:2021icn,Colangelo:2021nkr} and suffer from kinematic singularities~\cite{Colangelo:2021nkr}.

The basis dependence affects all single-particle intermediate states in the unitarity relation apart from the pseudoscalar contributions, as these do not contribute to the sum rules. The sum rules are exactly fulfilled by the pion box~\cite{Colangelo:2017fiz}. Scalar intermediate states or two-particle $S$-wave contributions are in general basis dependent, but they are not affected by spurious singularities, see Ref.~\cite{Danilkin:2021icn}. In the basis of Ref.~\cite{Colangelo:2017fiz}, axial-vector contributions are affected by singularities, but there exists an alternative basis, where this problem is absent, as discussed in Ref.~\cite{Colangelo:2021nkr}. By making use just of the minimal set of sum rules that are necessary to render the entire HLbL contribution basis independent, it is impossible to fully remove the spurious kinematic singularities from the contribution of tensor-meson resonances or two-particle $D$- and higher partial waves. Whether this can be achieved by making use of additional sum rules remains to be studied.

As long as no representation is available that is manifestly free of any kinematic singularities, the spurious singularities need to be subtracted as described in Ref.~\cite{Colangelo:2017fiz}. The same subtraction scheme needs to be applied in all contributions that are affected by the singularities. In the sum over all intermediate states, the subtraction again vanishes due to the sum rules. The subtraction scheme introduces an ambiguity in the contribution of individual intermediate states that is in addition to the general basis dependence. Due to these ambiguities, one cannot expect to obtain a meaningful result for these contributions unless the sum of included states fulfills the sum rules. This is one of the reasons why to date no evaluation of the tensor-resonance contributions within the dispersive framework is available~\cite{Aoyama:2020ynm}.

\subsection{Dispersing in the photon virtualities}

Instead of fixing the photon virtualities, writing dispersion relations in the Mandelstam variables, and finally taking the limit $q_4\to0$, here we propose to take a different approach: we first take the limit $q_4\to0$ and then write dispersion relations for the functions $\hat\Pi_i(q_1^2,q_2^2,q_3^2)$ entering the master formula, exploiting the analytic structure in the variables $q_i^2$. This alternative set of dispersion relations has been briefly discussed in Ref.~\cite{Colangelo:2019uex}. As explained there, these new dispersion relations have the disadvantage that the original cuts in the Mandelstam variables and in the photon virtualities are no longer separated. 

However, this approach has an important advantage over the dispersion relations for the four-point function: all the redundancies of the BTT set disappear in the $(g-2)_\mu$ limit. The functions $\hat\Pi_i(q_1^2,q_2^2,q_3^2)$ in the $(g-2)_\mu$ kinematic limit are free from any kinematic singularities. Working directly with them removes the problem of spurious divergences. Hence, this alternative is a promising approach for the model-independent evaluation of the contributions of $D$- and higher partial waves, or narrow tensor-meson resonances such as the $f_2(1270)$~\cite{Hoferichter:2019nlq}.

A potential pitfall is the fact that for the new dispersion relations in triangle kinematics, we need to reconstruct additional hadronic sub-processes, in particular $\gamma^*\gamma^*\gamma\to2\pi$ as well as $\pi\pi\to\pi\pi\gamma$. These sub-processes require their own tensor decomposition, which could potentially re-introduce the problem of redundancies and kinematic singularities. In the following, we derive the BTT tensor decompositions for these sub-processes and we show that in the limit of $(g-2)_\mu$ kinematics, all but a single redundancy in $\gamma^*\gamma^*\gamma\to2\pi$ disappear, which under the assumption of a uniform asymptotic behavior of the tensor amplitude can be traded for one kinematic constraint. This enables dispersion relations for scalar functions free of kinematic singularities.

In the case of dispersion relations in triangle kinematics, the objects under consideration are the functions $\hat\Pi_i$ in Eq.~\eqref{eq:HLbLTensorPiHatDecomposition}, which in $(g-2)_\mu$ kinematics depend on the three photon virtualities, $\hat\Pi_i(q_1^2,q_2^2,q_3^2)$. For the contribution to $(g-2)_\mu$, only a restricted domain of the three virtualities belongs to the physical region, as determined by the master formula~\eqref{eq:MasterFormula3Dim}, but analytic continuation allows us to treat the three virtualities as independent variables and to continue the function beyond the physical region. We start by writing a dispersion relation for $\hat\Pi_i(q_1^2,q_2^2,q_3^2)$ in $q_3^2$, while keeping the other two virtualities $q_{1,2}^2$ fixed:
\begin{align}
	\label{eq:PiHatDispersionRelation}
	\hat\Pi_i(q_1^2,q_2^2,q_3^2) = \frac{1}{\pi} \int_{s_0}^\infty ds' \frac{1}{s' - q_3^2 - i \epsilon} \Im\, \hat\Pi_i(q_1^2,q_2^2,s') \, ,
\end{align}
where the imaginary part is obtained from
\begin{align}
	\label{eq:ImPiHat}
	\Im\, \hat\Pi_i(q_1^2,q_2^2,s') = \frac{\hat\Pi_i(q_1^2,q_2^2,s'+i\epsilon) - \hat\Pi_i(q_1^2,q_2^2,s'-i\epsilon)}{2i}
\end{align}
and the lowest threshold is $s_0 = M_{\pi^0}^2$. We now demonstrate how to obtain this imaginary part by taking the appropriate limits of imaginary parts in four-point kinematics.

In Ref.~\cite{Colangelo:2017fiz}, a basis of 27 scalar functions $\check \Pi_i$ for HLbL scattering was derived, which applies to four-point kinematics at fixed $t=q_2^2$ and in the limit $q_4^2 = 0$. The 19 functions $\hat\Pi_i$ relevant for $(g-2)_\mu$ can be obtained from a subset of the $\check\Pi_i$ functions,
\begin{align}
	\check\Pi_i = \hat\Pi_{g_i} + (s-q_3^2) \bar\Delta_i + (s-q_3^2)^2 \bar{\bar\Delta}_i \, ,
\end{align}
i.e., by denoting the arguments as $\check\Pi_i(s;q_1^2,q_2^2,q_3^2)$, the limit of $(g-2)_\mu$ kinematics is given by
\begin{align}
	\hat\Pi_{g_i}(q_1^2,q_2^2,q_3^2) = \check\Pi_i( q_3^2; q_1^2,q_2^2,q_3^2) \, .
\end{align}
The imaginary part~\eqref{eq:ImPiHat} can be written as
\begin{align}
	\label{eq:ImPiHat2Disc}
	\Im\, \hat\Pi_{g_i}(q_1^2,q_2^2,s') &= \lim_{q_3^2\to s'} \frac{\check\Pi_i(s'+i\epsilon; q_1^2,q_2^2,q_3^2+i\epsilon) - \check\Pi_i(s'-i\epsilon; q_1^2,q_2^2,q_3^2-i\epsilon)}{2i} \nn
		&= \lim_{q_3^2\to s'} \begin{aligned}[t]
			&\bigg[ \frac{\check\Pi_i(s'+i\epsilon; q_1^2,q_2^2,q_3^2+i\epsilon) - \check\Pi_i(s'-i\epsilon; q_1^2,q_2^2,q_3^2+i\epsilon)}{2i} \\
			&+ \frac{\check\Pi_i(s'-i\epsilon; q_1^2,q_2^2,q_3^2+i\epsilon) - \check\Pi_i(s'-i\epsilon; q_1^2,q_2^2,q_3^2-i\epsilon)}{2i} \bigg] \end{aligned} \nn
		&= \lim_{q_3^2\to s'} \begin{aligned}[t]
			&\bigg[ \frac{\check\Pi_i(s'+i\epsilon; q_1^2,q_2^2,q_3^2+i\epsilon) - \check\Pi_i(s'-i\epsilon; q_1^2,q_2^2,q_3^2+i\epsilon)}{2i} \\
			&+ \left( \frac{\check\Pi_i(s'+i\epsilon; q_1^2,q_2^2,q_3^2+i\epsilon) - \check\Pi_i(s'+i\epsilon; q_1^2,q_2^2,q_3^2-i\epsilon)}{2i} \right)^* \bigg] \end{aligned} \nn
		&=: \lim_{q_3^2\to s'} \left[ \Im_s \check\Pi_i(s'; q_1^2,q_2^2,q_3^2+i\epsilon) + \left( \Im_3 \check\Pi_i(s'+i\epsilon; q_1^2,q_2^2,q_3^2) \right)^* \right]  \, .
\end{align}
Here, we denote by $\Im_s$ the $s$-channel discontinuity in four-point kinematics, analytically continued in the third photon virtuality to $q_3^2+i\epsilon$, while $\Im_3$ denotes the discontinuity in the variable $q_3^2$, again in four-point kinematics and now analytically continued in the Mandelstam variable to $s'+i\epsilon$. Due to the analytic continuation, these discontinuities $\Im_s$ and $\Im_3$ in general are complex quantities.

Before taking the limit $q_3^2\to s'$, the expression contains kinematic singularities of the form $1/(q_1^2+q_3^2)$ and $1/\lambda(q_1^2,q_2^2,q_3^2)$, which are present in the quantities $\bar\Delta_i$ and $\bar{\bar\Delta}_i$~\cite{Colangelo:2017fiz}, with $\lambda$ denoting the K\"all\'en triangle function. In the original $s$-channel dispersion relations~\cite{Colangelo:2017fiz}, the residue of these kinematic singularities vanishes due to the presence of sum rules for the scalar functions $\check\Pi_i$. Since individual partial waves or single narrow resonances violate the sum rules, the residues of the kinematic singularities had to be subtracted ``by hand.'' The basis $\check\Pi_i$ was then chosen in a way that leads to a simple form of kinematic singularities and at the same time optimizes the convergence of the partial-wave-expanded pion box. The alternative basis of $\check\Pi_i$ functions discussed in Ref.~\cite{Colangelo:2021nkr} contains singularities that are products of $1/q_i^2$ and $1/(q_1^2 + q_3^2)$.

We note that the limit of each of the two discontinuities~\eqref{eq:ImPiHat2Disc} may be singular: if the soft photon is emitted from an external leg of the sub-process, the limit $q_4\to0$ puts an internal propagator on shell. With the additional derivative in Eq.~\eqref{eq:BasicMasterFormula}, this potentially leads to a double-pole in $1/(s'-q_3^2)$. However, analogous poles exist in both discontinuities $\Im_s\check\Pi_i$ and $\Im_3\check\Pi_i$ and they are guaranteed to cancel in the sum of the two discontinuities because the HLbL tensor is free of such singularities. The two leading terms in the expansion around $q_4 = 0$ are related to the non-radiative process by Low's theorem~\cite{Low:1958sn} and the same is true for the poles in higher-orders in the expansion using dispersion relations as will be demonstrated in a future publication~\cite{TriangleDR4PiGamma}. Due to this absence of poles in the sum of the two discontinuities, the contributions from $\bar\Delta_i$ and $\bar{\bar\Delta}_i$ vanish in the limit $q_3^2\to s'$, so that $\Im \hat\Pi_{g_i}$ does not contain any kinematic singularities. For this reason, the contribution of a single partial wave or a narrow resonance to $(g-2)_\mu$ can be defined without relying on a sum rule that is violated by this particular contribution alone. The cancellation of soft singularities is illustrated for a simplified situation with single poles in scalar toy examples in App.~\ref{app:ScalarToy}. The cancellation in the realistic case of HLbL and its sub-processes will be presented in Ref.~\cite{TriangleDR4PiGamma}.

Writing a dispersion relation in $q_3^2$ and fixing $q_{1,2}^2$ in Eq.~\eqref{eq:PiHatDispersionRelation} is an arbitrary choice: crossing symmetry requires that dispersion relations in any of the other virtualities lead to the same result. In the final dispersive representation this symmetrization needs to be taken into account, in a way that avoids any double counting. In the present article, we will show how this is achieved for single-particle intermediate states: we add the crossed versions of the contributions that are generated from the first term in Eq.~\eqref{eq:ImPiHat2Disc}, corresponding to $t$- and $u$-channel discontinuities. These contributions have discontinuities in $q_3^2$, which accordingly must be excluded from the dispersion relation in $q_3^2$, in order to avoid a double counting. The explicit symmetrization for the complete vector-meson and two-pion contribution is more involved and is left for future work. The resulting representation will fulfill by construction all constraints from crossing symmetry. Furthermore, it will satisfy single-variable dispersion relations in any of the three virtualities and include the leading intermediate states in the unitarity relations of all channels.


\section{Unitarity relations}
\label{sec:URs}

\begin{figure}[t]
	\centering
	\begin{align*}
	\includegraphics[height=2cm,valign=c]{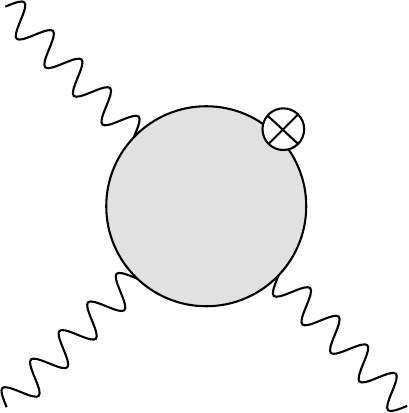}
	\quad = \quad 
	\includegraphics[height=2cm,valign=c]{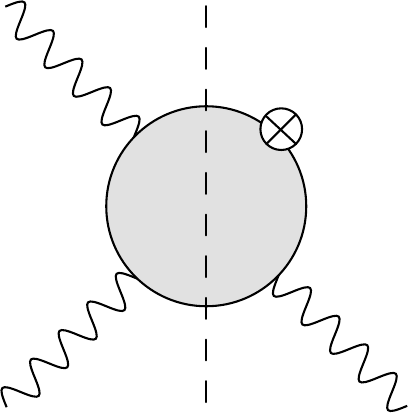}
	\quad + \quad
	\includegraphics[height=2cm,valign=c]{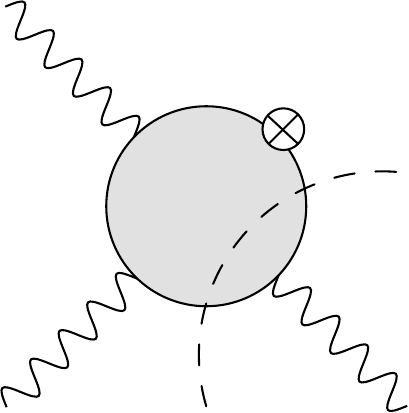}
	\end{align*}
	\caption{Unitarity cuts contributing to the discontinuity of HLbL with respect to $q_3^2$ in triangle kinematics. The static external electromagnetic field is denoted by a crossed circle.}
	\label{img:HLbLTwoCuts}
	\begin{align*}
	\begin{alignedat}{2}
	\includegraphics[height=1.5cm,valign=c]{images/HLbL-sCut}
	\quad &= \quad 
	\includegraphics[height=1.5cm,valign=c]{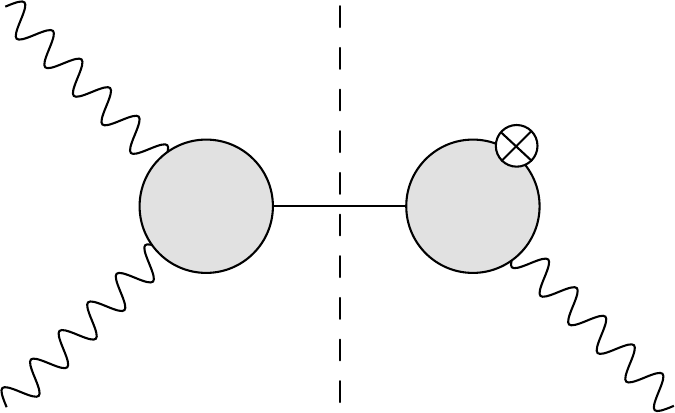}
	\quad + \quad
	&&\includegraphics[height=1.5cm,valign=c]{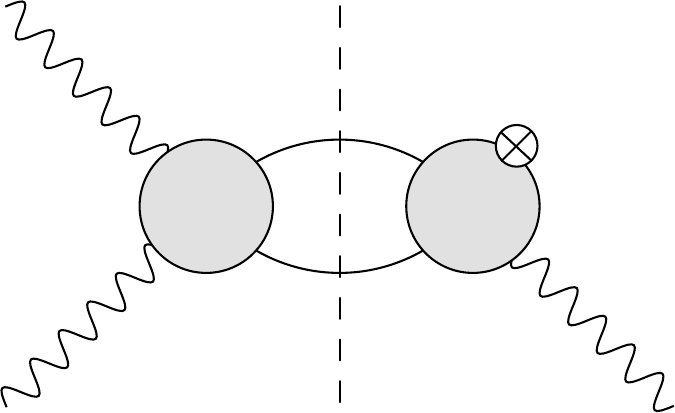}
	\quad + \quad
	\includegraphics[height=1.5cm,valign=c]{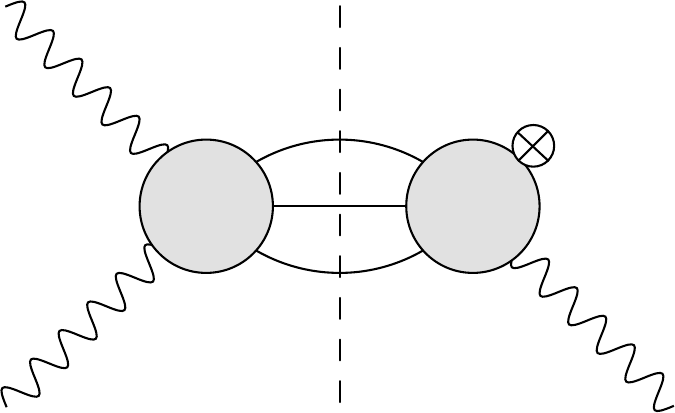}
	\quad + \quad \ldots \nn
	\nn
	\includegraphics[height=1.5cm,valign=c]{images/HLbL-q32Cut}
	\quad &= \quad 
	&&\includegraphics[height=1.5cm,valign=c]{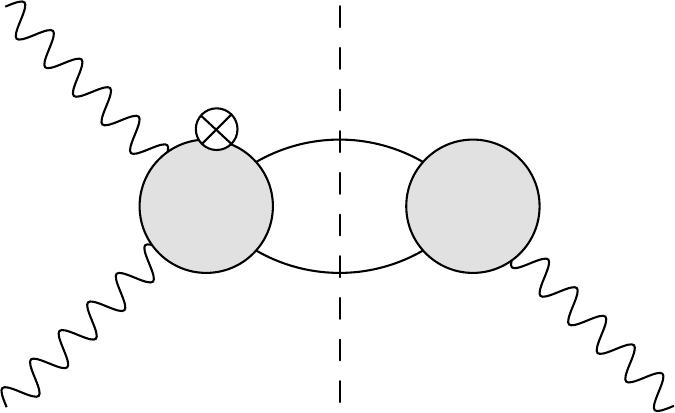}
	\quad + \quad
	\includegraphics[height=1.5cm,valign=c]{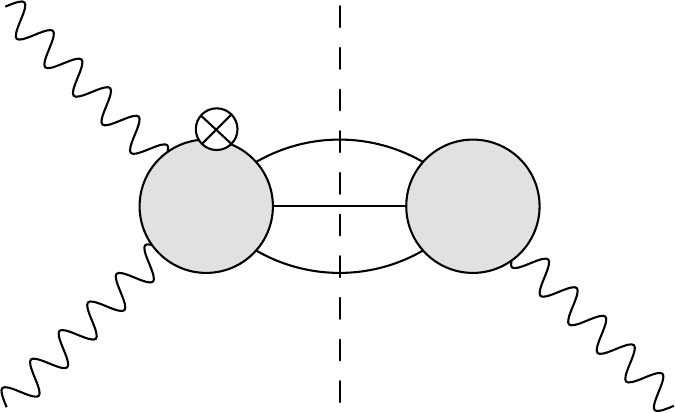}
	\quad + \quad \ldots
	\end{alignedat}
	\end{align*}
	\caption{The contribution of different intermediate states to the $s$-channel and $q_3^2$-channel discontinuities.}
	\label{img:HLbLIntermediateStates}
\end{figure}

According to Eq.~\eqref{eq:ImPiHat2Disc}, the relevant imaginary part that is needed in the new dispersion relations can be obtained from the sum of the discontinuities in the $s$-channel and the $q_3^2$-channel in four-point kinematics, illustrated in Fig.~\ref{img:HLbLTwoCuts}. Unitarity of the $S$-matrix provides these discontinuities in the form of two different relations. The first one is the $s$-channel unitarity relation that has already been employed in the established dispersive approach:
\begin{align}
	\label{eq:SChannelUnitarity}
	\Im_s &\left( e^4 (2\pi)^4 \delta^{(4)}(q_1 + q_2 + q_3 - q_4) H_{\lambda_1\lambda_2,\lambda_3\lambda_4} \right) \nn
		&= \sum_n \frac{1}{2S_n} \left( \prod_{i=1}^n \int \widetilde{dp_i} \right) \< n; \{p_i\} | \gamma^*(-q_3,\lambda_3) \gamma(q_4,\lambda_4) \>^* \< n; \{p_i\} | \gamma^*(q_1,\lambda_1) \gamma^*(q_2,\lambda_2) \> \, ,
\end{align}
where $S_n$ denotes the symmetry factor for the intermediate state $|n\>$. The Lorentz-invariant measure is abbreviated by $\widetilde{dp} := \frac{d^3p}{(2\pi)^3 2p^0}$.
Similarly, the discontinuity in the virtuality $q_3^2$ can be obtained from the unitarity relation, where the fourth photon is crossed to the initial state:
\begin{align}
	\label{eq:Q3ChannelUnitarity}
	\Im_3 &\left( e^4 (2\pi)^4 \delta^{(4)}(q_1 + q_2 + q_3 - q_4) H_{\lambda_1\lambda_2\lambda_4,\lambda_3} \right) \nn
		&= \sum_n \frac{1}{2S_n} \left( \prod_{i=1}^n \int \widetilde{dp_i} \right) \< n; \{p_i\} | \gamma^*(-q_3,\lambda_3) \>^* \< n; \{p_i\} | \gamma^*(q_1,\lambda_1) \gamma^*(q_2,\lambda_2) \gamma(-q_4,\lambda_4) \> \, .
\end{align}

The general strategy of the dispersive evaluation of the HLbL contribution to $(g-2)_\mu$ amounts to summing up individual contributions to the unitarity relations~\eqref{eq:SChannelUnitarity} and~\eqref{eq:Q3ChannelUnitarity}. Of course, in practice it is not possible to resum the whole tower of intermediate states and one needs to truncate the sum. The remainder is assumed to be small at low energies, where the lightest intermediate states dominate, but it becomes more important at higher energies and in the end it needs to be taken into account by a proper matching to asymptotic constraints~\cite{Melnikov:2003xd,Bijnens:2019ghy,Colangelo:2019uex,Colangelo:2019lpu,Leutgeb:2019gbz,Cappiello:2019hwh,Ludtke:2020moa,Aoyama:2020ynm,Bijnens:2020xnl,Bijnens:2021jqo,Colangelo:2021nkr,Bijnens:2022itw}.

The contributions of the lightest intermediate states to the unitarity relations in the $s$- and $q_3^2$-channels are illustrated in terms of unitarity diagrams in Fig.~\ref{img:HLbLIntermediateStates}. In order to evaluate the discontinuities, input for the sub-processes is required. In the case of the $s$-channel discontinuities, the input is identical to the one in the familiar dispersion relations, although evaluated for a different kinematic configuration: the evaluation of the one-particle intermediate state requires the pion (and $\eta$, $\eta'$) transition form factor as input. For two-pion intermediate states, the helicity partial waves for $\gamma^*\gamma^*\to2\pi$ are the required input. The formalism for the full inclusion of three-particle intermediate states in the $s$-channel is currently not available. This contribution contains axial-vector resonances, which are expected to be numerically relevant~\cite{Aoyama:2020ynm,Melnikov:2003xd,Jegerlehner:2017gek,Leutgeb:2019gbz,Cappiello:2019hwh,Masjuan:2020jsf}. In a first step, these effects can be described in a narrow-width approximation (NWA), replacing the three-pion intermediate state by a narrow resonance. The required input in this approximation are the axial-vector transition form factors, $\gamma^*\gamma^*\to A$. The effect of two-pion intermediate states gets enhanced close to scalar or tensor resonances. In this case, the two-pion unitarization can be compared to a NWA, which is used to include scalar and tensor resonances in different isospin channels~\cite{Danilkin:2021icn}. Again, the respective transition form factors are required as input.

The input for the discontinuities in the $q_3^2$-channel are given by the pion vector form factor (VFF) and $\gamma^*\gamma^*\gamma\to2\pi$ in the case of two-pion intermediate states. In the case of three-pion intermediate states, the sub-processes are $\gamma^*\to3\pi$ and $\gamma^*\gamma^*\gamma\to3\pi$, with potentially non-negligible effects due to the narrow vector resonances $\omega$ and $\phi$.\footnote{In the case of the pion pole in the dispersion relations in four-point kinematics, the three-pion cut is included in the dispersive treatment of the pion transition form factor~\cite{Hoferichter:2014vra,Hoferichter:2018kwz,Hoferichter:2018dmo}.} Therefore, compared to the established dispersion relations in four-point kinematics, the dispersion relations in triangle kinematics require the processes $\gamma^*\gamma^*\gamma\to2\pi$ and $\gamma^*\gamma^*\gamma\to V$ as new inputs, where $V$ denotes a vector resonance.

\begin{figure}[t]
	\centering
	\begin{align*}
	\includegraphics[height=1.5cm,valign=c]{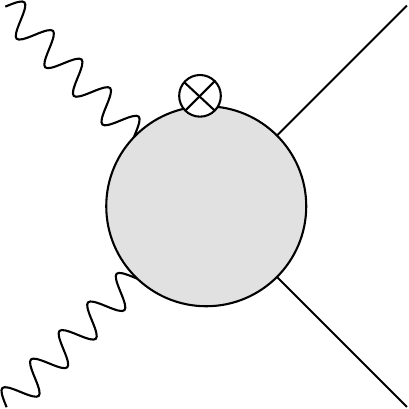}
	\quad = \quad 
	\includegraphics[height=1.5cm,valign=c]{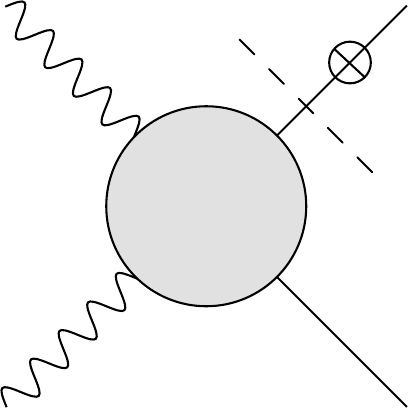}
	\quad + \quad
	\includegraphics[height=1.5cm,valign=c]{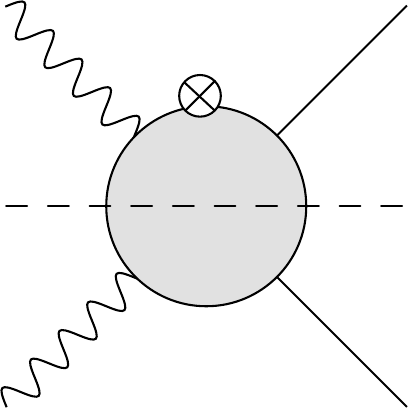}
	\quad + \quad
	\includegraphics[height=1.5cm,valign=c]{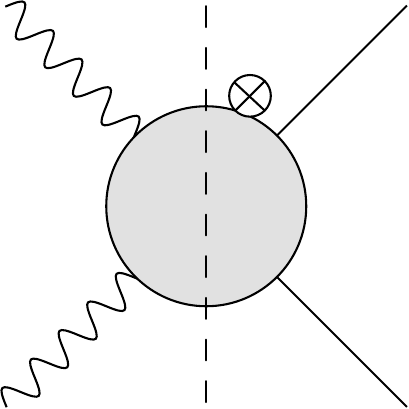}
	\quad + \quad
	\includegraphics[height=1.5cm,valign=c]{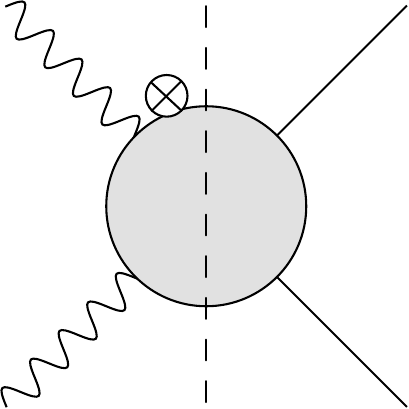}
	\end{align*}
	\caption{Unitarity cuts for $\gamma^*\gamma^*\gamma\to2\pi$ for a soft external on-shell photon. The first cut on the RHS denotes the soft divergence, the second diagram denotes the left-hand cut. The last two diagrams are the two $s$-channel cuts. Crossed diagrams are not shown.}
	\label{img:gggpipi}
	
	\begin{align*}
	&\begin{alignedat}{1}
	\includegraphics[height=1.5cm,valign=c]{images/gggpipi-LHC}
	\quad &= \quad
	\includegraphics[width=1.5cm,valign=c]{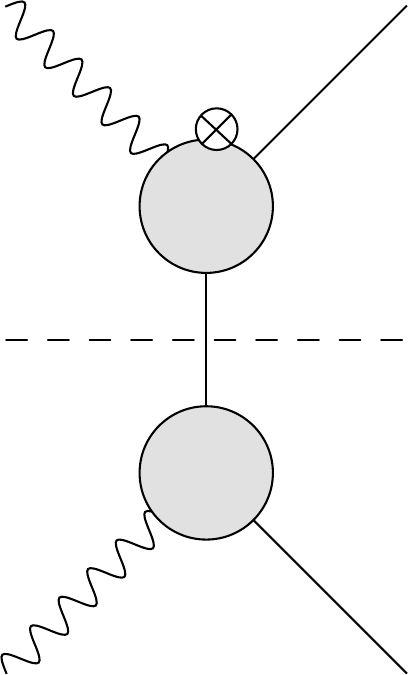}
	\quad + \quad
	\includegraphics[width=1.5cm,valign=c]{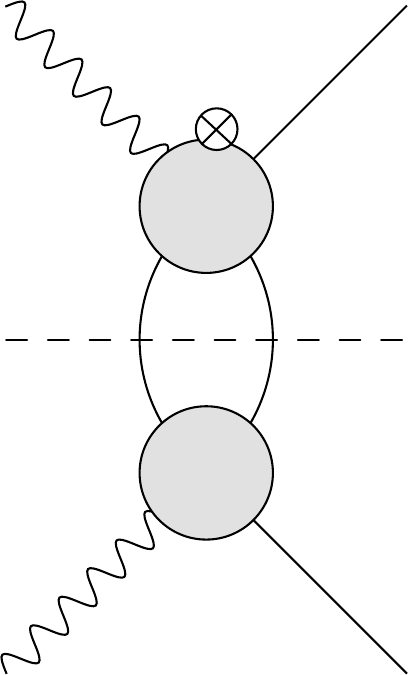}
	\quad + \quad
	\includegraphics[width=1.5cm,valign=c]{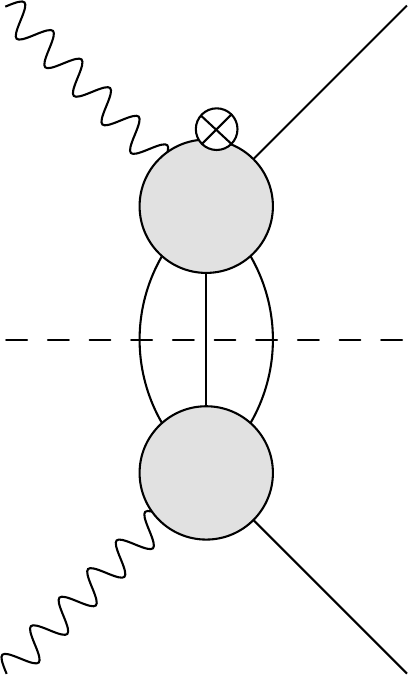}
	\quad + \quad \ldots
	\end{alignedat}
	\nn
	\nn
	&\begin{alignedat}{2}
	\includegraphics[height=1.5cm,valign=c]{images/gggpipi-Cut1}
	\quad &= \quad
	\includegraphics[height=1.5cm,valign=c]{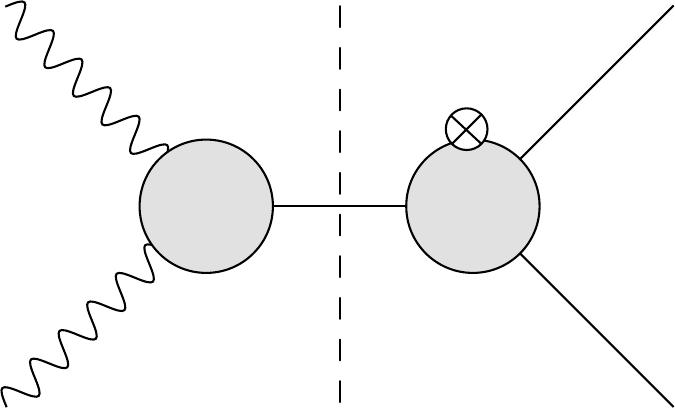}
	\quad + \quad
	&&\includegraphics[height=1.5cm,valign=c]{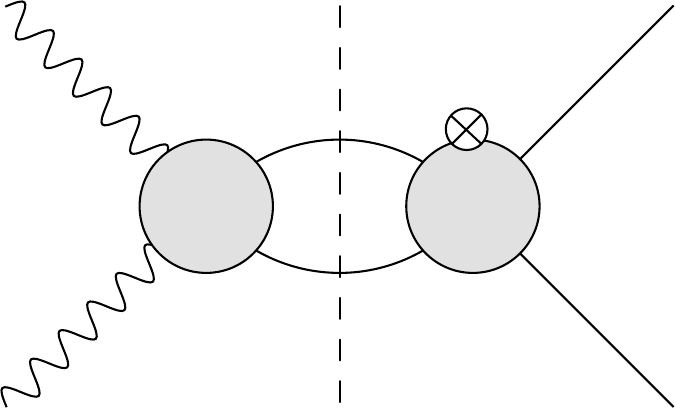}
	\quad + \quad
	\includegraphics[height=1.5cm,valign=c]{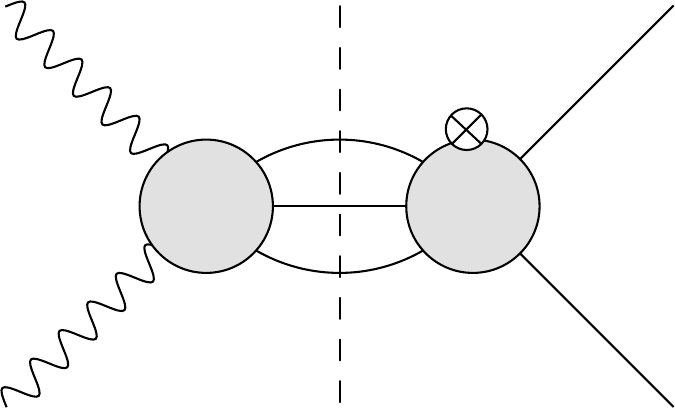}
	\quad + \quad \ldots \nn
	\nn
	\includegraphics[height=1.5cm,valign=c]{images/gggpipi-Cut2}
	\quad &= \quad
	&&\includegraphics[height=1.5cm,valign=c]{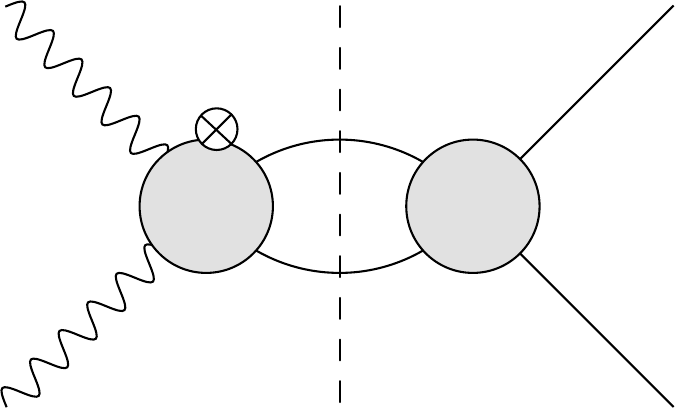}
	\quad + \quad \ldots
	\end{alignedat}
	\end{align*}
	\caption{The contribution of different intermediate states to the discontinuities in $\gamma^*\gamma^*\gamma\to2\pi$. The last diagram denotes $\pi\pi$ rescattering: the process $\gamma^*\gamma^*\gamma\to2\pi$ itself reappears as a sub-process.}
	\label{img:gggpipiIntermediateStates}
\end{figure}

The new inputs required for the dispersion relations in triangle kinematics should be reconstructed again dispersively. For $\gamma^*\gamma^*\gamma \to V$, much can be taken over directly from HLbL: in particular, this sub-process will be linked to the iso-scalar vector resonances in the pion transition form factor (TFF) reshuffled from the pion pole in the established dispersion relations~\cite{Hoferichter:2018kwz}. We discuss the tensor decomposition and kinematics in Sect.~\ref{sec:Q32ChannelVectorResonances}.

The second new input is the sub-process $\gamma^*\gamma^*\gamma\to2\pi$ needed for the two-pion intermediate state in the $q_3^2$-channel cut. The different unitarity cuts for $\gamma^*\gamma^*\gamma\to2\pi$ are shown in Fig.~\ref{img:gggpipi}. The soft limit is understood after taking the derivative with respect to $q_4$ in Eq.~\eqref{eq:BasicMasterFormula}. Only terms that are singular or finite in this limit are required. The singular terms can be expressed in terms of $\gamma^*\gamma^*\to2\pi$ via dispersion relations~\cite{TriangleDR4PiGamma}. The finite remainder is not directly determined by $\gamma^*\gamma^*\to2\pi$ and needs its own dispersion relation. The relevant intermediate states of the different unitarity cuts are illustrated in Fig.~\ref{img:gggpipiIntermediateStates}: if the photon virtualities are kept fixed, the original five-particle process reduces to four-point kinematics in the $(g-2)_\mu$ limit. The complexity again increases with the multiplicity of the intermediate states. The formalism for a fully dispersive reconstruction of the three-pion intermediate state is not available, but resonant contributions to the three-particle channel can be estimated in a NWA. Therefore, the main unknown sub-process is $\pi \pi \to \gamma \pi \pi$ for a soft photon.

\begin{table}[t]
	\centering
	\begin{tabular}{ c | c c c c c c }
	\toprule
	& \multicolumn{6}{c}{DR in four-point kinematics} \\
	\cmidrule{2-7}
	triangle-DR & $\pi^0, \eta, \eta'$ & $2\pi$ & $S$ & $A$ & $T$ & $\ldots$ \\
	\midrule
	\midrule
	$\pi^0, \eta, \eta'$
		& $\includegraphics[height=0.85cm,valign=c]{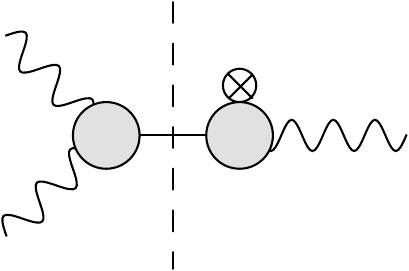}$
		& $\times$
		& $\times$
		& $\times$
		& $\times$
		& $\times$
		\\
	\midrule
		& $\times$
		& \includegraphics[height=0.85cm,valign=c]{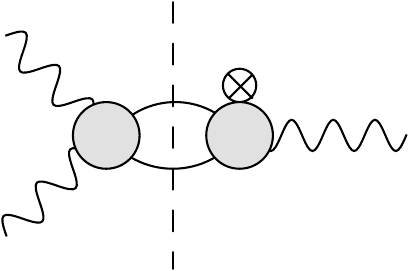}
		& $\times$
		& $\times$
		& $\times$
		& $\times$
		\\[0.5cm]
	$2\pi$
		& \includegraphics[height=0.85cm,valign=c]{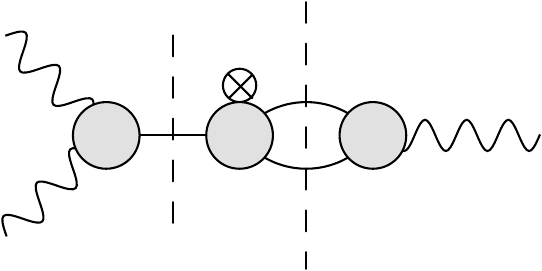}
		& \includegraphics[height=0.85cm,valign=c]{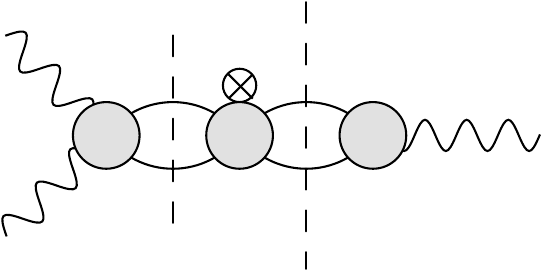}
		& \includegraphics[height=0.85cm,valign=c]{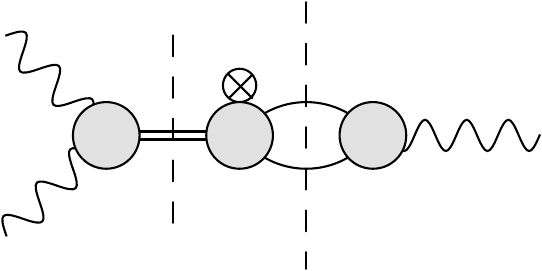}
		& \includegraphics[height=0.85cm,valign=c]{images/R-2pi}
		& \includegraphics[height=0.85cm,valign=c]{images/R-2pi}
		& \includegraphics[height=0.85cm,valign=c]{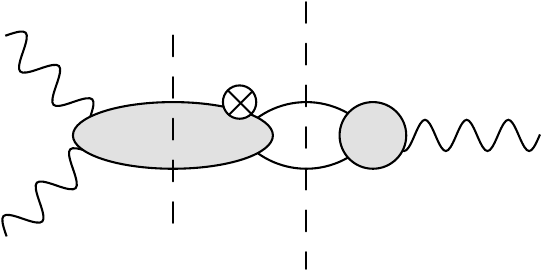}
		\\[0.5cm]
		&	\adjustbox{valign=c}{\begin{tikzpicture}[every node/.style={inner sep=0,outer sep=0}]
				\node{\includegraphics[height=0.85cm,valign=c]{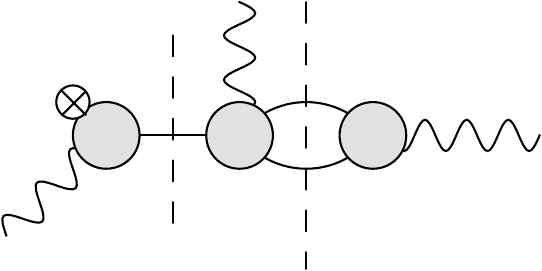}};
				\node[minimum height=1cm, minimum width=1.7cm,fill=white,opacity=0.7]{};
			\end{tikzpicture}}
		&	\adjustbox{valign=c}{\begin{tikzpicture}[every node/.style={inner sep=0,outer sep=0}]
				\node{\includegraphics[height=0.85cm,valign=c]{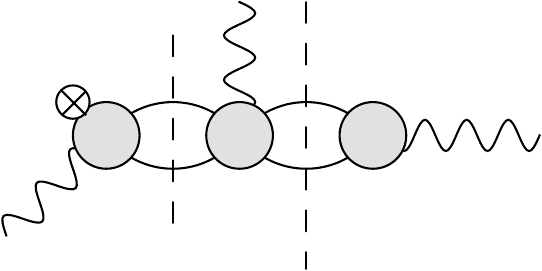}};
				\node[minimum height=1cm, minimum width=1.7cm,fill=white,opacity=0.7]{};
			\end{tikzpicture}}
		&	\adjustbox{valign=c}{\begin{tikzpicture}[every node/.style={inner sep=0,outer sep=0}]
				\node{\includegraphics[height=0.85cm,valign=c]{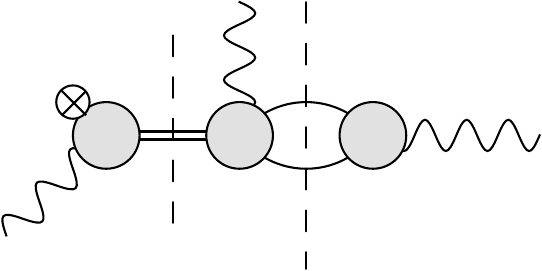}};
				\node[minimum height=1cm, minimum width=1.7cm,fill=white,opacity=0.7]{};
			\end{tikzpicture}}
		&	\adjustbox{valign=c}{\begin{tikzpicture}[every node/.style={inner sep=0,outer sep=0}]
				\node{\includegraphics[height=0.85cm,valign=c]{images/R-2pi-crossed}};
				\node[minimum height=1cm, minimum width=1.7cm,fill=white,opacity=0.7]{};
			\end{tikzpicture}}
		&	\adjustbox{valign=c}{\begin{tikzpicture}[every node/.style={inner sep=0,outer sep=0}]
				\node{\includegraphics[height=0.85cm,valign=c]{images/R-2pi-crossed}};
				\node[minimum height=1cm, minimum width=1.7cm,fill=white,opacity=0.7]{};
			\end{tikzpicture}}
		&	\adjustbox{valign=c}{\begin{tikzpicture}[every node/.style={inner sep=0,outer sep=0}]
				\node{\includegraphics[height=0.85cm,valign=c]{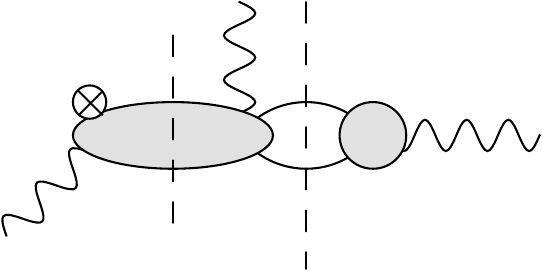}};
			\end{tikzpicture}}
		\\
	\midrule
	\multirow{2}{*}{$V$}
		&	\begin{tikzpicture}[every node/.style={inner sep=0,outer sep=0}]
				\node{\includegraphics[height=0.85cm,valign=c]{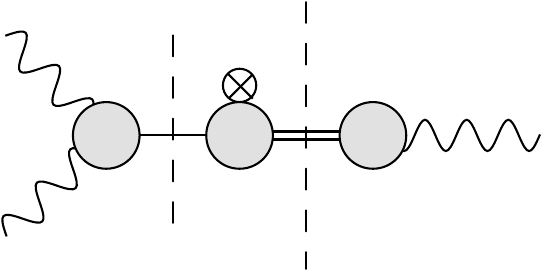}};
			\end{tikzpicture}
		&	\begin{tikzpicture}[every node/.style={inner sep=0,outer sep=0}]
				\node{\includegraphics[height=0.85cm,valign=c]{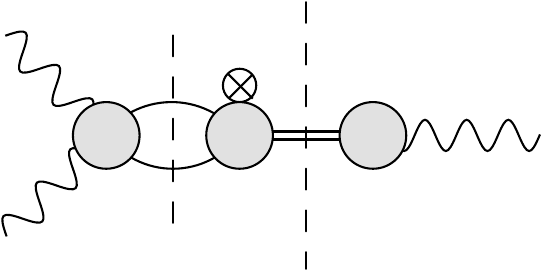}};
			\end{tikzpicture}
		&	\begin{tikzpicture}[every node/.style={inner sep=0,outer sep=0}]
				\node{\includegraphics[height=0.85cm,valign=c]{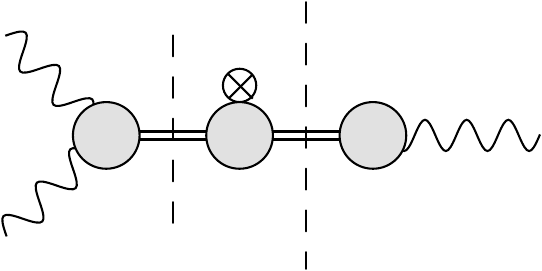}};
			\end{tikzpicture}
		&	\begin{tikzpicture}[every node/.style={inner sep=0,outer sep=0}]
				\node{\includegraphics[height=0.85cm,valign=c]{images/R-R}};
			\end{tikzpicture}
		&	\begin{tikzpicture}[every node/.style={inner sep=0,outer sep=0}]
				\node{\includegraphics[height=0.85cm,valign=c]{images/R-R}};
			\end{tikzpicture}
		&	\begin{tikzpicture}[every node/.style={inner sep=0,outer sep=0}]
				\node{\includegraphics[height=0.85cm,valign=c]{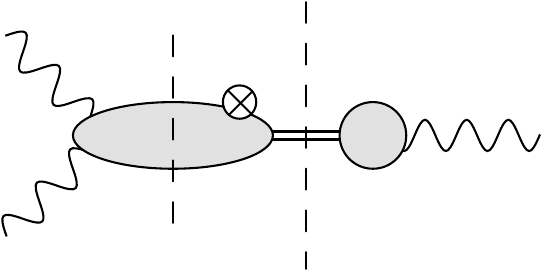}};
			\end{tikzpicture}
		\\
		&	\adjustbox{valign=c}{\begin{tikzpicture}[every node/.style={inner sep=0,outer sep=0}]
				\node{\includegraphics[height=0.85cm,valign=c]{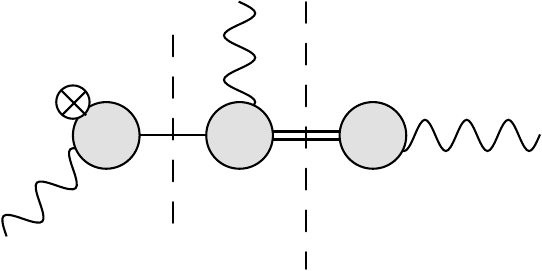}};
				\node[minimum height=1cm, minimum width=1.7cm,fill=white,opacity=0.7]{};
			\end{tikzpicture}}
		&	\adjustbox{valign=c}{\begin{tikzpicture}[every node/.style={inner sep=0,outer sep=0}]
				\node{\includegraphics[height=0.85cm,valign=c]{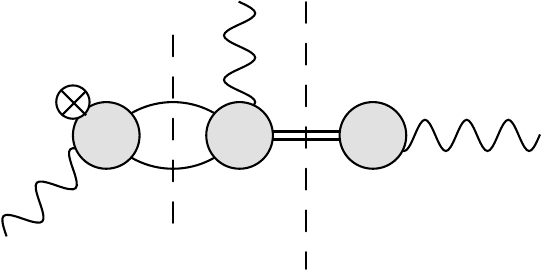}};
				\node[minimum height=1cm, minimum width=1.7cm,fill=white,opacity=0.7]{};
			\end{tikzpicture}}
		&	\adjustbox{valign=c}{\begin{tikzpicture}[every node/.style={inner sep=0,outer sep=0}]
				\node{\includegraphics[height=0.85cm,valign=c]{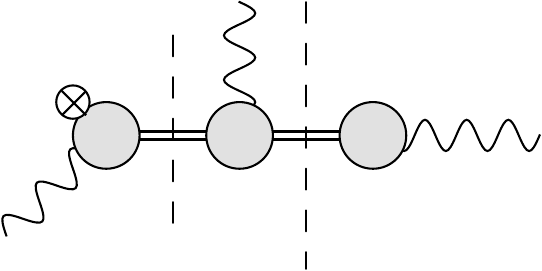}};
				\node[minimum height=1cm, minimum width=1.7cm,fill=white,opacity=0.7]{};
			\end{tikzpicture}}
		&	\adjustbox{valign=c}{\begin{tikzpicture}[every node/.style={inner sep=0,outer sep=0}]
				\node{\includegraphics[height=0.85cm,valign=c]{images/R-R-crossed}};
				\node[minimum height=1cm, minimum width=1.7cm,fill=white,opacity=0.7]{};
			\end{tikzpicture}}
		&	\adjustbox{valign=c}{\begin{tikzpicture}[every node/.style={inner sep=0,outer sep=0}]
				\node{\includegraphics[height=0.85cm,valign=c]{images/R-R-crossed}};
				\node[minimum height=1cm, minimum width=1.7cm,fill=white,opacity=0.7]{};
			\end{tikzpicture}}
		&	\adjustbox{valign=c}{\begin{tikzpicture}[every node/.style={inner sep=0,outer sep=0}]
				\node{\includegraphics[height=0.85cm,valign=c]{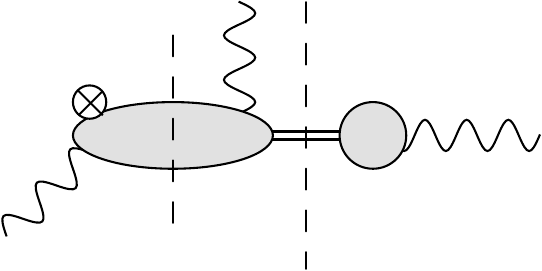}};
			\end{tikzpicture}}
		\\
	\midrule
	$S$
		& $\times$
		& $\times$
		& \includegraphics[height=0.85cm,valign=c]{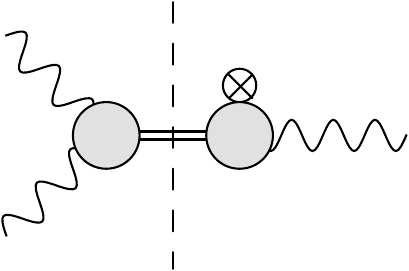}
		& $\times$
		& $\times$
		& $\times$
		\\
	\midrule
	$A$
		& $\times$
		& $\times$
		& $\times$
		& \includegraphics[height=0.85cm,valign=c]{images/R}
		& $\times$
		& $\times$
		\\
	\midrule
	$T$
		& $\times$
		& $\times$
		& $\times$
		& $\times$
		& \includegraphics[height=0.85cm,valign=c]{images/R}
		& $\times$
		\\
	\midrule
	\multirow{2}{*}{$\ldots$}
		&	\begin{tikzpicture}[every node/.style={inner sep=0,outer sep=0}]
				\node{\includegraphics[height=0.85cm,valign=c]{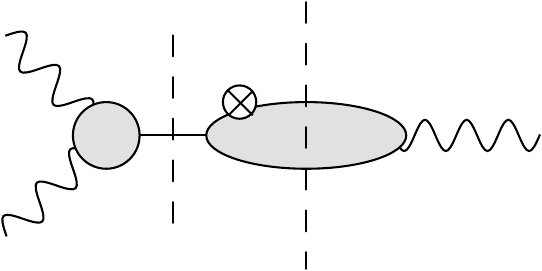}};
			\end{tikzpicture}
		&	\begin{tikzpicture}[every node/.style={inner sep=0,outer sep=0}]
				\node{\includegraphics[height=0.85cm,valign=c]{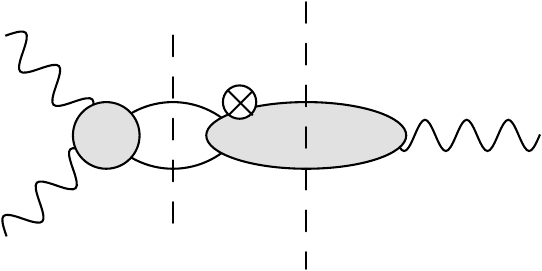}};
			\end{tikzpicture}
		&	\begin{tikzpicture}[every node/.style={inner sep=0,outer sep=0}]
				\node{\includegraphics[height=0.85cm,valign=c]{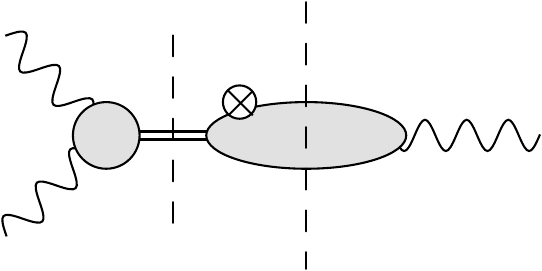}};
			\end{tikzpicture}
		&	\begin{tikzpicture}[every node/.style={inner sep=0,outer sep=0}]
				\node{\includegraphics[height=0.85cm,valign=c]{images/R-X}};
			\end{tikzpicture}
		&	\begin{tikzpicture}[every node/.style={inner sep=0,outer sep=0}]
				\node{\includegraphics[height=0.85cm,valign=c]{images/R-X}};
			\end{tikzpicture}
		& \multirow{2}{*}{$\ldots$}
		\\
		&	\begin{tikzpicture}[every node/.style={inner sep=0,outer sep=0}]
				\node{\includegraphics[height=0.85cm,valign=c]{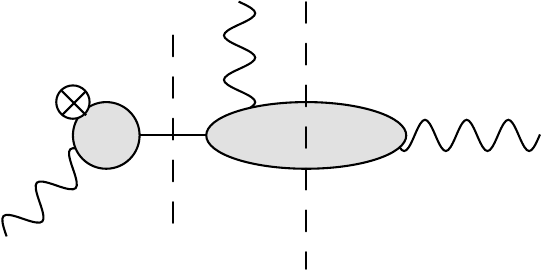}};
				\node[minimum height=1cm, minimum width=1.7cm,fill=white,opacity=0.7]{};
			\end{tikzpicture}
		&	\begin{tikzpicture}[every node/.style={inner sep=0,outer sep=0}]
				\node{\includegraphics[height=0.85cm,valign=c]{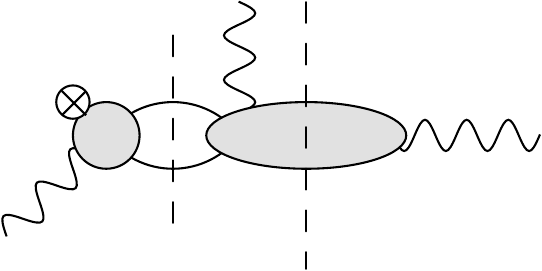}};
				\node[minimum height=1cm, minimum width=1.7cm,fill=white,opacity=0.7]{};
			\end{tikzpicture}
		&	\begin{tikzpicture}[every node/.style={inner sep=0,outer sep=0}]
				\node{\includegraphics[height=0.85cm,valign=c]{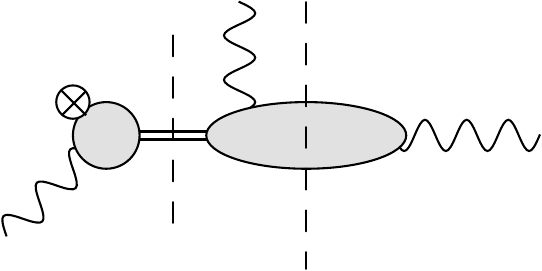}};
				\node[minimum height=1cm, minimum width=1.7cm,fill=white,opacity=0.7]{};
			\end{tikzpicture}
		&	\begin{tikzpicture}[every node/.style={inner sep=0,outer sep=0}]
				\node{\includegraphics[height=0.85cm,valign=c]{images/R-X-crossed}};
				\node[minimum height=1cm, minimum width=1.7cm,fill=white,opacity=0.7]{};
			\end{tikzpicture}
		&	\begin{tikzpicture}[every node/.style={inner sep=0,outer sep=0}]
				\node{\includegraphics[height=0.85cm,valign=c]{images/R-X-crossed}};
				\node[minimum height=1cm, minimum width=1.7cm,fill=white,opacity=0.7]{};
			\end{tikzpicture}
		\\
	\bottomrule
	\end{tabular}
	\caption{Comparison of different unitarity contributions in the established dispersive approach and the proposed dispersion relations in triangle kinematics. The soft external photon is denoted by a crossed circle. The longer dashed line is the primary cut in triangle-kinematics dispersion relations. Cuts through gray blobs denote even higher intermediate states that need to be covered via the implementation of asymptotic constraints. Some scalar and tensor resonances correspond to a NWA of two-pion contributions. Depending on the dispersion relation for the sub-processes, the diagrams in the first row of the $V$ intermediate state only contribute to normalizations. The light-gray diagrams are already taken into account by implementing crossing symmetry (which is not shown explicitly), hence these topologies should be excluded in order to avoid a double counting.}
	\label{tab:Reshuffling}
\end{table}

Any dispersion relation allows one to split up the entire HLbL contribution into a sum over intermediate states in the unitarity relation. However, the notion of the contribution of an individual intermediate state, obtained by inserting one term of the unitarity sum into the dispersion integral, depends on the dispersion relation under consideration. This is true for basis changes in the existing approach, as explained in Sect.~\ref{sec:ExistingApproach}, but also if one uses dispersion relations in a different kinematic variable. Only the result for the sum over all intermediate states is unique. In particular, this means that, e.g., the pion pole as defined in the established dispersive approach~\cite{Colangelo:2015ama,Hoferichter:2018kwz} does not coincide with the pion pole in triangle kinematics, as discussed in Ref.~\cite{Colangelo:2019uex}. When comparing the two approaches, one finds that a reshuffling happens between the contributions of different intermediate states. Since each dispersive approach requires some truncation of the unitarity sum, the correspondence is not exact, but the remainder needs to be covered by the uncertainties in the matching to an inclusive asymptotic contribution. We compare the two dispersion relations in Tab.~\ref{tab:Reshuffling}: the splitting by intermediate states in the established approach corresponds to columns, while the rows correspond to the contributions in the new dispersion relations in triangle kinematics. Therefore, if asymptotic constraints are included for the sub-processes, the established dispersion relations perform a resummation of columns, while the new approach would correspond to a resummation of rows. Crosses in the table denote the absence of a contribution. This sketch illustrates that the most promising strategy will be to combine the two approaches, which however requires some care in avoiding any double-counting. A detailed analysis of the reshuffling and the matching to asymptotic constraints is left for future work and will be illustrated for the simpler case of the VVA three-point function in a forthcoming publication~\cite{DRforVVA}.

In Sect.~\ref{sec:SingleParticle}, we consider single-particle intermediate states in triangle kinematics, while two-pion intermediate states will be discussed in Sect.~\ref{sec:TwoPionSubprocesses}.


\section{Single-particle intermediate states}
\label{sec:SingleParticle}

As shown in Fig.~\ref{img:HLbLIntermediateStates} and Tab.~\ref{tab:Reshuffling}, the $s$-channel cut receives single-particle contributions from pseudoscalar poles, as well as from resonances in the NWA. The $q_3^2$-channel discontinuity receives single-particle contributions only in the NWA due to vector-meson resonances.

In Sect.~\ref{sec:PionPole}, we work out the explicit expression for the pion-pole contribution in triangle kinematics and compare the result to the pion pole in the established dispersion relations in four-point kinematics. Similar results follow immediately for the other pseudoscalars $\eta$ and $\eta'$. In Sect.~\ref{sec:SChannelResonances}, we derive analogous expressions for resonance contributions in the NWA. In Sect.~\ref{sec:Q32ChannelVectorResonances}, we discuss vector-meson resonances in the $q_3^2$-channel.

\subsection{Pion pole}
\label{sec:PionPole}

The contribution of a single neutral pion in the $s$-channel unitarity relation is given by~\cite{Colangelo:2015ama}
\begin{align}
	\label{eq:SChannelUnitarityOnePion}
	\Im_s^\pi &\left( e^4 (2\pi)^4 \delta^{(4)}(q_1 + q_2 + q_3 - q_4) H_{\lambda_1\lambda_2,\lambda_3\lambda_4} \right) \nn
		&= \frac{1}{2} \int \widetilde{dp} \, \< \pi^0(p) | \gamma^*(-q_3,\lambda_3) \gamma(q_4,\lambda_4) \>^* \< \pi^0(p) | \gamma^*(q_1,\lambda_1) \gamma^*(q_2,\lambda_2) \> \, .
\end{align}
The matrix element of the sub-process is reduced according to
\begin{align}
	\label{eq:ggpiReduction}
	\< \pi^0(p) | \gamma^*(q_1,\lambda_1) \gamma^*(q_2,\lambda_2) \> &= - e^2 (2\pi)^4 \delta^{(4)}(p-q_1-q_2) \epsilon_\mu^{\lambda_1}(q_1) \epsilon_\nu^{\lambda_2}(q_2) \nn
		&\quad \times \int d^4x \, e^{-i q_1 \cdot x} \< \pi^0(p) | T \{ j^\mu_\mathrm{em}(x) j^\nu_\mathrm{em}(0) \} | 0 \> \, ,
\end{align}
relating it to the pion transition form factor (TFF)
\begin{align}
	i \int d^4x \, e^{-i q \cdot x} \< \pi^0(p) | T \{ j^\mu_\mathrm{em}(x) j^\nu_\mathrm{em}(0) \} | 0 \> = \epsilon^{\mu\nu\alpha\beta} q_\alpha p_\beta \F_{\pi^0\gamma^*\gamma^*}(q^2,(q-p)^2) \, ,
\end{align}
with $\epsilon^{0123} = +1$. It should be stressed that the TFF $\F_{\pi^0\gamma^*\gamma^*}$ is a scalar function of two independent scalar variables (the photon virtualities)---it does not explicitly depend on the four-vectors. E.g., momentum conservation is not part of the TFF but appears in the form of the delta function in Eq.~\eqref{eq:ggpiReduction}. Inserting the pion TFF into Eq.~\eqref{eq:SChannelUnitarityOnePion} allows one to perform the phase-space integral, leading to
\begin{align}
	\Im_s^\pi H_{\lambda_1\lambda_2,\lambda_3\lambda_4} &= - \pi \, \epsilon_\mu^{\lambda_1}(q_1) \epsilon_\nu^{\lambda_2}(q_2) {\epsilon_\lambda^{\lambda_3}}^*(-q_3) {\epsilon_\sigma^{\lambda_4}}^*(q_4) \nn
		&\quad \times \delta(s-\mpio^2) \epsilon^{\mu\nu\alpha\beta} \epsilon^{\lambda\sigma\gamma\delta} {q_1}_\alpha {q_2}_\beta {q_3}_\gamma {q_4}_\delta \F_{\pi^0\gamma^*\gamma^*}(q_1^2,q_2^2) \F_{\pi^0\gamma^*\gamma^*}(q_3^2,0) \, .
\end{align}
This expression can be evaluated for fixed-$t$ kinematics. In Ref.~\cite{Colangelo:2017fiz}, the basis change from $s$-channel helicity amplitudes to the fixed-$t$ singly-on-shell tensor coefficient functions is provided, leading to the single-pion discontinuity of the functions $\check\Pi_i$:
\begin{align}
	\Im_s^\pi \, \check\Pi_1(s';q_1^2,q_2^2,q_3^2) &= -\pi \delta(s'-\mpio^2) \F_{\pi^0\gamma^*\gamma^*}(q_1^2,q_2^2) \F_{\pi^0\gamma^*\gamma^*}(q_3^2,0) \, , \nn
	\Im_s^\pi \, \check\Pi_i(s';q_1^2,q_2^2,q_3^2) &= 0 \, , \quad i \ne 1 \, .
\end{align}
Since there is no one-pion intermediate state in the $q_3^2$-channel, the one-pion discontinuity of the $\hat\Pi_i$ functions follows by taking the limit~\eqref{eq:ImPiHat2Disc}:
\begin{align}
	\Im^\pi\, \hat\Pi_1(q_1^2,q_2^2,s') &= -\pi \delta(s'-\mpio^2) \F_{\pi^0\gamma^*\gamma^*}(q_1^2,q_2^2) \F_{\pi^0\gamma^*\gamma^*}(\mpio^2,0) \, , \nn
	\Im^\pi\, \hat\Pi_i(q_1^2,q_2^2,s') &= 0 \, , \quad i \in \{ 2, \ldots, 11,13,14,16,17,39,50,51,54\} \, .
\end{align}
Inserting this imaginary part into the dispersion relation~\eqref{eq:PiHatDispersionRelation} leads to a pion-pole contribution just in $\hat\Pi_1$ and no pole contribution in any other function. However, due to crossing symmetry it is clear that $\hat\Pi_{2,3}$ contain pion poles in the $q_2^2$- and $q_1^2$-channels, respectively. Those contributions could be reconstructed in the dispersion relation in $q_3^2$ from higher intermediate states due to cuts through the pion TFF, starting with a two-pion cut, which correspond to the light-gray diagrams of the first column of Tab.~\ref{tab:Reshuffling}:
\begin{align}
	\scalebox{-1}[1]{\includegraphics[height=1.2cm,valign=c]{images/Pole}} \; = \; \includegraphics[height=1.2cm,valign=c]{images/Pole-2pi-crossed} \; + \; \includegraphics[height=1.2cm,valign=c]{images/Pole-V-crossed} \; + \; \includegraphics[height=1.2cm,valign=c]{images/Pole-X-crossed} \; \, .
\end{align}
However, in practice it is simpler to directly include these contributions by imposing crossing symmetry, i.e., by adding the terms that appear in the crossed unitarity relations. These topologies then need to be omitted from the reconstruction of the cuts in $q_3^2$, in order to avoid a double counting. The same applies to the remaining topologies in the dispersion relation in $q_3^2$, which need to be symmetrized accordingly. This prescription leads to the final result for the pion-pole contribution in triangle kinematics:
\begin{align}
	\label{eq:PionPoleTriangleKinematics}
	\hat\Pi_1^\pi(q_1^2,q_2^2,q_3^2) &= \frac{ \F_{\pi^0\gamma^*\gamma^*}(q_1^2,q_2^2) \F_{\pi^0\gamma^*\gamma^*}(\mpio^2,0) }{q_3^2 - \mpio^2} \, , \nn
	\hat\Pi_i^\pi(q_1^2,q_2^2,q_3^2) &= 0 \, , \quad i \in \{ 4, 7, 17, 39, 54 \} \, ,
\end{align}
written in terms of the six representatives~\eqref{eq:PiHatFunctions}, while the remaining 13 functions follow from the crossing relations~\eqref{eq:CrossingRelationsPiHat}. Explicitly, they are given by
\begin{align}
	\hat\Pi_2^\pi(q_1^2,q_2^2,q_3^2) &= \frac{ \F_{\pi^0\gamma^*\gamma^*}(q_1^2,q_3^2) \F_{\pi^0\gamma^*\gamma^*}(\mpio^2,0) }{q_2^2 - \mpio^2} \, , \nn
	\hat\Pi_3^\pi(q_1^2,q_2^2,q_3^2) &= \frac{ \F_{\pi^0\gamma^*\gamma^*}(q_2^2,q_3^2) \F_{\pi^0\gamma^*\gamma^*}(\mpio^2,0) }{q_1^2 - \mpio^2} \, ,
\end{align}
and there are no further pion-pole contributions to the remaining functions. The pion-pole contribution fulfills the intrinsic crossing symmetries~\eqref{eq:InternalCrossingSymmetriesPiHat} due to the symmetry of the pion TFF, $\F_{\pi^0\gamma^*\gamma^*}(q_1^2,q_2^2) = \F_{\pi^0\gamma^*\gamma^*}(q_2^2,q_1^2)$.

The result for the pion pole~\eqref{eq:PionPoleTriangleKinematics} differs from the expression for the pion pole that follows from the dispersion relations in four-point kinematics~\cite{Colangelo:2015ama}, which is
\begin{align}
	\label{eq:PionPole4ptKin}
	\hat\Pi_1^{\pi^0\text{-pole}}(q_1^2,q_2^2,q_3^2) &= \frac{ \F_{\pi^0\gamma^*\gamma^*}(q_1^2,q_2^2) \F_{\pi^0\gamma^*\gamma^*}(q_3^2,0) }{q_3^2 - \mpio^2} \, , \nn
	\hat\Pi_i^{\pi^0\text{-pole}}(q_1^2,q_2^2,q_3^2) &= 0 \, , \quad i \in \{ 4, 7, 17, 39, 54 \} \, .
\end{align}
This mismatch led to some confusion in the literature~\cite{Melnikov:2019xkq,Knecht:2020xyr}, although the reason for it was already explained in Ref.~\cite{Colangelo:2019uex}: the different expressions~\eqref{eq:PionPoleTriangleKinematics} and~\eqref{eq:PionPole4ptKin} do not put the validity of either dispersion relation into question. Both, the dispersion relations in four-point kinematics and the ones in triangle kinematics can be used to describe the HLbL contribution to $(g-2)_\mu$. They reconstruct the same function if the tower of intermediate states in the unitarity relation is resummed. However, since the limit $q_4\to0$ changes the meaning of the kinematic invariants, writing dispersion relations before or after taking this limit does not lead to the same expressions. In particular, the contributions of one particular intermediate state do not need to agree in the two formalisms. As discussed in Ref.~\cite{Colangelo:2019uex}, the difference between Eqs.~\eqref{eq:PionPoleTriangleKinematics} and~\eqref{eq:PionPole4ptKin} is regular at $q_3^2 = \mpio^2$:
\begin{align}
	\label{eq:PionPoleDifference}
	\hat\Pi_1^{\pi^0\text{-pole}}(q_1^2,q_2^2,q_3^2) - \hat\Pi_1^\pi(q_1^2,q_2^2,q_3^2) = \F_{\pi^0\gamma^*\gamma^*}(q_1^2,q_2^2) \frac{ \F_{\pi^0\gamma^*\gamma^*}(q_3^2,0) - \F_{\pi^0\gamma^*\gamma^*}(\mpio^2,0)}{q_3^2 - \mpio^2} \, .
\end{align}
As shown in Tab.~\ref{tab:Reshuffling}, this quantity can be identified with contributions from higher intermediate states in the dispersion relation in triangle kinematics~\cite{Colangelo:2021nkr}, which needs to be considered when combining the two approaches: the $q_3^2$-channel pole $\hat\Pi_1^\pi(q_1^2,q_2^2,q_3^2)$ in Eq.~\eqref{eq:PionPoleTriangleKinematics} corresponds to the first row of the table (for the case of a $\pi^0$) and consists only of the upper left entry, while $\hat\Pi_1^{\pi^0\text{-pole}}(q_1^2,q_2^2,q_3^2)$ in Eq.~\eqref{eq:PionPole4ptKin} contains all the diagrams of the first column that are not shown in light-gray.\footnote{Note that this reshuffling of the pion pole is not related to the implementation of crossing symmetry discussed above: Eq.~\eqref{eq:PionPoleDifference} does not contain any pion pole in the $q_{1,2}^2$ channels.}

\subsection[Narrow resonances in the $s$-channel]{\boldmath Narrow resonances in the $s$-channel}
\label{sec:SChannelResonances}

In complete analogy to the single-pion (or, more generally, single-pseudoscalar) contribution in the $s$-channel, we can consider single-particle scalar, axial-vector, and tensor intermediate states, describing the contribution of resonances in the NWA.

\subsubsection{Scalar resonances}

The contribution of a scalar resonance to the $s$-channel unitarity relation is given by
\begin{align}
	\label{eq:SChannelUnitarityScalar}
	\Im_s^S &\left( e^4 (2\pi)^4 \delta^{(4)}(q_1 + q_2 + q_3 - q_4) H_{\lambda_1\lambda_2,\lambda_3\lambda_4} \right) \nn
		&= \frac{1}{2} \int \widetilde{dp} \, \< S(p) | \gamma^*(-q_3,\lambda_3) \gamma(q_4,\lambda_4) \>^* \< S(p) | \gamma^*(q_1,\lambda_1) \gamma^*(q_2,\lambda_2) \> \, .
\end{align}
The matrix element of the sub-process can be decomposed according to~\cite{Hoferichter:2020lap}
\begin{align}
	\label{eq:ggSReduction}
	\< S(p) | \gamma^*(q_1,\lambda_1) \gamma^*(q_2,\lambda_2) \> &= i e^2 (2\pi)^4 \delta^{(4)}(p-q_1-q_2) \epsilon_\mu^{\lambda_1}(q_1) \epsilon_\nu^{\lambda_2}(q_2) \M^{\mu\nu}(q_1,q_2\to p) \, , \nn
	\M^{\mu\nu}(q_1,q_2\to p) &= i \int d^4x \, e^{-i q_1 \cdot x} \< S(p) | T \{ j^\mu_\mathrm{em}(x) j^\nu_\mathrm{em}(0) \} | 0 \> \nn
		&=\frac{\F_1^S(q_1^2,q_2^2)}{m_S} T_1^{\mu\nu}  + \frac{\F_2^S(q_1^2,q_2^2)}{m_S^3} T_2^{\mu\nu} \, ,
\end{align}
where the Lorentz structures are given by
\begin{align}
	\label{eq:BTTScalar}
	T_1^{\mu\nu} &= q_1 \cdot q_2 g^{\mu\nu} - q_2^\mu q_1^\nu \, \nn
	T_2^{\mu\nu} &= q_1^2 q_2^2 g^{\mu\nu} + q_1 \cdot q_2 q_1^\mu q_2^\nu - q_1^2 q_2^\mu q_2^\nu - q_2^2 q_1^\mu q_1^\nu \, ,
\end{align}
and $\F_1^S$ and $\F_2^S$ are the scalar TFFs. For fixed-$t$ kinematics, one obtains the following discontinuity of the functions $\check\Pi_i$~\cite{Danilkin:2021icn}:
\begin{align}
	\Im_s^S \, \check\Pi_4(s';q_1^2,q_2^2,q_3^2) &= -\pi \delta(s'-m_S^2) \left( \frac{\F_1^S(q_1^2,q_2^2)}{m_S^2} - \frac{(s'+q_1^2+q_2^2)}{2m_S^4} \F_2^S(q_1^2,q_2^2) \right) \F_1^S(q_3^2,0)  \, , \nn
	\Im_s^S \, \check\Pi_{15}(s';q_1^2,q_2^2,q_3^2) &= -\pi \delta(s'-m_S^2) \frac{\F_2^S(q_1^2,q_2^2) \F_1^S(q_3^2,0)}{m_S^4} \, , \nn
	\Im_s^S \, \check\Pi_i(s';q_1^2,q_2^2,q_3^2) &= 0 \, , \quad i \notin \{4,15\} \, .
\end{align}
The discontinuity of the $\hat\Pi_i$ functions follows as:
\begin{align}
	\Im^S \, \hat\Pi_4(q_1^2,q_2^2,s') &= -\pi \delta(s'-m_S^2) \left( \frac{\F_1^S(q_1^2,q_2^2)}{m_S^2} - \frac{(m_S^2+q_1^2+q_2^2)}{2m_S^4} \F_2^S(q_1^2,q_2^2) \right) \F_1^S(m_S^2,0)  \, , \nn
	\Im^S \, \hat\Pi_{17}(q_1^2,q_2^2,s') &= -\pi \delta(s'-m_S^2) \frac{\F_2^S(q_1^2,q_2^2) \F_1^S(m_S^2,0)}{m_S^4} \, , \nn
	\Im^S \, \hat\Pi_i(q_1^2,q_2^2,s') &= 0 \, , \quad i \in \{1,2,3,5,\ldots,11,13,14,16,39,50,51,54\} \, .
\end{align}
Inserting this imaginary part into the dispersion relation~\eqref{eq:PiHatDispersionRelation} leads to
\begin{align}
	\label{eq:ScalarResonanceTriangleKinematics}
	\hat\Pi_4^S(q_1^2,q_2^2,q_3^2) &= \frac{\F_1^S(m_S^2,0)}{q_3^2 - m_S^2} \left( \frac{\F_1^S(q_1^2,q_2^2)}{m_S^2} - \frac{(m_S^2+q_1^2+q_2^2)}{2m_S^4} \F_2^S(q_1^2,q_2^2) \right) \, , \nn
	\hat\Pi_{17}^S(q_1^2,q_2^2,q_3^2) &= \frac{\F_1^S(m_S^2,0)}{q_3^2 - m_S^2} \frac{\F_2^S(q_1^2,q_2^2)}{m_S^4} \, , \nn
	\hat\Pi_i^S(q_1^2,q_2^2,q_3^2) &= 0 \, , \quad i \in \{1,7,39,54\} \, .
\end{align}
The six representative functions do not contain any scalar-meson poles in the crossed channels. However, the implementation of crossing symmetry in analogy to the pion pole implies that some of the remaining scalar functions contain scalar-meson poles in the crossed channels, as follows directly from Eq.~\eqref{eq:ScalarResonanceTriangleKinematics} and the crossing relations~\eqref{eq:CrossingRelationsPiHat}. E.g., the contribution to $\hat\Pi_5$ reads
\begin{align}
	\label{eq:ScalarResonancePihat5}
	\hat\Pi_5^S(q_1^2,q_2^2,q_3^2) &= \frac{\F_1^S(m_S^2,0)}{q_2^2 - m_S^2} \left( \frac{\F_1^S(q_1^2,q_3^2)}{m_S^2} - \frac{(m_S^2+q_1^2+q_3^2)}{2m_S^4} \F_2^S(q_1^2,q_3^2) \right) \, ,
\end{align}
and a double counting needs to be avoided by omitting from the $q_3^2$-dispersion relation the light-gray diagrams in the $S$ column, which correspond, e.g., to two-pion and vector-meson intermediate states in the scalar-meson TFF in Eq.~\eqref{eq:ScalarResonancePihat5}.
The result~\eqref{eq:ScalarResonanceTriangleKinematics} can be obtained from the scalar contribution in four-point kinematics~\cite{Danilkin:2021icn} by keeping only the pure pole in $q_3^2$. It differs from it by a piece regular at $q_3^2 = m_S^2$, which corresponds to the entries in the ``$S$'' column of Tab.~\ref{tab:Reshuffling} that do not belong to the ``$S$'' row and are not light-gray diagrams.

\subsubsection{Axial-vector resonances}

Next, we consider the contribution of an axial-vector resonance to the $s$-channel:
\begin{align}
	\label{eq:SChannelUnitarityAxial}
	\Im_s^A &\left( e^4 (2\pi)^4 \delta^{(4)}(q_1 + q_2 + q_3 - q_4) H_{\lambda_1\lambda_2,\lambda_3\lambda_4} \right) \nn
		&= \frac{1}{2} \sum_{\lambda_A} \int \widetilde{dp} \, \< A(p,\lambda_A) | \gamma^*(-q_3,\lambda_3) \gamma(q_4,\lambda_4) \>^* \< A(p,\lambda_A) | \gamma^*(q_1,\lambda_1) \gamma^*(q_2,\lambda_2) \> \, .
\end{align}
The matrix element of the sub-process can be decomposed into Lorentz structures according to
\begin{align}
	\label{eq:ggAReduction}
	\< A(p,\lambda_A) | \gamma^*(q_1,\lambda_1) \gamma^*(q_2,\lambda_2) \> &= i e^2 (2\pi)^4 \delta^{(4)}(p-q_1-q_2) \epsilon_\mu^{\lambda_1}(q_1) \epsilon_\nu^{\lambda_2}(q_2) \M^{\mu\nu}(q_1,q_2\to \{p,\lambda_A\}) \, , \nn
	\M^{\mu\nu}(q_1,q_2\to \{p,\lambda_A\}) &= i \int d^4x \, e^{-i q_1 \cdot x} \< A(p,\lambda_A) | T \{ j^\mu_\mathrm{em}(x) j^\nu_\mathrm{em}(0) \} | 0 \> \nn
		&=: {\epsilon_\alpha^{\lambda_A}}^*(p) \M^{\mu\nu\alpha}(-q_1,-q_2) \, , \nn
	\M^{\mu\nu\alpha}(q_1,q_2) &= -\M^{\mu\nu\alpha}(-q_1,-q_2) \nn
		&= \frac{i}{m_A^2} \sum_{i=1}^3 T_i^{\mu\nu\alpha} \F_i^A(q_1^2,q_2^2) \, ,
\end{align}
where the Lorentz structures are given by~\cite{Hoferichter:2020lap}
\begin{align}
	\label{eq:BTTAxial}
	T_1^{\mu\nu\alpha} &= \epsilon^{\mu\nu\beta\gamma} {q_1}_\beta {q_2}_\gamma ( q_1^\alpha - q_2^\alpha ) \, , \nn
	T_2^{\mu\nu\alpha} &= \epsilon^{\alpha\nu\beta\gamma} {q_1}_\beta {q_2}_\gamma q_1^\mu + \epsilon^{\alpha\mu\nu\beta} {q_2}_\beta q_1^2 \, , \nn
	T_3^{\mu\nu\alpha} &= \epsilon^{\alpha\mu\beta\gamma} {q_1}_\beta {q_2}_\gamma q_2^\nu + \epsilon^{\alpha\mu\nu\beta} {q_1}_\beta q_2^2 \, ,
\end{align}
and $\F_i^A$ are the axial-vector TFFs. For fixed-$t$ kinematics, the discontinuity of the functions $\check\Pi_i$ in the basis of Ref.~\cite{Colangelo:2017fiz} is rather complicated and contains kinematic singularities proportional to $s'-q_3^2$, which drop out when we take the limit $q_3^2\to s'$. This leads to
\begin{align}
	\Im^A \, \hat\Pi_5(q_1^2,q_2^2,s') &= \pi \delta(s'-m_A^2) \frac{m_A^2-q_1^2-q_2^2}{2m_A^4} \left( 2 \F_1^A(q_1^2,q_2^2) + \F_3^A(q_1^2,q_2^2) \right) \F_2^A(m_A^2,0)  \, , \nn
	\Im^A \, \hat\Pi_6(q_1^2,q_2^2,s') &= -\pi \delta(s'-m_A^2) \frac{m_A^2-q_1^2-q_2^2}{2m_A^4} \left( 2 \F_1^A(q_1^2,q_2^2) + \F_2^A(q_1^2,q_2^2) \right) \F_2^A(m_A^2,0)  \, , \nn
	\Im^A \, \hat\Pi_{9}(q_1^2,q_2^2,s') &= - \Im^A \, \hat\Pi_{13}(q_1^2,q_2^2,s') \nn*
		& = -\pi \delta(s'-m_A^2) \frac{1}{m_A^4} \left( 2 \F_1^A(q_1^2,q_2^2) + \F_2^A(q_1^2,q_2^2) + \F_3^A(q_1^2,q_2^2) \right) \F_2^A(m_A^2,0)  \, , \nn
	\Im^A \, \hat\Pi_{10}(q_1^2,q_2^2,s') &= \pi \delta(s'-m_A^2) \frac{1}{m_A^4} \left( 2 \F_1^A(q_1^2,q_2^2) + \F_3^A(q_1^2,q_2^2) \right) \F_2^A(m_A^2,0)  \, , \nn
	\Im^A \, \hat\Pi_{11,54}(q_1^2,q_2^2,s') &= - \Im^A \, \hat\Pi_{16}(q_1^2,q_2^2,s') \nn
		& = -\pi \delta(s'-m_A^2) \frac{1}{2m_A^4} \left( 4 \F_1^A(q_1^2,q_2^2) + \F_2^A(q_1^2,q_2^2) + \F_3^A(q_1^2,q_2^2) \right) \F_2^A(m_A^2,0)  \, , \nn
	\Im^A \, \hat\Pi_{14}(q_1^2,q_2^2,s') &= -\pi \delta(s'-m_A^2) \frac{1}{m_A^4} \left( 2 \F_1^A(q_1^2,q_2^2) + \F_2^A(q_1^2,q_2^2) \right) \F_2^A(m_A^2,0)  \, , \nn
	\Im^A \, \hat\Pi_{17,39,50,51}(q_1^2,q_2^2,s') &= \pi \delta(s'-m_A^2) \frac{1}{2m_A^4} \left( \F_2^A(q_1^2,q_2^2) - \F_3^A(q_1^2,q_2^2) \right) \F_2^A(m_A^2,0)  \, , \nn
	\Im^A \, \hat\Pi_i(q_1^2,q_2^2,s') &= 0 \, , \quad i \in \{1,2,3,4,7,8\} \, .
\end{align}
Plugging these imaginary parts into dispersion relations in $q_3^2$ leads to axial-vector contributions in only the $q_3^2$-channel. In analogy to the pion-pole and scalar-resonance contributions, crossing symmetry implies that there are also axial-vector poles in the $q_1^2$- and $q_2^2$-channels. Therefore, we write the full axial-vector contribution as
\begin{align}
	\label{eq:AxialResonanceTriangleKinematicsFull}
	\hat\Pi_i^A = \hat\Pi_i^{A,1} + \hat\Pi_i^{A,2} + \hat\Pi_i^{A,3} \, ,
\end{align}
which fulfills all constraints of crossing symmetry. Explicitly, it is given in terms of the six representative functions~\eqref{eq:PiHatFunctions} as
\begin{align}
	\label{eq:AxialResonanceTriangleKinematicsQ32}
	\hat\Pi_{17}^{A,3}(q_1^2,q_2^2,q_3^2) &= \hat\Pi_{39}^{A,3}(q_1^2,q_2^2,q_3^2) = \frac{\F_2^A(m_A^2,0)}{q_3^2 - m_A^2} \frac{\F_3^A(q_1^2,q_2^2) - \F_2^A(q_1^2,q_2^2)}{2m_A^4} \, , \nn
	\hat\Pi_{54}^{A,3}(q_1^2,q_2^2,q_3^2) &= \frac{\F_2^A(m_A^2,0)}{q_3^2 - m_A^2} \frac{4 \F_1^A(q_1^2,q_2^2) + \F_2^A(q_1^2,q_2^2) + \F_3^A(q_1^2,q_2^2)}{2m_A^4} \, , \nn
	\hat\Pi_i^{A,3}(q_1^2,q_2^2,q_3^2) &= 0 \, , \quad i \in \{1,4,7\} \, ,
\end{align}
for the axial-vector contributions in the $q_3^2$-channels, while the contributions in the other two channels are
\begin{align}
	\label{eq:AxialResonanceTriangleKinematicsQ12}
	\hat\Pi_1^{A,1}(q_1^2,q_2^2,q_3^2) &= 0 \, , \nn
	\hat\Pi_{4}^{A,1}(q_1^2,q_2^2,q_3^2) &= -\left(m_A^2 - q_2^2 - q_3^2\right) \frac{\F_2^A(m_A^2,0)}{q_1^2 - m_A^2} \frac{2 \F_1^A(q_2^2,q_3^2) + \F_3^A(q_2^2,q_3^2) }{2m_A^4} \, , \nn
	\hat\Pi_{7}^{A,1}(q_1^2,q_2^2,q_3^2) &= \frac{\F_2^A(m_A^2,0)}{q_1^2 - m_A^2} \frac{2\F_1^A(q_2^2,q_3^2) + \F_2^A(q_2^2,q_3^2) + \F_3^A(q_2^2,q_3^2)}{m_A^4} \, , \nn
	\hat\Pi_{17}^{A,1}(q_1^2,q_2^2,q_3^2) &= -\frac{\F_2^A(m_A^2,0)}{q_1^2 - m_A^2} \frac{4\F_1^A(q_2^2,q_3^2) + \F_2^A(q_2^2,q_3^2) + \F_3^A(q_2^2,q_3^2)}{2m_A^4} \, , \nn
	\hat\Pi_{39}^{A,1}(q_1^2,q_2^2,q_3^2) &= \hat\Pi_{54}^{A,1}(q_1^2,q_2^2,q_3^2) = \frac{\F_2^A(m_A^2,0)}{q_1^2 - m_A^2} \frac{\F_3^A(q_2^2,q_3^2) - \F_2^A(q_2^2,q_3^2)}{2m_A^4} \, ,
\end{align}
as well as
\begin{align}
	\label{eq:AxialResonanceTriangleKinematicsQ22}
	\hat\Pi_1^{A,2}(q_1^2,q_2^2,q_3^2) &= 0 \, , \nn
	\hat\Pi_{4}^{A,2}(q_1^2,q_2^2,q_3^2) &= -\left(m_A^2 - q_1^2 - q_3^2\right) \frac{\F_2^A(m_A^2,0)}{q_2^2 - m_A^2} \frac{2 \F_1^A(q_1^2,q_3^2) + \F_3^A(q_1^2,q_3^2) }{2m_A^4} \, , \nn
	\hat\Pi_{7}^{A,2}(q_1^2,q_2^2,q_3^2) &= -\frac{\F_2^A(m_A^2,0)}{q_2^2 - m_A^2} \frac{2\F_1^A(q_1^2,q_3^2) + \F_3^A(q_1^2,q_3^2)}{m_A^4} \, , \nn
	\hat\Pi_{17}^{A,2}(q_1^2,q_2^2,q_3^2) &= -\frac{\F_2^A(m_A^2,0)}{q_2^2 - m_A^2} \frac{4\F_1^A(q_1^2,q_3^2) + \F_2^A(q_1^2,q_3^2) + \F_3^A(q_1^2,q_3^2)}{2m_A^4} \, , \nn
	\hat\Pi_{39}^{A,2}(q_1^2,q_2^2,q_3^2) &= -\hat\Pi_{54}^{A,2}(q_1^2,q_2^2,q_3^2) = \frac{\F_2^A(m_A^2,0)}{q_2^2 - m_A^2} \frac{\F_3^A(q_1^2,q_3^2) - \F_2^A(q_1^2,q_3^2)}{2m_A^4} \, .
\end{align}
The axial-vector contribution to the remaining 13 functions follows from the crossing relations~\eqref{eq:CrossingRelationsPiHat}. This implementation of crossing symmetry is analogous to the pion pole or scalar resonances, but in the case of axial-vector resonances most of the scalar functions receive contributions from multiple channels. We also note that due to the symmetries of the axial-vector TFFs~\cite{Hoferichter:2020lap}, the intrinsic crossing symmetries~\eqref{eq:InternalCrossingSymmetriesPiHat} are manifestly fulfilled.
Similarly to the pion pole and scalar contribution, the added crossed-channel axial-vector contributions contain two-pion, vector-meson, and higher intermediate states in the $q_3^2$-channel, due to the singularity structure of the TFFs that depend on $q_3^2$. This needs to be considered when taking into account these cuts in $q_3^2$, in order to avoid a double counting, again in complete analogy to the pion-pole or scalar-meson contributions.

With the modified basis of $\check\Pi_i$ functions discussed in Ref.~\cite{Colangelo:2021nkr}, axial-vector contributions can be taken into account in the dispersion relation in four-point kinematics without introducing spurious kinematic singularities. The results~\eqref{eq:AxialResonanceTriangleKinematicsQ32}, \eqref{eq:AxialResonanceTriangleKinematicsQ12}, and \eqref{eq:AxialResonanceTriangleKinematicsQ22} differ from the results in four-point kinematics~\cite{Danilkin:2021icn} only by non-pole pieces. As before, this difference is given by the entries in the ``$A$'' column of Tab.~\ref{tab:Reshuffling} that do not belong to the ``$A$'' row and are not light-gray diagrams.

\subsubsection{Tensor resonances}

We finally consider the contribution of a tensor resonance to the $s$-channel:
\begin{align}
	\label{eq:SChannelUnitarityTensor}
	\Im_s^T &\left( e^4 (2\pi)^4 \delta^{(4)}(q_1 + q_2 + q_3 - q_4) H_{\lambda_1\lambda_2,\lambda_3\lambda_4} \right) \nn
		&= \frac{1}{2} \sum_{\lambda_T} \int \widetilde{dp} \, \< T(p,\lambda_T) | \gamma^*(-q_3,\lambda_3) \gamma(q_4,\lambda_4) \>^* \< T(p,\lambda_T) | \gamma^*(q_1,\lambda_1) \gamma^*(q_2,\lambda_2) \> \, .
\end{align}
The matrix element of the sub-process can be decomposed into Lorentz structures according to
\begin{align}
	\label{eq:ggTReduction}
	\< T(p,\lambda_T) | \gamma^*(q_1,\lambda_1) \gamma^*(q_2,\lambda_2) \> &= i e^2 (2\pi)^4 \delta^{(4)}(p-q_1-q_2) \epsilon_\mu^{\lambda_1}(q_1) \epsilon_\nu^{\lambda_2}(q_2) \M^{\mu\nu}(q_1,q_2\to \{p,\lambda_T\}) \, , \nn
	\M^{\mu\nu}(q_1,q_2\to \{p,\lambda_T\}) &= i \int d^4x \, e^{-i q_1 \cdot x} \< T(p,\lambda_T) | T \{ j^\mu_\mathrm{em}(x) j^\nu_\mathrm{em}(0) \} | 0 \> \nn
		&=: {\epsilon_{\alpha\beta}^{\lambda_T}}^*(p) \M^{\mu\nu\alpha\beta}(-q_1,-q_2) \, , \nn
	\M^{\mu\nu\alpha\beta}(q_1,q_2) &= \M^{\mu\nu\alpha\beta}(-q_1,-q_2) \nn
		&= \sum_{i=1}^5 T_i^{\mu\nu\alpha\beta} \frac{1}{m_T^{n_i}} \F_i^T(q_1^2,q_2^2) \, ,
\end{align}
with $n_1=1$ and the other $n_i=3$ and where the Lorentz structures are given in Ref.~\cite{Hoferichter:2020lap}. The polarization sum is
\begin{align}
	\label{eq:TensorPolarisationSum}
	s^T_{\alpha\beta\alpha'\beta'}(p) := \sum_{\lambda_T} \epsilon_{\alpha\beta}^{\lambda_T}(p) {\epsilon_{\alpha'\beta'}^{\lambda_T}}^*(p) = \frac{1}{2} \left( s_{\alpha\beta'} s_{\alpha'\beta} + s_{\alpha\alpha'} s_{\beta\beta'} \right) - \frac{1}{3} s_{\alpha\beta} s_{\alpha'\beta'} ,
\end{align}
where
\begin{align}
	s_{\alpha\alpha'} := - \left( g_{\alpha\alpha'} - \frac{p_\alpha p_{\alpha'}}{m_T^2} \right) .
\end{align}
The projection onto the functions $\hat\Pi_i$ leads to the following imaginary parts:
\begin{align}
	\Im^T \, \hat\Pi_i(q_1^2,q_2^2,s') &= \pi \delta(s' - m_T^2) \sum_{j=1}^5 t_{i,j}(q_1^2,q_2^2) \frac{\F_j^T(q_1^2,q_2^2)}{m_T^6} ( \F_1^T(m_T^2,0) + \F_5^T(m_T^2,0) ) \, ,
\end{align}
where the coefficients $t_{i,j}$ are defined in App.~\ref{app:TensorMesons}. In analogy to the axial-vector contributions, we combine dispersion relations in all three virtualities in order to arrive at a tensor-meson contribution that respects crossing symmetry. The full tensor-meson contributions is given by
\begin{align}
	\hat\Pi_i^T = \hat\Pi_i^{T,1} + \hat\Pi_i^{T,2} + \hat\Pi_i^{T,3} \, ,
\end{align}
which fulfills all constraints of crossing symmetry and can be defined in terms of the six representative functions~\eqref{eq:PiHatFunctions}:
\begin{align}
	\label{eq:TensorResonanceTriangleKinematicsQ32}
	\hat\Pi_i^{T,3}(q_1^2,q_2^2,q_3^2) &= - \sum_{j=1}^5 t_{i,j}(q_1^2,q_2^2) \frac{\F_j^T(q_1^2,q_2^2)}{m_T^6} \frac{\F_1^T(m_T^2,0) + \F_5^T(m_T^2,0)}{q_3^2 - m_T^2} \, , \quad i \in \{ 4, 7, 17, 39, 54 \} \, , \nn
	\hat\Pi_1^{T,3}(q_1^2,q_2^2,q_3^2) &= 0 \, .
\end{align}
The crossed-channel contributions to the six representative functions are obtained as
\begin{align}
	\label{eq:TensorResonanceTriangleKinematicsQ12}
	\hat\Pi_1^{T,1}(q_1^2,q_2^2,q_3^2) &= 0 \, , \nn
	\hat\Pi_4^{T,1}(q_1^2,q_2^2,q_3^2) &= - \sum_{j=1}^5 t_{5,j}(q_2^2,q_3^2) \frac{\F_j^T(q_2^2,q_3^2)}{m_T^6} \frac{\F_1^T(m_T^2,0) + \F_5^T(m_T^2,0)}{q_1^2 - m_T^2} \, , \nn 
	\hat\Pi_7^{T,1}(q_1^2,q_2^2,q_3^2) &= - \sum_{j=1}^5 t_{9,j}(q_2^2,q_3^2) \frac{\F_j^T(q_2^2,q_3^2)}{m_T^6} \frac{\F_1^T(m_T^2,0) + \F_5^T(m_T^2,0)}{q_1^2 - m_T^2} \, , \nn 
	\hat\Pi_{17}^{T,1}(q_1^2,q_2^2,q_3^2) &= - \sum_{j=1}^5 t_{16,j}(q_2^2,q_3^2) \frac{\F_j^T(q_2^2,q_3^2)}{m_T^6} \frac{\F_1^T(m_T^2,0) + \F_5^T(m_T^2,0)}{q_1^2 - m_T^2} \, , \nn 
	\hat\Pi_{39}^{T,1}(q_1^2,q_2^2,q_3^2) &= - \sum_{j=1}^5 t_{39,j}(q_2^2,q_3^2) \frac{\F_j^T(q_2^2,q_3^2)}{m_T^6} \frac{\F_1^T(m_T^2,0) + \F_5^T(m_T^2,0)}{q_1^2 - m_T^2} \, , \nn 
	\hat\Pi_{54}^{T,1}(q_1^2,q_2^2,q_3^2) &= - \sum_{j=1}^5 t_{50,j}(q_2^2,q_3^2) \frac{\F_j^T(q_2^2,q_3^2)}{m_T^6} \frac{\F_1^T(m_T^2,0) + \F_5^T(m_T^2,0)}{q_1^2 - m_T^2} 
\end{align}
and
\begin{align}
	\label{eq:TensorResonanceTriangleKinematicsQ22}
	\hat\Pi_1^{T,2}(q_1^2,q_2^2,q_3^2) &= 0 \, , \nn
	\hat\Pi_4^{T,2}(q_1^2,q_2^2,q_3^2) &= - \sum_{j=1}^5 t_{5,j}(q_1^2,q_3^2) \frac{\F_j^T(q_1^2,q_3^2)}{m_T^6} \frac{\F_1^T(m_T^2,0) + \F_5^T(m_T^2,0)}{q_2^2 - m_T^2} \, , \nn 
	\hat\Pi_7^{T,2}(q_1^2,q_2^2,q_3^2) &= - \sum_{j=1}^5 t_{10,j}(q_1^2,q_3^2) \frac{\F_j^T(q_1^2,q_3^2)}{m_T^6} \frac{\F_1^T(m_T^2,0) + \F_5^T(m_T^2,0)}{q_2^2 - m_T^2} \, , \nn 
	\hat\Pi_{17}^{T,2}(q_1^2,q_2^2,q_3^2) &= - \sum_{j=1}^5 t_{16,j}(q_1^2,q_3^2) \frac{\F_j^T(q_1^2,q_3^2)}{m_T^6} \frac{\F_1^T(m_T^2,0) + \F_5^T(m_T^2,0)}{q_2^2 - m_T^2} \, , \nn 
	\hat\Pi_{39}^{T,2}(q_1^2,q_2^2,q_3^2) &= - \sum_{j=1}^5 t_{39,j}(q_1^2,q_3^2) \frac{\F_j^T(q_1^2,q_3^2)}{m_T^6} \frac{\F_1^T(m_T^2,0) + \F_5^T(m_T^2,0)}{q_2^2 - m_T^2} \, , \nn 
	\hat\Pi_{54}^{T,2}(q_1^2,q_2^2,q_3^2) &= \sum_{j=1}^5 t_{50,j}(q_1^2,q_3^2) \frac{\F_j^T(q_1^2,q_3^2)}{m_T^6} \frac{\F_1^T(m_T^2,0) + \F_5^T(m_T^2,0)}{q_2^2 - m_T^2} \, ,
\end{align}
while the contribution to the remaining 13 functions again follows directly from the crossing relations~\eqref{eq:CrossingRelationsPiHat}. The same comment regarding double counting with two-pion, vector-meson, and higher cuts in $q_3^2$ applies as for the other resonances.

To the best of our knowledge, there is no alternative basis of $\check\Pi_i$ functions that would allow dispersion relations in four-point kinematics for the tensor-meson contributions that are manifestly free from spurious kinematic singularities if no additional sum rules compared to the ones of Ref.~\cite{Colangelo:2017fiz} are invoked. The modified basis discussed in Ref.~\cite{Colangelo:2021nkr} reduces the spurious kinematic singularities in the tensor-meson contribution to simple poles of the type $1/q_1^2$ for fixed-$t$ kinematics.

\subsection[Vector resonances in the $q_3^2$-channel]{\boldmath Vector resonances in the $q_3^2$-channel}
\label{sec:Q32ChannelVectorResonances}

In the $q_3^2$-channel, single-particle intermediate states only appear in the NWA, in particular the iso-scalar vector-meson resonances $\omega$ and $\phi$ (the prominent iso-vector $\rho$ resonance is best described in terms of two-pion $P$-wave rescattering). The unitarity relation reads
\begin{align}
	\Im_3^V &\left( e^4 (2\pi)^4 \delta^{(4)}(q_1 + q_2 + q_3 - q_4) H_{\lambda_1\lambda_2\lambda_4,\lambda_3} \right) \nn
		&= \frac{1}{2} \sum_{\lambda_V} \int \widetilde{dp} \, \< V(p,\lambda_V) | \gamma^*(-q_3,\lambda_3) \>^* \< V(p,\lambda_V) | \gamma^*(q_1,\lambda_1) \gamma^*(q_2,\lambda_2) \gamma(-q_4,\lambda_4) \> \, .
\end{align}
The matrix element of the first sub-process is simply given by
\begin{align}
	\< V(p,\lambda_V) | \gamma^*(-q_3,\lambda_3) \> = - i e \, \epsilon_\mu^{\lambda_3}(-q_3) \epsilon^\mu_{\lambda_V}(p)^* (2\pi)^4 \delta^{(4)}(p + q_3) m_V f_V \, ,
\end{align}
where the vector-meson decay constant $f_V$ is defined as
\begin{align}
	\< 0 | j_\mathrm{em}^\mu(x) | V(p,\lambda_V) \> = m_V f_V \epsilon^\mu_{\lambda_V}(p) e^{- i p \cdot x} \, .
\end{align}
For the matrix element of the second sub-process $\gamma^*\gamma^*\gamma\to V$, we define
\begin{align}
	\< V(p,\lambda_V) | \gamma^*(q_1,\lambda_1) \gamma^*(q_2,\lambda_2) \gamma(-q_4,\lambda_4) \> &= i (2\pi)^4 \delta^{(4)}(q_1+q_2-q_4-p) e^3 \nn
		&\quad \times \epsilon_\mu^{\lambda_1}(q_1) \epsilon_\nu^{\lambda_2}(q_2) \epsilon_\sigma^{\lambda_4}(-q_4) \epsilon_\lambda^{\lambda_V}(p)^* \, \Pi_V^{\mu\nu\lambda\sigma} \, ,
\end{align}
where
\begin{align}
	\epsilon_\lambda^{\lambda_V}(p)^* \, \Pi_V^{\mu\nu\lambda\sigma}(q_1,q_2,p) = \int d^4x d^4y \, e^{-i(q_1 \cdot x + q_2 \cdot y)} \, \< V(p,\lambda_V) | T \{ j_\mathrm{em}^\mu(x) j_\mathrm{em}^\nu(y) j_\mathrm{em}^\sigma(0) \} | 0 \> \, ,
\end{align}
and we perform the BTT tensor decomposition~\cite{Bardeen:1969aw,Tarrach:1975tu} for $\Pi_V^{\mu\nu\lambda\sigma}(q_1,q_2,p)$ in close analogy to the case of HLbL scattering~\cite{Colangelo:2015ama}. We first impose transversality for the three photons by making use of gauge projectors
\begin{align}
	I_{12}^{\mu\nu} &= g^{\mu\nu} - \frac{q_2^\mu q_1^\nu}{q_1 \cdot q_2} \, , \quad I_{4}^{\sigma\sigma'} = g^{\sigma\sigma'} - \frac{q_4^\sigma q_4^{\sigma'}}{q_4^2} \, ,
\end{align}
and we remove kinematic singularities in the projected tensor structures according to the BTT recipe. This leads to a highly redundant generating set of 72 tensor structures. For the dispersion relations in triangle kinematics, we can immediately take the derivative with respect to the external photon momentum and put $q_4\to0$. After this step, only 26 linear combinations of tensor structures are non-vanishing. In a final step, we note that in any observable (in particular in $(g-2)_\mu$) the tensor $\Pi_V^{\mu\nu\lambda\sigma}$ appears contracted with the vector-meson polarization sum
\begin{align}
	\sum_{\lambda_V} \epsilon_\lambda^{\lambda_V}(p) \epsilon_{\lambda'}^{\lambda_V}(p)^* = - \left( g_{\lambda\lambda'} - \frac{p_\lambda p_{\lambda'}}{m_V^2} \right) \, .
\end{align}
This implies that out of the 26 derivative tensor structures, only 19 linear combinations enter $(g-2)_\mu$, which can be chosen to be identical to the HLbL tensor structures $\hat T_i^{\mu\nu\lambda\sigma;\rho}$ in Eq.~\eqref{eq:HLbLThatTensorStructures}: the contraction with the polarization sum has the same effect as imposing the QED Ward identity, with the difference that factors of $m_V^2$ in the denominator should not be regarded as kinematic singularities. Choosing the HLbL structures instead of the ones that naturally come out of the BTT construction with a vector meson only amounts to a basis change that does not introduce kinematic singularities but involves factors of $1/m_V^2$. The ideal basis for a dispersive reconstruction of the scalar functions depends on the asymptotic behavior, which will require a dedicated analysis. Here, we decompose the tensor as
\begin{align}
	\frac{\p}{\p q_{4\rho}} \Pi_V^{\mu\nu\lambda\sigma}(q_1,q_2,p)  \bigg|_{q_4=0} = \sum_{i=1}^{19} \hat T_i^{\mu\nu\lambda\sigma;\rho}(q_1,q_2) \F^V_i(q_1^2,q_2^2) \, ,
\end{align}
dropping directly the unphysical contributions that vanish upon contraction with the polarization sum. Hence, the unitarity relation leads to
\begin{align}
	\Im^V \hat\Pi_{g_i}(q_1^2,q_2^2,s') = \pi \delta(s' - m_V^2) m_V f_V \F_i^V(q_1^2,q_2^2) \, ,
\end{align}
and therefore
\begin{align}
	\hat\Pi_{g_i}^{V,3}(q_1^2,q_2^2,q_3^2) = - \F_i^V(q_1^2,q_2^2) \frac{m_V f_V}{q_3^2 - m_V^2} \, .
\end{align}
Analogous expressions hold for the contributions in the two crossed channels. Again, when writing a representation that is manifestly crossing symmetric, a double counting must be avoided. E.g., the crossed pion-pole contributions already contain the vector-resonance contribution that corresponds to the pion pole in $\F_i^V$ in the $q_1^2$- and $q_2^2$-channels.


\section{Tensor decomposition for two-pion sub-processes}
\label{sec:TwoPionSubprocesses}

Apart from single-particle intermediate states, we are mainly interested in two-pion contributions in the new formalism: in the $D$-wave of $\pi\pi$ scattering, we find the $f_2(1270)$ resonance. In order to compare the description of this resonance in terms of a NWA with the two-pion representation (in analogy to the comparison for scalar resonances performed in Ref.~\cite{Danilkin:2021icn}), we need to reconstruct the two-pion sub-processes appearing in the unitarity relations for HLbL. The main missing input is the five-particle process $\gamma^*\gamma^*\gamma\to2\pi$, up to the first non-trivial order in the soft-photon expansion. As a nested sub-process, the process $\pi\pi\to\pi\pi\gamma$ appears. In the following subsections, we present the Lorentz decomposition for these sub-processes, which are key to set up a dispersive treatment. The dispersive reconstruction itself, which solves two-pion unitarity, will be the subject of a future publication~\cite{TriangleDR4PiGamma}.

\subsection[The process $\pi\pi\to\pi\pi\gamma$]{\boldmath The process $\pi\pi\to\pi\pi\gamma$}

\subsubsection{Kinematics and matrix element}

We consider $\pi\pi$ scattering with the emission of an additional soft photon, $\pi\pi\to\pi\pi\gamma$, with polarization $\lambda$. We define the process via the matrix element
\begin{align}
	&\< \pi^c(-p_3) \pi^d(-p_4) \gamma(-q,\lambda) | \pi^a(p_1) \pi^b(p_2) \> \nn
		&= -i e {\epsilon_\mu^{\lambda}}^*(-q) \int d^4x\, e^{-i q \cdot x} \< \pi^c(-p_3) \pi^d(-p_4) | j^\mu_\mathrm{em}(x) | \pi^a(p_1) \pi^b(p_2) \> \nn
		&= -i e (2\pi)^4 \delta^{(4)}(p_1+p_2+p_3+p_4+q) {\epsilon_\mu^{\lambda}}^*(-q) \< \pi^c(-p_3) \pi^d(-p_4) | j^\mu_\mathrm{em}(0) | \pi^a(p_1) \pi^b(p_2) \> \nn
		&=: -i e (2\pi)^4 \delta^{(4)}(p_1+p_2+p_3+p_4+q) {\epsilon_\mu^{\lambda}}^*(-q) \M^{\mu}(p_1,p_2,p_3,p_4) \, .
\end{align}
In the end, we will be interested in the limit of a soft on-shell photon. Via Low's theorem~\cite{Low:1958sn}, the first two terms, i.e., the divergent and finite pieces in an expansion in the soft-photon momentum are determined in terms of $\pi\pi$ scattering. These terms need to be defined in a gauge-invariant way that does not introduce kinematic singularities and such that the remainder is non-singular in the soft-photon limit, see also Ref.~\cite{Moussallam:2013una} for a related discussion. The part of the remainder that is linear in the soft-photon momentum still contributes to $(g-2)_\mu$. It is not fixed by Low's theorem and needs to be reconstructed dispersively~\cite{TriangleDR4PiGamma}. Possible input could also be provided by lattice QCD~\cite{Baroni:2018iau,Briceno:2019nns,Briceno:2022omu}. In the following, we will derive the Lorentz decomposition for this contribution.

\subsubsection{BTT decomposition}

In a first step, we consider the decomposition of the matrix element into gauge-invariant Lorentz structures. Applying the BTT~\cite{Bardeen:1969aw,Tarrach:1975tu} recipe to the matrix element $\M^\mu(p_1,p_2,p_3,p_4)$ is a trivial exercise. One starts with four independent four-vectors and applies gauge projectors, which leave three independent structures. However, these structures become degenerate in certain kinematic limits, requiring the introduction of three redundant Tarrach structures. This is equivalent to including the crossed Lorentz structures. The decomposition then reads
\begin{align}
	\M^\mu(p_1,p_2,p_3,p_4) &= \sum_{i=1}^6 T^\mu_i \M_i \, ,
\end{align}
where
\begin{align}
	T_1^\mu = p_1^\mu (p_2 \cdot q) - p_2^\mu (p_1 \cdot q)
\end{align}
and the remaining structures are related by crossing:
\begin{align}
	T_2^\mu &= \Cr{23}{T_1^\mu} \, , \quad 
	T_3^\mu = \Cr{24}{T_1^\mu} \, , \quad 
	T_4^\mu = -\Cr{13}{T_1^\mu} \, , \quad 
	T_5^\mu = -\Cr{14}{T_1^\mu} \, , \quad 
	T_6^\mu = \Cr{13}{\Cr{24}{T_1^\mu}} \, .
\end{align}
Here, we define the crossing operators $\Cr{ij}{}$ to exchange momenta (and isospin indices) of the pions $i$ and $j$. There is one internal crossing symmetry,
\begin{align}
	T_1^\mu = - \Cr{12}{T_1^\mu} \, .
\end{align}
Crossing symmetry of the full amplitude implies that the scalar functions $\M_i$ fulfill the same crossing relations as the Lorentz structures.

Gauge invariance is manifestly fulfilled by the Lorentz structures,
\begin{align}
	q_\mu T_i^\mu = 0 \, ,
\end{align}
and at the same time the scalar functions $\M_i$ are free of kinematic singularities.
The three Tarrach redundancies read
\begin{align}
	(p_3 \cdot q) T_1^\mu - (p_2 \cdot q) T_2^\mu + (p_1 \cdot q) T_4^\mu &= 0 \, , \nn
	(p_4 \cdot q) T_1^\mu - (p_2 \cdot q) T_3^\mu + (p_1 \cdot q) T_5^\mu &= 0 \, , \nn
	(p_4 \cdot q) T_2^\mu - (p_3 \cdot q) T_3^\mu + (p_1 \cdot q) T_6^\mu &= 0 \, .
	\label{eq:gppppTarrach}
\end{align}
Eliminating redundant structures introduces kinematic singularities into the scalar coefficient functions.

Finally, we perform a basis change
\begin{align}
	\M^\mu(p_1,p_2,p_3,p_4) &= \sum_{i=1}^6 T^\mu_i \M_i = \sum_{i=1}^6 \hat T^\mu_i \hat \M_i \, ,
	\label{eq:gppppBTT}
\end{align}
where
\begin{align}
	\hat T_1^\mu &= T_4^\mu \, , \quad \hat T_2^\mu = -T_2^\mu \, , \quad \hat T_3^\mu = T_1^\mu \, , \quad
	\hat T_4^\mu = T_1^\mu + T_2^\mu + T_3^\mu \, , \nn
	\hat T_5^\mu &= - T_1^\mu + T_4^\mu + T_5^\mu \, , \quad \hat T_6^\mu = -T_2^\mu - T_4^\mu + T_6^\mu \, ,
\end{align}
explicitly
\begin{align}
	\begin{alignedat}{3}
		\hat T_1^\mu &= p_2^\mu (p_3 \cdot q) - p_3^\mu (p_2 \cdot q) \, , \quad &
		\hat T_2^\mu &= p_3^\mu (p_1 \cdot q) - p_1^\mu (p_3 \cdot q) \, , \quad &
		\hat T_3^\mu &= p_1^\mu (p_2 \cdot q) - p_2^\mu (p_1 \cdot q) \, , \\
		\hat T_4^\mu &= q^\mu (p_1 \cdot q) - p_1^\mu q^2 \, , \quad &
		\hat T_5^\mu &= q^\mu (p_2 \cdot q) - p_2^\mu q^2 \, , \quad &
		\hat T_6^\mu &= q^\mu (p_3 \cdot q) - p_3^\mu q^2 \, .
	\end{alignedat}
\end{align}
The Tarrach redundancies Eq.~\eqref{eq:gppppTarrach} imply that the shifts
\begin{align}
	\hat{\M}_1 &\mapsto \hat{\M}_1 + (p_1 \cdot q) \Delta_1 \,, \nn
	\hat{\M}_2 &\mapsto \hat{\M}_2 + (p_2 \cdot q) \Delta_1 + q^2 \Delta_3 \,, \nn
	\hat{\M}_3 &\mapsto \hat{\M}_3 + (p_3 \cdot q) \Delta_1 + q^2 \Delta_2 \,, \nn
	\hat{\M}_4 &\mapsto \hat{\M}_4 + (p_2 \cdot q) \Delta_2 - (p_3 \cdot q) \Delta_3 \,, \nn
	\hat{\M}_5 &\mapsto \hat{\M}_5 - (p_1 \cdot q) \Delta_2 \,, \nn
	\hat{\M}_6 &\mapsto \hat{\M}_6 + (p_1 \cdot q) \Delta_3
	\label{eq:gppppTarrachHat}
\end{align}
with arbitrary non-singular $\Delta_i$ leave the amplitude unchanged.

\subsubsection{Soft-photon limit}

In the soft-photon limit, $q\to0$, the scalar coefficient functions $\M_i$ contain double and single poles, which are determined by $\pi\pi$ scattering alone. We assume a gauge-invariant separation of these soft singularities that respects crossing symmetries and leaves a regular remainder, which can be achieved by using a dispersive definition as will be discussed in Ref.~\cite{TriangleDR4PiGamma}:
\begin{align}
	\M_i = \M_i^\text{poles} + \M_i^\text{non-pole} \, .
	\label{gppppSeparation}
\end{align}
We are only interested in the leading non-pole term, i.e., in the limit $\lim\limits_{q\to0} \M_i^\text{non-pole}$. Defining
\begin{align}
	\M^\mu_\text{reg}(p_1,p_2,p_3,p_4) &= \sum_{i=1}^6 \hat T^\mu_i \hat \M_i^\text{non-pole} \, ,
\end{align}
we obtain the desired contribution by taking the following derivative:
\begin{align}
	\left. \frac{\p}{\p q_\nu} \M^\mu_\text{reg}(p_1,p_2,p_3,p_4) \right|_{q=0} &= \sum_{i=1}^6 \left.\left(\frac{\p}{\p q_\nu} \hat T^\mu_i \right)\right|_{q=0} \hat \M_i^\text{non-pole}(q=0) \nn
		&= \sum_{i,j,k=1}^3 \epsilon_{ijk} p_i^\mu p_j^\nu \hat\M_k^\text{non-pole}(q=0) \, ,
		\label{eq:gppppBTTq=0}
\end{align}
with the antisymmetric tensor $\epsilon_{123}=1$. The first three coefficient functions $\hat\M_i$ contain a Tarrach redundancy of the form
\begin{align}
	\hat\M_i \mapsto \hat\M_i + (p_i \cdot q) \Delta \, , \quad i = 1,2,3 \, ,
\end{align}
with arbitrary non-singular $\Delta$, which, however, drops out in the limit $q\to0$.

In the following, we will focus on the mixed-charge channel $\pi^0 \pi^0 \to \pi^+ \pi^- \gamma$: in the isospin limit, the fully charged process $\pi^+ \pi^- \to \pi^+ \pi^- \gamma$ can be related to this amplitude~\cite{Kuhn:1998rh,Ecker:2002cw}. Bose symmetry implies
\begin{align}
	\M^\mu = \Cr{12}{\M^\mu} = -\Cr{34}{\M^\mu} = -  \Cr{12}{\Cr{34}{\M^\mu}} \, .
\end{align}
Since the scalar coefficient functions only depend on the scalar invariants, in the limit $q\to0$ they are invariant under simultaneous crossing of the two neutral and the two charged pions:
\begin{align}
	\hat\M_i(q=0) = \Cr{12}{\Cr{34}{\hat\M_i(q=0)}} \, .
\end{align}
Assuming that the definition of $\M_i^\text{poles}$ respects crossing symmetry, not all three scalar functions are independent at $q=0$, but one finds
\begin{align}
	\hat{\M}_2^\text{non-pole}(q=0) = -\hat{\M}_1^\text{non-pole}(q=0) \,.
\end{align}
Thus, we define the new tensor decomposition
\begin{align}
	\label{eq:gppppTensDecompDerivative}
	\left. \frac{\p}{\p q_\nu} \M^\mu_\text{reg}(p_1,p_2,p_3,p_4) \right|_{q=0} = (p_3^\mu p_4^\nu - p_4^\mu p_3^\nu) \bar{\M}_1 + (p_1^\mu p_2^\nu - p_2^\mu p_1^\nu) \bar{\M}_2
\end{align}
where
\begin{align}
	\bar{\M}_1 = \hat{\M}_1^\text{non-pole}(q=0) \,, \quad \bar{\M}_2 = \hat{\M}_3^\text{non-pole}(q=0) \, .
\end{align}
In the limit $q\to0$, the five-particle process reduces to four-point kinematics. The scalar functions $\bar\M_1$ and $\bar\M_2$ are functions of the Mandelstam variables $s=(p_1+p_2)^2$, $t=(p_1+p_3)^2$, and $u=(p_1+p_4)^2$, fulfilling $s+t+u=4M_\pi^2$.
Crossing symmetry further implies that $\bar\M_1$ is symmetric and $\bar\M_2$ is antisymmetric under $t \leftrightarrow u$. These functions will be reconstructed dispersively in Ref.~\cite{TriangleDR4PiGamma}.

\subsection[The process $\gamma^*\gamma^*\gamma\to2\pi$]{\boldmath The process $\gamma^*\gamma^*\gamma\to2\pi$}

\subsubsection{Kinematics and matrix element}

As a sub-process in the new dispersion relations in triangle kinematics, we require the (unphysical) process $\gamma^*\gamma^*\gamma\to2\pi$ with two off-shell photons as input. $C$-symmetry of the strong interaction implies that the two-pion state is odd under charge conjugation and hence pure isospin $I=1$, i.e., only charged pions contribute. We define the process via the matrix element
\begin{align}
	\label{eq:3g2piAmplitude}
	\< \pi^+(p_1) &\pi^-(p_2) | \gamma^*(q_1,\lambda_1) \gamma^*(q_2,\lambda_2) \gamma(q_3,\lambda_3) \> \nn
		&= i e^3 (2\pi)^4 \delta^{(4)}(p_1+p_2-q_1-q_2-q_3) \epsilon_\mu^{\lambda_1}(q_1) \epsilon_\nu^{\lambda_2}(q_2) \epsilon_\lambda^{\lambda_3}(q_3) \nn
			&\qquad \times \int d^4x\, d^4y\, e^{-i(q_1 \cdot x + q_2 \cdot y)} \< \pi^+(p_1) \pi^-(p_2) | T \{ j^\mu_\mathrm{em}(x) j^\nu_\mathrm{em}(y) j^\lambda_\mathrm{em}(0) \} | 0 \> \nn
		&=: i e^3 (2\pi)^4 \delta^{(4)}(p_1+p_2-q_1-q_2-q_3) \epsilon_\mu^{\lambda_1}(q_1) \epsilon_\nu^{\lambda_2}(q_2) \epsilon_\lambda^{\lambda_3}(q_3) \M^{\mu\nu\lambda}(p_1,p_2,q_1,q_2) \, .
\end{align}
We are interested in the case where the on-shell photon with $q_3^2=0$ is soft and we will need terms up to linear order in $q_3$.

\subsubsection{BTT decomposition}

We start from the BTT decomposition~\cite{Bardeen:1969aw,Tarrach:1975tu} for the process $\gamma^*\gamma^*\gamma^*\to\pi^+\pi^-$ with three off-shell photons. The construction is of considerable complexity, since the rank-3 tensor structures depend on four independent four-momenta. In the construction, we keep photon-crossing symmetry manifest. We define the momenta $q_4 := p_1 + p_2$, $q_5 := p_1 - p_2$, hence $q_1 + q_2 + q_3 = q_4$. The BTT construction starts off with the following 76 naive rank-3 tensor structures:\footnote{Although for five-point kinematics, the 64 structures that do not contain the metric tensor already form a basis in 4 space-time dimensions~\cite{Peraro:2020sfm}, we keep all structures as we are interested in the degenerate soft limit, which corresponds to four-point kinematics.}
\begin{align}
	\{ L_i^{\mu\nu\lambda} \} = \left\{ q_i^\mu q_j^\nu q_k^\lambda , \, q_i^\mu g^{\nu\lambda} , \, q_j^\nu g^{\mu\lambda} , \, q_k^\lambda g^{\mu\nu} \right\}_{ i \in \{ 2,3,4,5 \} , \, j \in \{ 1,3,4,5 \} , \, k \in \{ 1,2,4,5\} } \, .
\end{align}
The application of projectors
\begin{align}
	I_{12}^{\mu\nu} &= g^{\mu\nu} - \frac{q_2^{\mu} q_1^\nu}{q_1 \cdot q_2} \, , \quad
	I_{23}^{\nu\lambda} = g^{\nu\lambda} - \frac{q_3^{\nu} q_2^\lambda}{q_2 \cdot q_3} \, , \quad
	I_{31}^{\lambda\mu} = g^{\lambda\mu} - \frac{q_1^{\lambda} q_3^\mu}{q_1 \cdot q_3}
\end{align}
maps 40 structures directly to zero. The remaining structures map to structures with kinematic singularities, which are removed following the BTT recipe. In the end, the set of structures has to be enlarged again to account for Tarrach degeneracies. We end up with a highly redundant set of 74 off-shell structures, split into 20 distinct equivalence classes under photon crossing:
\begin{align}
	\M^{\mu\nu\lambda}(p_1,p_2,q_1,q_2) = \sum_{i=1}^{74} T_i^{\mu\nu\lambda} \A_i \, .
\end{align}
The 20 photon-crossing classes of tensor structures are defined in App.~\ref{app:BTT3g2pi}.
There are 38 Tarrach redundancies, leading to 36 independent structures in $D$ dimensions. In 4 space-time dimensions 9 additional relations due to the Schouten identity reduce the basis to 27 elements. This agrees with the number of helicity amplitudes for three off-shell photons, $3^3 = 27$. In contrast to $n$-particle processes with $n\le4$, such as $\gamma^*\gamma^*\to\pi\pi$ or HLbL~\cite{Colangelo:2015ama,Colangelo:2017fiz}, parity does not reduce the number of independent helicity amplitudes for a five-particle process~\cite{Peraro:2020sfm}: in this case, the contraction of the tensor structures with polarization vectors leads to angular dependences of the helicity amplitudes that can be expressed as a non-trivial dependence on the parity-odd invariant $\epsilon_{\mu\nu\lambda\sigma} q_1^\mu q_2^\nu q_3^\lambda q_5^\sigma$.

\subsubsection{Soft-photon limit}

In analogy to $\pi\pi\to\pi\pi\gamma$, we assume an appropriate gauge-invariant splitting of the amplitude into soft-singular and regular pieces,
\begin{align}
	\A_i = \A_i^\text{pole} + \A_i^\text{non-pole} \, ,
\end{align}
where the scalar functions $\A_i^\text{pole}$ contain double and single poles in the soft-photon limit $q_3\to0$ and can be expressed in terms of $\gamma^*\gamma^*\to\pi^+\pi^-$, whereas $\A_i^\text{non-pole}$ are regular in the limit $q_3\to0$. Defining
\begin{align}
	\label{eq:BTT3g2pi}
	\M_\text{reg}^{\mu\nu\lambda}(p_1,p_2,q_1,q_2) = \sum_{i=1}^{74} T_i^{\mu\nu\lambda} \A_i^\text{non-pole} \, ,
\end{align}
we are interested only in the leading non-pole term, which is obtained from
\begin{align}
	\frac{\p}{\p q_{3\sigma}} \M_\text{reg}^{\mu\nu\lambda}(p_1,p_2,q_1,q_2)\bigg|_{q_3=0} &= \sum_{i=1}^{74} \left( \frac{\p}{\p q_{3\sigma}} T_i^{\mu\nu\lambda} \right) \bigg|_{q_3=0} \A_i^\text{non-pole}(q_3=0) \nn
		&=: \sum_{i=1}^{74} T_i^{\mu\nu\lambda;\sigma}(q_1,q_2,q_5) \A_i^\text{non-pole}(q_3=0) \, .
\end{align}
With a basis change that does not introduce any kinematic singularities, it is possible to express the soft-photon limit of the regular part in terms of 34 structures
\begin{align}
	\frac{\p}{\p q_{3\sigma}} \M_\text{reg}^{\mu\nu\lambda}(p_1,p_2,q_1,q_2)\bigg|_{q_3=0} &= \sum_{i=1}^{34} \hat T_i^{\mu\nu\lambda;\sigma}(q_1,q_2,q_5) \hat\A_i \, .
\end{align}
The soft-photon limit of the five-particle process corresponds to four-point kinematics and we define Mandelstam variables
\begin{align}
	s = (q_1+q_2)^2 \, , \quad t = (q_1 - p_1)^2 \, , \quad u = (q_1 - p_2)^2 \, ,
\end{align}
fulfilling $s+t+u = q_1^2 + q_2^2 + 2 M_\pi^2$. The matrix
\begin{align}
	C_{ij}(s,t-u,q_1^2,q_2^2) := \hat T_i^{\mu\nu\lambda;\sigma}(q_1,q_2,q_5) \hat T^j_{\mu\nu\lambda;\sigma}(q_1,q_2,q_5)
\end{align}
has rank 27: the set of structures $\hat T_i^{\mu\nu\lambda;\sigma}$ still contains six Tarrach redundancies and the Schouten identity implies one additional linear relation in 4 space-time dimensions.

In order to further reduce the redundancies, we consider crossing symmetry, in analogy to the case of $\gamma^*\gamma^*\to\pi\pi$~\cite{Drechsel:1997xv,Colangelo:2015ama}. We define $\Cr{12}{}$ as the crossing operator exchanging the two off-shell photons and $\Cr{5}{}$ as the crossing operator for the two pions, exchanging $p_1$ and $p_2$ or, equivalently, $q_5 \mapsto -q_5$. The amplitude is even under photon crossing, but odd under pion crossing, since the two pions are in the isospin $I=1$ state:
\begin{align}
	\M_\text{reg}^{\mu\nu\lambda} = \Cr{12}{\M_\text{reg}^{\mu\nu\lambda}} = - \Cr{5}{\M_\text{reg}^{\mu\nu\lambda}} = - \Cr{5}{\Cr{12}{\M_\text{reg}^{\mu\nu\lambda}}} \, .
\end{align}
In the limit $q_3\to0$, the crossing operations act on the scalar functions as
\begin{align}
	\Cr{5}{\hat\A_i(s,t-u,q_1^2,q_2^2)} &= \hat\A_i(s,u-t,q_1^2,q_2^2) \, , \nn
	\Cr{5}{\Cr{12}{\hat\A_i(s,t-u,q_1^2,q_2^2)}} &= \hat\A_i(s,t-u,q_2^2,q_1^2) \, .
\end{align}
It is possible to choose the basis $\hat T_i^{\mu\nu\lambda;\sigma}$ in such a way that all the elements have definite crossing properties:
\begin{align}
	\hat T_i^{\mu\nu\lambda;\sigma} &= -\Cr{5}{\hat T_i^{\mu\nu\lambda;\sigma}} = -\Cr{5}{\Cr{12}{\hat T_i^{\mu\nu\lambda;\sigma}}} \quad \text{for } i \in \{ 1, \ldots, 9 \} \, , \nn
	\hat T_i^{\mu\nu\lambda;\sigma} &= -\Cr{5}{\hat T_i^{\mu\nu\lambda;\sigma}} = \Cr{5}{\Cr{12}{\hat T_i^{\mu\nu\lambda;\sigma}}} \quad \text{for } i \in \{ 10, \ldots, 18 \} \, , \nn
	\hat T_i^{\mu\nu\lambda;\sigma} &= \Cr{5}{\hat T_i^{\mu\nu\lambda;\sigma}} = -\Cr{5}{\Cr{12}{\hat T_i^{\mu\nu\lambda;\sigma}}} \quad \text{for } i \in \{ 19, \ldots, 28 \} \, , \nn
	\hat T_i^{\mu\nu\lambda;\sigma} &= \Cr{5}{\hat T_i^{\mu\nu\lambda;\sigma}} = \Cr{5}{\Cr{12}{\hat T_i^{\mu\nu\lambda;\sigma}}} \quad \text{for } i \in \{ 29, \ldots, 34 \} \, .
\end{align}
This implies that the scalar coefficient functions contain kinematic zeros of the following form:
\begin{align}
	\hat\A_i(s, t-u, q_1^2, q_2^2) &= \kappa_i \, \hat{\hat\A}_i(s, t-u, q_1^2, q_2^2) \, ,
\end{align}
where
\begin{align}
	\kappa_i &= 1 \quad \text{for } i \in \{ 1, \ldots, 9 \} \, , \nn
	\kappa_i &= q_1^2 - q_2^2 \quad \text{for } i \in \{ 10, \ldots, 18 \} \, , \nn
	\kappa_i &= t-u \quad \text{for } i \in \{ 19, \ldots, 28 \} \, , \nn
	\kappa_i &= (q_1^2-q_2^2)(t-u) \quad \text{for } i \in \{ 29, \ldots, 34 \} \, ,
\end{align}
and the functions $\hat{\hat\A}_i$ are still free from kinematic singularities. The kinematic zeros that follow from the crossing symmetries allow us to remove all but a single redundancy in the tensor basis, hence we only need to keep a subset of 28 structures $\kappa_i \hat T_i^{\mu\nu\lambda;\sigma}$ and write
\begin{align}
	\label{eq:3g2piSoftLimit}
	\frac{\p}{\p q_{3\sigma}} \M_\text{reg}^{\mu\nu\lambda}(p_1,p_2,q_1,q_2)\bigg|_{q_3=0} &= \sum_{i=1}^{28} \tilde T_i^{\mu\nu\lambda;\sigma}(q_1,q_2,q_5) \tilde\A_i(s, t-u, q_1^2, q_2^2) \, ,
\end{align}
where the tensor structures $\tilde T_i^{\mu\nu\lambda;\sigma}$ have mass dimensions between 4 and 10. They are given in App.~\ref{app:BTT3g2piSoftPhoton}. We express the scalar functions $\tilde \A_i$ in terms of the 74 coefficient functions $\A_i^\text{non-pole}(q_3=0)$:
\begin{align}
	\tilde\A_i(s, t-u, q_1^2, q_2^2) &= \sum_{j=1}^{74} a_{ij} \A_j^\text{non-pole}(q_3=0) \, ,
\end{align}
where the matrix $a_{ij}$ contains poles of the form $1/(q_1^2-q_2^2)$ and $1/(t-u)$, which are cancelled in $\tilde\A_i$ by kinematic zeros due to crossing symmetry. The matrix $a_{ij}$ is provided as supplementary material, together with a Mathematica notebook that makes use of \texttt{FeynCalc}~\cite{Mertig:1990an,Shtabovenko:2016sxi,Shtabovenko:2020gxv}.

The remaining redundancy only involves structures of dimension 8 and 10 and has the form
\begin{align}
	0 = (t-u)^2 \tilde T_{22}^{\mu\nu\lambda;\sigma} - (q_1^2 - q_2^2 - s)(q_1^2 - q_2^2 + s) \tilde T_{27}^{\mu\nu\lambda;\sigma} + (q_1^2 + q_2^2 - s) \tilde T_{28}^{\mu\nu\lambda;\sigma} \, .
\end{align}
We can write the soft-photon limit in terms of a basis:
\begin{align}
	\label{eq:3g2piSoftLimitBasis}
	\frac{\p}{\p q_{3\sigma}} \M_\text{reg}^{\mu\nu\lambda}(p_1,p_2,q_1,q_2)\bigg|_{q_3=0} &= \sum_{i=1}^{27} \B_i^{\mu\nu\lambda;\sigma}(q_1,q_2,q_5) \bar\A_i(s, t-u, q_1^2, q_2^2) \, ,
\end{align}
where
\begin{align}
	\label{eq:3g2piBasisTensors}
	\B_{22}^{\mu\nu\lambda;\sigma} &= \frac{1}{q_1^2+q_2^2-s} \tilde T_{22}^{\mu\nu\lambda;\sigma} \, , \quad
	\B_{27}^{\mu\nu\lambda;\sigma} = \frac{1}{q_1^2+q_2^2-s} \tilde T_{27}^{\mu\nu\lambda;\sigma} \, , \nn
	\B_i^{\mu\nu\lambda;\sigma} &= \tilde T_i^{\mu\nu\lambda;\sigma} \quad \text{for } i \in \{1, \ldots, 21, 23, \ldots, 26 \} \, .
\end{align}
The redundancy is traded for the following kinematic constraint:
\begin{align}
	\bar\A_{22} &= (q_1^2 + q_2^2 - s) \tilde\A_{22} - (t-u)^2 \tilde\A_{28} \, , \nn
	\bar\A_{27} &= (q_1^2 + q_2^2 - s) \tilde\A_{27} + (q_1^2 - q_2^2 - s)(q_1^2 - q_2^2 + s) \tilde\A_{28} \, , \nn
	\bar\A_i &= \tilde\A_i \quad \text{for } i \in \{1, \ldots, 21, 23, \ldots, 26 \} \, ,
\end{align}
which ensures that the spurious kinematic singularities in Eq.~\eqref{eq:3g2piBasisTensors} drop out in Eq.~\eqref{eq:3g2piSoftLimitBasis}. We note that in contrast to HLbL in four-point kinematics, the assumption of unsubtracted dispersion relations for the basis coefficient functions $\bar\A_i$ does not require sum rules for the functions $\tilde\A_i$ beyond the ones that guarantee basis independence for the regular tensor in the soft-photon limit. In addition, the spurious singularities in Eq.~\eqref{eq:3g2piBasisTensors} are harmless: their cancellation in the imaginary part of the HLbL scalar functions $\hat\Pi_i$ is numerically uncritical, because the imaginary part in the dispersion relation~\eqref{eq:PiHatDispersionRelation} is evaluated only for $q_{1,2}^2 \le 0$ and $s\ge4M_\pi^2$.

This shows that dispersion relations in triangle kinematics, together with the derived tensor decompositions for the sub-processes, open up a path towards the dispersive evaluation of two-pion contributions beyond $S$-waves, including tensor-meson resonances in the $D$-wave: the new formalism is not affected by ambiguous singular subtractions that require the exact fulfillment of sum rules by the HLbL tensor.


\section{Conclusions and outlook}
\label{sec:Conclusion}

In this paper, we introduced a novel dispersive framework for HLbL which directly applies in the kinematic limit relevant for $a_\mu$. We showed in detail how this allows us to overcome issues with kinematic singularities that affect intermediate states of spin two and higher present in the established dispersive approach in four-point kinematics. In the new framework, a reshuffling of intermediate-state contributions takes place and further sub-processes enter the two-pion unitarity relations. These can be dispersively reconstructed without introducing kinematic singularities nor ambiguities.
Our results pave the way for a first complete data-driven evaluation of all contributions to HLbL that are described in terms of exclusive hadronic intermediate states and that are required to reach an accuracy matching the final precision goal of the E989 experiment at Fermilab.

We stress that the goal of our new dispersive formalism is not to replace the established one but rather to extend and complement it. The dispersive reconstruction of the sub-processes needed to solve two-pion unitarity in triangle kinematics will allow us to obtain numerical results for two-pion contributions to HLbL beyond the $S$-wave, including the $f_2(1270)$ resonance in the $D$-wave. Once a description of all relevant sub-processes is available, a detailed analysis of the reshuffling of intermediate-state contributions with respect to the established approach will be possible yielding more robust estimates of suppressed effects. Therefore, while the new formalism offers a path towards a dispersive treatment of higher-spin resonances, we expect that the detailed comparison of the two approaches will be very useful even for contributions that can be included in the established approach, such as scalar and axial-vector contributions.
 Moreover, the new approach provides new perspectives on the matching onto asymptotic constraints~\cite{Colangelo:2019uex,TriangleDRAsymptotics}. A suitable combination of the two dispersive approaches to HLbL will enable a precise data-driven determination of this contribution with reliable uncertainties, compatible with all theoretical and experimental constraints.

	
	\section*{Acknowledgements}
	\addcontentsline{toc}{section}{\numberline{}Acknowledgements}

	We thank Gilberto Colangelo and Martin Hoferichter for valuable discussions and useful comments on the manuscript.
	Financial support by the Swiss National Science Foundation (Project No.~PCEFP2\_194272) is gratefully acknowledged. 	J.L. is supported by the FWF-DACH Grant I~3845-N27 and by the FWF doctoral program Particles and Interactions, project no. W1252-N27.

	
	\appendix
	

\section{Scalar toy examples}
\label{app:ScalarToy}

\subsection{Triangle diagram}

As an illustration of the cancellation of soft singularities between $s$- and $q_3^2$-channel unitarity cuts in HLbL, we consider the simple example of a scalar three-point function. We define the triangle function by
\begin{align}
	C_0(q_1^2,q_2^2,q_3^2) = \int \frac{d^4l}{(2\pi)^4} \frac{i}{l^2 - M^2} \frac{i}{(l-q_2)^2 - M^2} \frac{i}{(l+q_3)^2 - M^2} \, .
\end{align}
The Feynman parametrization reads
\begin{align}
	C_0(q_1^2,q_2^2,q_3^2) &= - \frac{1}{16\pi^2} \int_0^1 dx dy dz \frac{\delta(1-x-y-z)}{\Delta_{123}} \, , \nn
	\Delta_{ijk} &= M^2 - xy q_i^2 - xz q_j^2 - yz q_k^2 \, .
\end{align}
For values of $q_{2,3}^2$ that avoid anomalous thresholds~\cite{Hoferichter:2013ama}, the $C_0$ function fulfills the following dispersion relation:
\begin{align}
	\label{eq:C0DispRel}
	C_0(q_1^2,q_2^2,q_3^2) &= \frac{1}{\pi} \int_{4M^2}^\infty ds \frac{\Delta_1 C_0(s,q_2^2,q_3^2)}{s-q_1^2-i\epsilon} \, , \nn
	\Delta_1 C_0(s,q_2^2,q_3^2) &= \frac{C_0(s+i\epsilon,q_2^2,q_3^2)  - C_0(s-i\epsilon,q_2^2,q_3^2) }{2i} \nn
		&= \frac{1}{16\pi} \frac{1}{\lambda^{1/2}_{23}(s)} \log\left( \frac{ - s(s-q_2^2-q_3^2) + \sqrt{s(s-4M^2)} \lambda_{23}^{1/2}(s)}{- s(s-q_2^2-q_3^2) - \sqrt{s(s-4M^2)} \lambda_{23}^{1/2}(s)} \right) \, , \nn
	\lambda_{ij}(s) &= \lambda(s,q_i^2,q_j^2) \, .
\end{align}
Consider now the scalar triangle diagram for degenerate kinematics $q_3\to0$:
\begin{align}
	C_0(q_1^2,q_1^2,0) = \frac{1}{i} \int \frac{d^4l}{(2\pi)^4} \frac{1}{l^2 - M^2} \frac{1}{(l+q_1)^2 - M^2} \frac{1}{l^2 - M^2} \, .
\end{align}
Explicit evaluation of the integral gives
\begin{align}
	C_0(q_1^2,q_1^2,0) = -\frac{1}{16\pi^2} \frac{1}{\sqrt{q_1^2(q_1^2-4M^2)}} \log\left( \frac{\sqrt{q_1^2(q_1^2-4M^2)} + 2M^2-q_1^2}{2M^2} \right) \, .
\end{align}
This function satisfies a dispersion relation
\begin{align}
	C_0(q_1^2,q_1^2,0) &= \frac{1}{\pi} \int_{4M^2}^\infty ds \frac{\Delta C_0(s)}{s-q_1^2-i\epsilon} \, , \nn
	\Delta C_0(s) &= \frac{C_0(s+i\epsilon,s+i\epsilon,0)-C_0(s-i\epsilon,s-i\epsilon,0)}{2i} = -\frac{1}{16\pi} \frac{1}{\sqrt{s(s-4M^2)}} \, .
\end{align}
This representation can also be obtained by starting from the dispersion relation~\eqref{eq:C0DispRel}:
\begin{align}
	C_0(q_1^2,q_1^2,0) &= \frac{1}{\pi} \int_{4M^2}^\infty ds \frac{\Delta_1 C_0(s,q_1^2,0)}{s-q_1^2-i\epsilon} \, , \nn
	\Delta_1 C_0(s,q_1^2,0) &= \frac{1}{16\pi} \frac{1}{s-q_1^2} \log\left( \frac{ s - \sqrt{s(s-4M^2)}}{s + \sqrt{s(s-4M^2)}} \right) \, .
\end{align}
The double pole can be written as a derivative of the Cauchy kernel:
\begin{align}
	C_0(q_1^2,q_1^2,0) &= \frac{1}{\pi} \int_{4M^2}^\infty ds \frac{\p}{\p s}\left( \frac{1}{s-q_1^2} \right) \frac{-1}{16\pi} \log\left( \frac{ s - \sqrt{s(s-4M^2)}}{s + \sqrt{s(s-4M^2)}} \right) \nn
		&= \frac{1}{\pi} \int_{4M^2}^\infty ds \frac{1}{s-q_1^2}  \frac{\p}{\p s} \frac{1}{16\pi} \log\left( \frac{ s - \sqrt{s(s-4M^2)}}{s + \sqrt{s(s-4M^2)}} \right) \nn
		&= \frac{1}{\pi} \int_{4M^2}^\infty ds \frac{1}{s-q_1^2} \frac{-1}{16\pi} \frac{1}{\sqrt{s(s-4M^2)}} \, ,
\end{align}
where we integrated by parts. However, this trick only works because $q_1^2$ appears in $\Delta_1 C_0(s,q_1^2,0)$ as a pure pole. More generally, we can derive the imaginary part in analogy to Eq.~\eqref{eq:ImPiHat2Disc}:
\begin{align}
	\Delta C_0(s) &= \frac{C_0(s+i\epsilon,s+i\epsilon,0)-C_0(s-i\epsilon,s-i\epsilon,0)}{2i} \nn
		&= \lim_{t\to s} \left[ \frac{C_0(s+i\epsilon,t+i\epsilon,0)-C_0(s-i\epsilon,t-i\epsilon,0)}{2i} \right] \nn
		&= \lim_{t\to s} \begin{aligned}[t]
			&\bigg[ \frac{C_0(s+i\epsilon,t+i\epsilon,0)-C_0(s-i\epsilon,t+i\epsilon,0)}{2i} \nn
			& + \frac{C_0(s-i\epsilon,t+i\epsilon,0)-C_0(s-i\epsilon,t-i\epsilon,0)}{2i} \bigg] \end{aligned} \nn
		&= \lim_{t\to s} \begin{aligned}[t]
			&\bigg[ \frac{C_0(s+i\epsilon,t+i\epsilon,0)-C_0(s-i\epsilon,t+i\epsilon,0)}{2i} \nn
			& + \bigg( \frac{C_0(s+i\epsilon,t+i\epsilon,0)-C_0(s+i\epsilon,t-i\epsilon,0)}{2i} \bigg)^* \bigg] \end{aligned} \nn
		&= \lim_{t\to s} \begin{aligned}[t]
			&\bigg[ \Delta_1 C_0(s,t+i\epsilon,0) + \big( \Delta_2 C_0(s+i\epsilon,t,0) \big)^* \bigg] \, , \end{aligned}
\end{align}
where we made use of the Schwarz reflection principle. Note that the limits of the individual terms in the bracket do not exist due to the soft singularities in the discontinuities. However, these poles cancel in the sum of the two discontinuities. The explicit calculation gives
\begin{align}
	\Delta C_0(s) &= \lim_{t\to s} \bigg[ \Delta_1 C_0(s,t+i\epsilon,0) + \Delta_2 C_0(s+i\epsilon,t,0)^* \bigg] \nn
		&= \frac{1}{16\pi}  \lim_{t\to s} \bigg[ \frac{1}{s-t} f(s) + \frac{1}{t-s} f(t) \bigg] = \frac{1}{16\pi} f'(s) \, ,
\end{align}
where
\begin{align}
	f(s) = \log\left( \frac{ s - \sqrt{s(s-4M^2)}}{s + \sqrt{s(s-4M^2)}} \right) \, ,
\end{align}
i.e., indeed
\begin{align}
	\Delta C_0(s) =  -\frac{1}{16\pi} \frac{1}{\sqrt{s(s-4M^2)}} \, .
\end{align}

\subsection{Box diagram}

Similarly to the triangle diagram, we consider the scalar box integral with equal internal masses, defined by
\begin{align}
	D_0(q_1^2,q_2^2,q_3^2,q_4^2,s,t) = \frac{1}{i} \int \frac{d^4l}{(2\pi)^4} \frac{1}{l^2 - M^2} \frac{1}{(l+q_1)^2 - M^2} \frac{1}{(l+q_1+q_2)^2 - M^2} \frac{1}{(l-q_4)^2 - M^2} \, ,
\end{align}
where $s=(q_1+q_2)^2$ and $t=(q_2+q_3)^2$. This loop function satisfies the following dispersion relation in the Mandelstam variable $s$:
\begin{align}
	D_0(q_1^2,q_2^2,q_3^2,q_4^2,s,t) &= \frac{1}{\pi} \int_{4M^2}^\infty ds' \frac{\Delta_s D_0(q_1^2,q_2^2,q_3^2,q_4^2,s',t)}{s'-s-i\epsilon} \, , \nn
	\Delta_s D_0(q_1^2,q_2^2,q_3^2,q_4^2,s,t) &= \frac{D_0(q_1^2,q_2^2,q_3^2,q_4^2,s+i\epsilon,t)-D_0(q_1^2,q_2^2,q_3^2,q_4^2,s-i\epsilon,t)}{2i} \nn
		&= \frac{1}{64\pi^2} \sqrt{1-\frac{4M^2}{s}} \int d\Omega_l \frac{1}{(l+q_1)^2-M^2} \frac{1}{(l-q_4)^2-M^2} \, ,
\end{align}
where in the phase-space integral, $l^2 = M^2$, $|\vec l| = \frac{\sqrt{s-4M^2}}{2}$. The phase-space integral can be evaluated explicitly by using a Feynman parametrization for the two propagators. It can also be converted into a second dispersion integral, which leads to the double-spectral representation~\cite{Barut1967,Colangelo:2015ama}.

Alternatively, the $D_0$ function satisfies a dispersion relation in the virtuality $q_3^2$ for fixed Mandelstam variables:
\begin{align}
	D_0(q_1^2,q_2^2,q_3^2,q_4^2,s,t) &= \frac{1}{\pi} \int_{4M^2}^\infty dw \frac{\Delta_3 D_0(q_1^2,q_2^2,w,q_4^2,s,t)}{w-q_3^2-i\epsilon} \, , \nn
	\Delta_3 D_0(q_1^2,q_2^2,q_3^2,q_4^2,s,t) &= \frac{D_0(q_1^2,q_2^2,q_3^2+i\epsilon,q_4^2,s,t)-D_0(q_1^2,q_2^2,q_3^2-i\epsilon,q_4^2,s,t)}{2i} \nn
		&= \frac{1}{64\pi^2} \sqrt{1-\frac{4M^2}{q_3^2}} \int d\Omega_l \frac{1}{(l-q_1-q_4)^2-M^2} \frac{1}{(l-q_4)^2-M^2} \, ,
\end{align}
where $l^2 = M^2$, $|\vec l| = \frac{\sqrt{q_3^2-4M^2}}{2}$.

We now consider the box diagram in the kinematic limit $q_4\to0$:
\begin{align}
	D_0(q_1^2,q_2^2,q_3^2,0,q_3^2,q_1^2) = \frac{1}{i} \int \frac{d^4l}{(2\pi)^4} \frac{1}{l^2 - M^2} \frac{1}{(l+q_1)^2 - M^2} \frac{1}{(l+q_1+q_2)^2 - M^2} \frac{1}{l^2 - M^2} \, .
\end{align}
This function satisfies the following dispersion relation in $q_3^2$:
\begin{align}
	\label{eq:D0DegenerateDiscontinuity}
	D_0(q_1^2,q_2^2,q_3^2,0,q_3^2,q_1^2) &= \frac{1}{\pi} \int_{4M^2}^\infty ds \frac{\Delta D_0(q_1^2,q_2^2,s)}{s-q_3^2-i\epsilon} \, , \nn
	\Delta D_0(q_1^2,q_2^2,s) &= \frac{D_0(q_1^2,q_2^2,s+i\epsilon,0,s+i\epsilon,q_1^2)-D_0(q_1^2,q_2^2,s-i\epsilon,0,s-i\epsilon,q_1^2)}{2i} \nn
		&= \frac{1}{16\pi} \frac{1}{\sqrt{s(s-4M^2)}} \frac{2M^2(s-q_1^2+q_2^2) - s q_2^2}{M^2 \lambda_{12}(s)+s q_1^2 q_2^2} \, .
\end{align}
Similarly to the case of the triangle diagram, this discontinuity can be obtained from the discontinuities for non-degenerate kinematics:
\begin{align}
	\Delta D_0(q_1^2,q_2^2,s) &= \lim_{w \to s} \begin{aligned}[t]
		&\bigg[ \frac{D_0(q_1^2,q_2^2,s+i\epsilon,0,w+i\epsilon,q_1^2)-D_0(q_1^2,q_2^2,s+i\epsilon,0,w-i\epsilon,q_1^2)}{2i} \nn
		& +  \frac{D_0(q_1^2,q_2^2,s+i\epsilon,0,w-i\epsilon,q_1^2)-D_0(q_1^2,q_2^2,s-i\epsilon,0,w-i\epsilon,q_1^2)}{2i} \bigg] \end{aligned} \nn
	&= \lim_{w \to s} \begin{aligned}[t]
		&\bigg[ \frac{D_0(q_1^2,q_2^2,s+i\epsilon,0,w+i\epsilon,q_1^2)-D_0(q_1^2,q_2^2,s+i\epsilon,0,w-i\epsilon,q_1^2)}{2i} \nn
		& + \left( \frac{D_0(q_1^2,q_2^2,s+i\epsilon,0,w+i\epsilon,q_1^2) - D_0(q_1^2,q_2^2,s-i\epsilon,0,w+i\epsilon,q_1^2)}{2i}\right)^* \bigg] \end{aligned} \nn
	&= \lim_{w \to s} \bigg[ \Delta_s D_0(q_1^2,q_2^2,s+i\epsilon,0,w,q_1^2) + \Delta_3 D_0(q_1^2,q_2^2,s,0,w+i\epsilon,q_1^2)^* \bigg] \, ,
\end{align}
where again the limit of the individual terms does not exist due to soft singularities in the discontinuities, which cancel in the sum. By explicitly calculating the discontinuities $\Delta_s D_0$ and $\Delta_3 D_0$ using Feynman parameters, we indeed reproduce~\eqref{eq:D0DegenerateDiscontinuity}.


\section{Tensor-meson contributions}
\label{app:TensorMesons}

The contribution of tensor mesons in the NWA  to the $s$-channel unitarity relation is given by Eqs.~\eqref{eq:TensorResonanceTriangleKinematicsQ32}, \eqref{eq:TensorResonanceTriangleKinematicsQ12}, and \eqref{eq:TensorResonanceTriangleKinematicsQ22}, with the following coefficients:
\begin{align}
	t_{4,1}(q_1^2,q_2^2) &= \frac{8 m_T^4}{3} \, , \nn
	t_{4,2}(q_1^2,q_2^2) &= \frac{2}{3} \left(m_T^4+(q_1^2+q_2^2) m_T^2-2 (q_1^2-q_2^2)^2\right) \, , \nn
	t_{4,3}(q_1^2,q_2^2) &= \frac{2}{3} \left(m_T^4-(q_1^2+q_2^2) m_T^2-2 (q_1^2-q_2^2)^2\right) \, , \nn
	t_{4,4}(q_1^2,q_2^2) &= t_{4,5}(q_2^2,q_1^2) = -\frac{4}{3} \left(m_T^4-q_2^2 m_T^2-(q_1^2-q_2^2)^2\right) \, , \nn
	t_{5,1}(q_1^2,q_2^2) &= t_{6,1}(q_1^2,q_2^2) = -m_T^2 (m_T^2-q_1^2-q_2^2) \, , \nn
	t_{5,3}(q_1^2,q_2^2) &= -t_{5,5}(q_1^2,q_2^2) = t_{6,3}(q_2^2,q_1^2) = -t_{6,4}(q_2^2,q_1^2) = (m_T^2+q_2^2) (m_T^2-q_1^2-q_2^2) \, , \nn
	t_{5,4}(q_1^2,q_2^2) &= t_{6,5}(q_1^2,q_2^2) = \frac{1}{2} (m_T^2-q_1^2-q_2^2)^2 \, , \nn
	t_{7,4}(q_1^2,q_2^2) &= t_{7,5}(q_1^2,q_2^2) = -t_{7,2}(q_1^2,q_2^2) = -t_{7,3}(q_1^2,q_2^2) \nn
		&= t_{8,4}(q_2^2,q_1^2) = t_{8,5}(q_2^2,q_1^2) = -t_{8,2}(q_2^2,q_1^2) = -t_{8,3}(q_2^2,q_1^2) = 2 (m_T^2-q_1^2+q_2^2) \, , \nn
	t_{9,3}(q_1^2,q_2^2) &= t_{10,1}(q_1^2,q_2^2) = t_{13,3}(q_1^2,q_2^2) = t_{14,1}(q_1^2,q_2^2) \nn
		&= t_{50,1}(q_1^2,q_2^2) = t_{51,1}(q_1^2,q_2^2) = -t_{39,1}(q_1^2,q_2^2) = -2 m_T^2 \, , \nn
	t_{9,5}(q_1^2,q_2^2) &= t_{10,4}(q_1^2,q_2^2) = t_{13,4}(q_1^2,q_2^2) = t_{14,5}(q_1^2,q_2^2) \nn
		&= -t_{9,4}(q_1^2,q_2^2) = -t_{13,5}(q_1^2,q_2^2) = m_T^2-q_1^2-q_2^2 \, , \nn
	t_{10,3}(q_1^2,q_2^2) &= -t_{10,5}(q_1^2,q_2^2) = t_{14,3}(q_2^2,q_1^2) = -t_{14,4}(q_2^2,q_1^2) = 2 (m_T^2+q_2^2) \, , \nn
	t_{11,3}(q_1^2,q_2^2) &= t_{54,3}(q_1^2,q_2^2) = t_{16,3}(q_2^2,q_1^2) = q_1^2-q_2^2 \, , \nn
	t_{11,4}(q_1^2,q_2^2) &= t_{54,4}(q_1^2,q_2^2) = -t_{16,4}(q_1^2,q_2^2) = t_{16,5}(q_2^2,q_1^2) \nn
		&= -t_{11,5}(q_2^2,q_1^2) = -t_{54,5}(q_2^2,q_1^2) = \frac{1}{2} (m_T^2-q_1^2+q_2^2) \, , \nn
	t_{17,1}(q_1^2,q_2^2) &= -\frac{10 m_T^2}{3} \, , \nn
	t_{17,3}(q_1^2,q_2^2) &= \frac{1}{3} (2 m_T^2+5 (q_1^2+q_2^2)) \, , \nn
	t_{17,4}(q_1^2,q_2^2) &= t_{17,5}(q_2^2,q_1^2) = \frac{1}{6} (7 m_T^2-7 q_1^2-13 q_2^2) \, , \nn
	t_{39,3}(q_1^2,q_2^2) &= 2 m_T^2-q_1^2-q_2^2 \, , \nn
	t_{39,4}(q_1^2,q_2^2) &= t_{39,5}(q_2^2,q_1^2) = \frac{1}{2} (-3 m_T^2+3 q_1^2+q_2^2) \, , \nn
	t_{50,3}(q_1^2,q_2^2) &= t_{51,3}(q_1^2,q_2^2) = q_1^2+q_2^2 \, , \nn
	t_{50,4}(q_1^2,q_2^2) &= t_{51,4}(q_1^2,q_2^2) = t_{50,5}(q_2^2,q_1^2) = t_{51,5}(q_2^2,q_1^2) = \frac{1}{2} (m_T^2-q_1^2-3 q_2^2)
\end{align}
and all other $t_{i,j}$ vanish.


\section{\boldmath Tensor decomposition for $\gamma^*\gamma^*\gamma\to2\pi$}

\subsection{Off-shell tensor structures}
\label{app:BTT3g2pi}

The 20 photon-crossing classes of tensor structures for the off-shell process $\gamma^*\gamma^*\gamma^*\to2\pi$ are defined by the following representative elements:
\begin{align}
	T_1^{\mu\nu\lambda} &= (q_1\cdot q_2)\bigl(q_3^{\mu} g^{\lambda\nu} - q_3^{\nu} g^{\lambda\mu}\bigr) + (q_1\cdot q_3)\bigl(q_2^{\lambda} g^{\mu\nu} - q_2^{\mu} g^{\lambda\nu}\bigr) \nn
		&\quad + q_1^{\nu}\bigl((q_2\cdot q_3) g^{\lambda\mu} - q_2^{\lambda} q_3^{\mu}\bigr) + q_1^{\lambda}\bigl(q_2^{\mu} q_3^{\nu} - (q_2\cdot q_3) g^{\mu\nu}\bigr) \, , \nn
	T_2^{\mu\nu\lambda} &= \bigl(q_3^{\mu}(q_1\cdot q_2) - q_2^{\mu}(q_1\cdot q_3)\bigr)\bigl(q_2^{\lambda} q_3^{\nu} - (q_2\cdot q_3) g^{\lambda\nu}\bigr) \, , \nn
	T_5^{\mu\nu\lambda} &= \bigl(q_4^{\mu}(q_1\cdot q_2) - q_2^{\mu}(q_1\cdot q_4)\bigr)\bigl(q_2^{\lambda} q_3^{\nu} - (q_2\cdot q_3) g^{\lambda\nu}\bigr) \, , \nn
	T_{11}^{\mu\nu\lambda} &= \bigl(q_5^{\mu}(q_1\cdot q_2) - q_2^{\mu}(q_1\cdot q_5)\bigr)\bigl(q_2^{\lambda} q_3^{\nu} - (q_2\cdot q_3) g^{\lambda\nu}\bigr) \, , \nn
	T_{17}^{\mu\nu\lambda} &= \bigl(q_5^{\mu}(q_1\cdot q_4) - q_4^{\mu}(q_1\cdot q_5)\bigr)\bigl(q_2^{\lambda} q_3^{\nu} - (q_2\cdot q_3) g^{\lambda\nu}\bigr) \, , \nn
	T_{20}^{\mu\nu\lambda} &= q_4^{\nu}\left[(q_2\cdot q_3)\bigl((q_1\cdot q_5) g^{\lambda\mu} - q_1^{\lambda} q_5^{\mu}\bigr) + q_2^{\lambda}\bigl(q_5^{\mu}(q_1\cdot q_3) - q_3^{\mu}(q_1\cdot q_5)\bigr)\right] \nn
		&\quad + (q_2\cdot q_4)\left[q_5^{\mu}\bigl(q_1^{\lambda} q_3^{\nu} - (q_1\cdot q_3) g^{\lambda\nu}\bigr) + (q_1\cdot q_5)\bigl(q_3^{\mu} g^{\lambda\nu} - q_3^{\nu} g^{\lambda\mu}\bigr)\right] \, , \nn
	T_{26}^{\mu\nu\lambda} &= q_4^{\lambda} \left[(q_1\cdot q_3) \bigl(q_2^{\mu} q_4^{\nu} - (q_2\cdot q_4) g^{\mu \nu}\bigr) + q_3^{\mu} \bigl(q_1^{\nu} (q_2\cdot q_4) - q_4^{\nu} (q_1\cdot q_2)\bigr)\right] \nn
 		&\quad + (q_3\cdot q_4) \left[ q_4^{\nu} \bigl((q_1\cdot q_2) g^{\lambda \mu} - q_1^{\lambda} q_2^{\mu}\bigr) + (q_2\cdot q_4) \bigl(q_1^{\lambda} g^{\mu \nu} - q_1^{\nu} g^{\lambda \mu}\bigr)\right] \, , \nn
	T_{29}^{\mu\nu\lambda} &= q_5^{\lambda} \left[ (q_1\cdot q_3) \bigl( q_2^{\mu} q_5^{\nu} - (q_2\cdot q_5) g^{\mu\nu}\bigr)+q_3^{\mu} \bigl(q_1^{\nu} (q_2\cdot q_5)-q_5^{\nu} (q_1\cdot q_2)\bigr) \right] \nn
		&\quad + (q_3\cdot q_5) \left[q_5^{\nu} \bigl( (q_1\cdot q_2) g^{\lambda\mu}-q_1^{\lambda} q_2^{\mu} \bigr) + (q_2\cdot q_5) \bigl( q_1^{\lambda} g^{\mu\nu} - q_1^{\nu} g^{\lambda\mu}\bigr) \right] \, , \nn
	T_{32}^{\mu\nu\lambda} &= 2 (q_1\cdot q_2) (q_1\cdot q_3) q_4^{\mu} g^{\lambda \nu} - (q_1\cdot q_4) \left[(q_1\cdot q_2) \bigl(q_3^{\mu} g^{\lambda \nu} - q_3^{\nu} g^{\lambda \mu}\bigr) + (q_1\cdot q_3) \bigl(q_2^{\mu} g^{\lambda \nu} - q_2^{\lambda} g^{\mu \nu}\bigr)\right] \nn
 		&\quad - q_1^{\nu} \left[(q_1\cdot q_4) (q_2\cdot q_3) g^{\lambda \mu} + q_2^{\lambda} \bigl(2 q_4^{\mu} (q_1\cdot q_3) - q_3^{\mu} (q_1\cdot q_4)\bigr)\right] \nn
 		&\quad - q_1^{\lambda} \left[(q_2\cdot q_3) \bigl((q_1\cdot q_4) g^{\mu \nu} - 2 q_4^{\mu} q_1^{\nu}\bigr) + q_3^{\nu} \bigl(2 q_4^{\mu} (q_1\cdot q_2) - q_2^{\mu} (q_1\cdot q_4)\bigr)\right] \, , \nn
	T_{35}^{\mu\nu\lambda} &= 2 (q_1\cdot q_2) (q_1\cdot q_3) q_5^{\mu} g^{\lambda \nu} + (q_1\cdot q_5) \left[(q_1\cdot q_2) \bigl(q_3^{\nu} g^{\lambda \mu} - q_3^{\mu} g^{\lambda \nu}\bigr) + (q_1\cdot q_3) \bigl(q_2^{\lambda} g^{\mu \nu} - q_2^{\mu} g^{\lambda \nu}\bigr)\right] \nn*
		&\quad + q_1^{\nu} \left[q_2^{\lambda} \bigl(q_3^{\mu} (q_1\cdot q_5) - 2 q_5^{\mu} (q_1\cdot q_3)\bigr) - (q_1\cdot q_5) (q_2\cdot q_3) g^{\lambda \mu}\right] \nn*
		&\quad + q_1^{\lambda} \left[(q_2\cdot q_3) \bigl(2 q_5^{\mu} q_1^{\nu} - (q_1\cdot q_5) g^{\mu \nu}\bigr) + q_3^{\nu} \bigl(q_2^{\mu} (q_1\cdot q_5) - 2 q_5^{\mu} (q_1\cdot q_2)\bigr)\right] \, , \nn
	T_{38}^{\mu\nu\lambda} &= \bigl(q_4^{\mu}\left(q_1\cdot q_3\right) - q_3^{\mu}\left(q_1\cdot q_4\right)\bigr)\bigl[\left(q_3\cdot q_4\right)\bigl(\left(q_2\cdot q_5\right) g^{\lambda\nu} - q_2^{\lambda} q_5^{\nu}\bigr) + q_4^{\lambda}\bigl(q_5^{\nu}\left(q_2\cdot q_3\right) - q_3^{\nu}\left(q_2\cdot q_5\right)\bigr)\bigr] \, , \nn
	T_{44}^{\mu\nu\lambda} &= \bigl(q_4^{\mu}\left(q_1\cdot q_3\right) - q_3^{\mu}\left(q_1\cdot q_4\right)\bigr)\bigl[\left(q_3\cdot q_5\right)\bigl(\left(q_2\cdot q_5\right) g^{\lambda\nu} - q_2^{\lambda} q_5^{\nu}\bigr) + q_5^{\lambda}\bigl(q_5^{\nu}\left(q_2\cdot q_3\right) - q_3^{\nu}\left(q_2\cdot q_5\right)\bigr)\bigr] \, , \nn
	T_{50}^{\mu\nu\lambda} &= \bigl(q_5^{\mu}\left(q_1\cdot q_4\right) - q_4^{\mu}\left(q_1\cdot q_5\right)\bigr)\bigl[\left(q_2\cdot q_4\right)\bigl(\left(q_3\cdot q_5\right) g^{\lambda\nu} - q_5^{\lambda} q_3^{\nu}\bigr) + q_4^{\nu}\bigl(q_5^{\lambda}\left(q_2\cdot q_3\right) - q_2^{\lambda}\left(q_3\cdot q_5\right)\bigr)\bigr] \, , \nn
	T_{56}^{\mu\nu\lambda} &= \bigl(q_5^{\mu}\left(q_1\cdot q_4\right) - q_4^{\mu}\left(q_1\cdot q_5\right)\bigr)\bigl[\left(q_3\cdot q_5\right)\bigl(\left(q_2\cdot q_5\right) g^{\lambda\nu} - q_2^{\lambda} q_5^{\nu}\bigr) + q_5^{\lambda}\bigl(q_5^{\nu}\left(q_2\cdot q_3\right) - q_3^{\nu}\left(q_2\cdot q_5\right)\bigr)\bigr] \, , \nn
	T_{59}^{\mu\nu\lambda} &= \bigl(q_5^{\mu}\left(q_1\cdot q_4\right) - q_4^{\mu}\left(q_1\cdot q_5\right)\bigr)\bigl[\left(q_3\cdot q_4\right)\bigl(\left(q_2\cdot q_4\right) g^{\lambda\nu} - q_2^{\lambda} q_4^{\nu}\bigr) + q_4^{\lambda}\bigl(q_4^{\nu}\left(q_2\cdot q_3\right) - q_3^{\nu}\left(q_2\cdot q_4\right)\bigr)\bigr] \, , \nn
	T_{62}^{\mu\nu\lambda} &= q_4^{\lambda} \Big[ (q_1\cdot q_4) \left[ (q_2\cdot q_4) \bigl((q_1\cdot q_3) g^{\mu \nu} + q_3^{\mu} q_1^{\nu}\bigr) - q_4^{\nu} \bigl(q_3^{\mu} (q_1\cdot q_2) + q_2^{\mu} (q_1\cdot q_3)\bigr)\right] \nn
		&\qquad + 2 q_4^{\mu} (q_1\cdot q_3) \bigl(q_4^{\nu} (q_1\cdot q_2) - q_1^{\nu} (q_2\cdot q_4)\bigr)\Big] \nn
 		&\quad + (q_3\cdot q_4) \Big[ q_4^{\nu} \left[ (q_1\cdot q_4) \bigl((q_1\cdot q_2) g^{\lambda \mu} + q_1^{\lambda} q_2^{\mu}\bigr) - 2 q_1^{\lambda} q_4^{\mu} (q_1\cdot q_2)\right] \nn
		&\qquad\qquad\quad + (q_2\cdot q_4) \left[ 2 q_1^{\lambda} q_4^{\mu} q_1^{\nu} - (q_1\cdot q_4) \bigl(q_1^{\nu} g^{\lambda \mu} + q_1^{\lambda} g^{\mu \nu}\bigr)\right]\Big] \, , \nn
	T_{65}^{\mu\nu\lambda} &= q_4^{\lambda} \Big[(q_2\cdot q_4) \left[ (q_1\cdot q_5) \bigl((q_1\cdot q_3) g^{\mu \nu} + q_3^{\mu} q_1^{\nu}\bigr) - 2 q_5^{\mu} q_1^{\nu} (q_1\cdot q_3)\right] \nn
		&\qquad + q_4^{\nu} \left[ 2 q_5^{\mu} (q_1\cdot q_2) (q_1\cdot q_3) - (q_1\cdot q_5) \bigl(q_3^{\mu} (q_1\cdot q_2) + q_2^{\mu} (q_1\cdot q_3)\bigr)\right]\Big] \nn
		&\quad + (q_3\cdot q_4) \Big[ q_4^{\nu} \left[ (q_1\cdot q_5) \bigl((q_1\cdot q_2) g^{\lambda \mu} + q_1^{\lambda} q_2^{\mu}\bigr) - 2 q_1^{\lambda} q_5^{\mu} (q_1\cdot q_2)\right] \nn
		&\qquad\qquad\quad + (q_2\cdot q_4) \left(2 q_1^{\lambda} q_5^{\mu} q_1^{\nu} - (q_1\cdot q_5) \left(q_1^{\nu} g^{\lambda \mu} + q_1^{\lambda} g^{\mu \nu}\right)\right)\Big] \, , \nn
	T_{68}^{\mu\nu\lambda} &= (q_1\cdot q_4) \Big[ q_5^{\lambda} \left[(q_2\cdot q_5) \bigl((q_1\cdot q_3) g^{\mu \nu} + q_3^{\mu} q_1^{\nu}\bigr) - q_5^{\nu} \bigl(q_3^{\mu} (q_1\cdot q_2) + q_2^{\mu} (q_1\cdot q_3)\bigr)\right] \nn
		&\qquad\qquad\quad + (q_3\cdot q_5) \left[q_5^{\nu} \bigl((q_1\cdot q_2) g^{\lambda \mu} + q_1^{\lambda} q_2^{\mu}\bigr) - (q_2\cdot q_5) \bigl(q_1^{\nu} g^{\lambda \mu} + q_1^{\lambda} g^{\mu \nu}\bigr)\right]\Big] \nn
 		&\quad + 2 q_4^{\mu} \bigl(q_5^{\lambda} (q_1\cdot q_3) - q_1^{\lambda} (q_3\cdot q_5)\bigr) \bigl(q_5^{\nu} (q_1\cdot q_2) - q_1^{\nu} (q_2\cdot q_5)\bigr) \, , \nn
	T_{71}^{\mu\nu\lambda} &= q_5^{\lambda} \Big[ (q_1\cdot q_5) \left[ (q_2\cdot q_5) \bigl((q_1\cdot q_3) g^{\mu \nu} + q_3^{\mu} q_1^{\nu}\bigr) - q_5^{\nu} \bigl(q_3^{\mu} (q_1\cdot q_2) + q_2^{\mu} (q_1\cdot q_3)\bigr)\right] \nn
		&\qquad + 2 q_5^{\mu} (q_1\cdot q_3) \bigl(q_5^{\nu} (q_1\cdot q_2) - q_1^{\nu} (q_2\cdot q_5)\bigr)\Big] \nn
 		&\quad + (q_3\cdot q_5) \Big[ q_5^{\nu} \left[(q_1\cdot q_5) \bigl((q_1\cdot q_2) g^{\lambda \mu} + q_1^{\lambda} q_2^{\mu}\bigr) - 2 q_1^{\lambda} q_5^{\mu} (q_1\cdot q_2)\right] \nn
		&\qquad\qquad\quad + (q_2\cdot q_5) \left[2 q_1^{\lambda} q_5^{\mu} q_1^{\nu} - (q_1\cdot q_5) \bigl(q_1^{\nu} g^{\lambda \mu} + q_1^{\lambda} g^{\mu \nu}\bigr)\right]\Big] \, , \nn
	T_{74}^{\mu\nu\lambda} &= \bigl(q_5^{\lambda}\left(q_3\cdot q_4\right) - q_4^{\lambda}\left(q_3\cdot q_5\right)\bigr) \bigl(q_5^{\mu}\left(q_1\cdot q_4\right) - q_4^{\mu}\left(q_1\cdot q_5\right)\bigr) \bigl(q_5^{\nu}\left(q_2\cdot q_4\right) - q_4^{\nu}\left(q_2\cdot q_5\right)\bigr) \, .
\end{align}
The remaining 54 structures in Eq.~\eqref{eq:BTT3g2pi} can be obtained from the given ones by applying the following photon-crossing operations:
\begin{align}
	T_3^{\mu\nu\lambda} &= \Cr{12}{T_2^{\mu\nu\lambda}} \, , \; T_4^{\mu\nu\lambda} = \Cr{13}{T_2^{\mu\nu\lambda}} \, , \nn
	T_6^{\mu\nu\lambda} &= \Cr{12}{T_5^{\mu\nu\lambda}} \, , \; T_7^{\mu\nu\lambda} = \Cr{13}{T_5^{\mu\nu\lambda}} \, , \; T_8^{\mu\nu\lambda} = \Cr{23}{T_5^{\mu\nu\lambda}} \, , \; T_9^{\mu\nu\lambda} = \Cr{12}{T_8^{\mu\nu\lambda}} \, , \; T_{10}^{\mu\nu\lambda} = \Cr{12}{T_7^{\mu\nu\lambda}} \, , \nn
	T_{12}^{\mu\nu\lambda} &= \Cr{12}{T_{11}^{\mu\nu\lambda}} \, , \; T_{13}^{\mu\nu\lambda} = \Cr{13}{T_{11}^{\mu\nu\lambda}} \, , \; T_{14}^{\mu\nu\lambda} = \Cr{23}{T_{11}^{\mu\nu\lambda}} \, , \; T_{15}^{\mu\nu\lambda} = \Cr{12}{T_{14}^{\mu\nu\lambda}} \, , \; T_{16}^{\mu\nu\lambda} = \Cr{12}{T_{13}^{\mu\nu\lambda}} \, , \nn
	T_{18}^{\mu\nu\lambda} &= \Cr{12}{T_{17}^{\mu\nu\lambda}} \, , \; T_{19}^{\mu\nu\lambda} = \Cr{13}{T_{17}^{\mu\nu\lambda}} \, , \nn
	T_{21}^{\mu\nu\lambda} &= \Cr{12}{T_{20}^{\mu\nu\lambda}} \, , \; T_{22}^{\mu\nu\lambda} = \Cr{13}{T_{20}^{\mu\nu\lambda}} \, , \; T_{23}^{\mu\nu\lambda} = \Cr{23}{T_{20}^{\mu\nu\lambda}} \, , \; T_{24}^{\mu\nu\lambda} = \Cr{12}{T_{23}^{\mu\nu\lambda}} \, , \; T_{25}^{\mu\nu\lambda} = \Cr{12}{T_{22}^{\mu\nu\lambda}} \, , \nn
	T_{27}^{\mu\nu\lambda} &= \Cr{12}{T_{26}^{\mu\nu\lambda}} \, , \; T_{28}^{\mu\nu\lambda} = \Cr{13}{T_{26}^{\mu\nu\lambda}} \, , \nn
	T_{30}^{\mu\nu\lambda} &= \Cr{12}{T_{29}^{\mu\nu\lambda}} \, , \; T_{31}^{\mu\nu\lambda} = \Cr{13}{T_{29}^{\mu\nu\lambda}} \, , \nn
	T_{33}^{\mu\nu\lambda} &= \Cr{12}{T_{32}^{\mu\nu\lambda}} \, , \; T_{34}^{\mu\nu\lambda} = \Cr{13}{T_{32}^{\mu\nu\lambda}} \, , \nn
	T_{36}^{\mu\nu\lambda} &= \Cr{12}{T_{35}^{\mu\nu\lambda}} \, , \; T_{37}^{\mu\nu\lambda} = \Cr{13}{T_{35}^{\mu\nu\lambda}} \, , \nn
	T_{39}^{\mu\nu\lambda} &= \Cr{12}{T_{38}^{\mu\nu\lambda}} \, , \; T_{40}^{\mu\nu\lambda} = \Cr{13}{T_{38}^{\mu\nu\lambda}} \, , \; T_{41}^{\mu\nu\lambda} = \Cr{23}{T_{38}^{\mu\nu\lambda}} \, , \; T_{42}^{\mu\nu\lambda} = \Cr{12}{T_{41}^{\mu\nu\lambda}} \, , \; T_{43}^{\mu\nu\lambda} = \Cr{12}{T_{40}^{\mu\nu\lambda}} \, , \nn
	T_{45}^{\mu\nu\lambda} &= \Cr{12}{T_{44}^{\mu\nu\lambda}} \, , \; T_{46}^{\mu\nu\lambda} = \Cr{13}{T_{44}^{\mu\nu\lambda}} \, , \; T_{47}^{\mu\nu\lambda} = \Cr{23}{T_{44}^{\mu\nu\lambda}} \, , \; T_{48}^{\mu\nu\lambda} = \Cr{12}{T_{47}^{\mu\nu\lambda}} \, , \; T_{49}^{\mu\nu\lambda} = \Cr{12}{T_{46}^{\mu\nu\lambda}} \, , \nn
	T_{51}^{\mu\nu\lambda} &= \Cr{12}{T_{50}^{\mu\nu\lambda}} \, , \; T_{52}^{\mu\nu\lambda} = \Cr{13}{T_{50}^{\mu\nu\lambda}} \, , \; T_{53}^{\mu\nu\lambda} = \Cr{23}{T_{50}^{\mu\nu\lambda}} \, , \; T_{54}^{\mu\nu\lambda} = \Cr{12}{T_{53}^{\mu\nu\lambda}} \, , \; T_{55}^{\mu\nu\lambda} = \Cr{12}{T_{52}^{\mu\nu\lambda}} \, , \nn
	T_{57}^{\mu\nu\lambda} &= \Cr{12}{T_{56}^{\mu\nu\lambda}} \, , \; T_{58}^{\mu\nu\lambda} = \Cr{13}{T_{56}^{\mu\nu\lambda}} \, , \nn
	T_{60}^{\mu\nu\lambda} &= \Cr{12}{T_{59}^{\mu\nu\lambda}} \, , \; T_{61}^{\mu\nu\lambda} = \Cr{13}{T_{59}^{\mu\nu\lambda}} \, , \nn
	T_{63}^{\mu\nu\lambda} &= \Cr{12}{T_{62}^{\mu\nu\lambda}} \, , \; T_{64}^{\mu\nu\lambda} = \Cr{13}{T_{62}^{\mu\nu\lambda}} \, , \nn
	T_{66}^{\mu\nu\lambda} &= \Cr{12}{T_{65}^{\mu\nu\lambda}} \, , \; T_{67}^{\mu\nu\lambda} = \Cr{13}{T_{65}^{\mu\nu\lambda}} \, , \nn
	T_{69}^{\mu\nu\lambda} &= \Cr{12}{T_{68}^{\mu\nu\lambda}} \, , \; T_{70}^{\mu\nu\lambda} = \Cr{13}{T_{68}^{\mu\nu\lambda}} \, , \nn
	T_{72}^{\mu\nu\lambda} &= \Cr{12}{T_{71}^{\mu\nu\lambda}} \, , \; T_{73}^{\mu\nu\lambda} = \Cr{13}{T_{71}^{\mu\nu\lambda}} \, .
\end{align}
The tensor structures $T_i^{\mu\nu\lambda}$ can also be found in the supplementary material.

\subsection{Tensor structures for the soft-photon limit}
\label{app:BTT3g2piSoftPhoton}

We express the 28 tensor structures in the soft-photon limit in terms of the derivatives of the 74 off-shell structures defined in App.~\ref{app:BTT3g2pi}:
\begin{align}
	T_i^{\mu\nu\lambda;\sigma}(q_1,q_2,q_5) = \left( \frac{\p}{\p q_{3\sigma}} T_i^{\mu\nu\lambda} \right) \bigg|_{q_3=0} \, .
\end{align}
They are given by
\begin{align}
	\tilde T_1^{\mu\nu\lambda;\sigma} &= T_{11}^{\mu\nu\lambda;\sigma} + T_{12}^{\mu\nu\lambda;\sigma} \, , \quad
	\tilde T_2^{\mu\nu\lambda;\sigma} = T_{13}^{\mu\nu\lambda;\sigma} + T_{16}^{\mu\nu\lambda;\sigma} \, , \quad
	\tilde T_3^{\mu\nu\lambda;\sigma} = T_{17}^{\mu\nu\lambda;\sigma} + T_{18}^{\mu\nu\lambda;\sigma} \, , \nn
	\tilde T_4^{\mu\nu\lambda;\sigma} &= T_{20}^{\mu\nu\lambda;\sigma} + T_{21}^{\mu\nu\lambda;\sigma} \, , \quad
	\tilde T_5^{\mu\nu\lambda;\sigma} = T_{22}^{\mu\nu\lambda;\sigma} + T_{25}^{\mu\nu\lambda;\sigma} \, , \quad
	\tilde T_6^{\mu\nu\lambda;\sigma} = T_{23}^{\mu\nu\lambda;\sigma} + T_{24}^{\mu\nu\lambda;\sigma} \, , \nn
	\tilde T_7^{\mu\nu\lambda;\sigma} &= (t-u) T_{1}^{\mu\nu\lambda;\sigma} \, , \nn
	\tilde T_8^{\mu\nu\lambda;\sigma} &= T_{41}^{\mu\nu\lambda;\sigma} + T_{42}^{\mu\nu\lambda;\sigma} \, , \quad
	\tilde T_9^{\mu\nu\lambda;\sigma} = T_{56}^{\mu\nu\lambda;\sigma} + T_{57}^{\mu\nu\lambda;\sigma} \, , \quad
	\tilde T_{10}^{\mu\nu\lambda;\sigma} = T_{58}^{\mu\nu\lambda;\sigma} \, , \nn
	\tilde T_{11}^{\mu\nu\lambda;\sigma} &= (q_1^2-q_2^2) \left( T_{11}^{\mu\nu\lambda;\sigma} - T_{12}^{\mu\nu\lambda;\sigma} \right) \, , \quad
	\tilde T_{12}^{\mu\nu\lambda;\sigma} = (q_1^2-q_2^2) \left( T_{13}^{\mu\nu\lambda;\sigma} - T_{16}^{\mu\nu\lambda;\sigma} \right) \, , \nn
	\tilde T_{13}^{\mu\nu\lambda;\sigma} &= (q_1^2-q_2^2) \left( T_{17}^{\mu\nu\lambda;\sigma} - T_{18}^{\mu\nu\lambda;\sigma} \right) \, , \quad
	\tilde T_{14}^{\mu\nu\lambda;\sigma} = (q_1^2-q_2^2) \left( T_{20}^{\mu\nu\lambda;\sigma} - T_{21}^{\mu\nu\lambda;\sigma} \right) \, , \nn
	\tilde T_{15}^{\mu\nu\lambda;\sigma} &= (q_1^2-q_2^2) \left( T_{23}^{\mu\nu\lambda;\sigma} - T_{24}^{\mu\nu\lambda;\sigma} \right) \, , \quad
	\tilde T_{16}^{\mu\nu\lambda;\sigma} = (q_1^2-q_2^2) \left( T_{22}^{\mu\nu\lambda;\sigma} - T_{25}^{\mu\nu\lambda;\sigma} \right) \, , \nn
	\tilde T_{17}^{\mu\nu\lambda;\sigma} &= (t-u) T_{4}^{\mu\nu\lambda;\sigma} \, , \quad
	\tilde T_{18}^{\mu\nu\lambda;\sigma} = (t-u) \left( T_{26}^{\mu\nu\lambda;\sigma} - T_{27}^{\mu\nu\lambda;\sigma} \right) \, , \nn
	\tilde T_{19}^{\mu\nu\lambda;\sigma} &= (t-u) T_{28}^{\mu\nu\lambda;\sigma} \, , \quad
	\tilde T_{20}^{\mu\nu\lambda;\sigma} = (t-u) \left( T_{29}^{\mu\nu\lambda;\sigma} - T_{30}^{\mu\nu\lambda;\sigma} \right) \, , \quad
	\tilde T_{21}^{\mu\nu\lambda;\sigma} = (t-u) T_{31}^{\mu\nu\lambda;\sigma} \, , \nn
	\tilde T_{22}^{\mu\nu\lambda;\sigma} &= (q_1^2-q_2^2-s) T_{41}^{\mu\nu\lambda;\sigma} - (q_1^2-q_2^2+s) T_{42}^{\mu\nu\lambda;\sigma} \, , \quad
	\tilde T_{23}^{\mu\nu\lambda;\sigma} = (q_1^2-q_2^2) \left( T_{56}^{\mu\nu\lambda;\sigma} - T_{57}^{\mu\nu\lambda;\sigma} \right) \, , \nn
	\tilde T_{24}^{\mu\nu\lambda;\sigma} &= (q_1^2-q_2^2) \left( \frac{t-u}{2} \, T_{30}^{\mu\nu\lambda;\sigma} + T_{58}^{\mu\nu\lambda;\sigma} - T_{72}^{\mu\nu\lambda;\sigma} \right) \, , \quad
	\tilde T_{25}^{\mu\nu\lambda;\sigma} = (t-u) \left( T_{50}^{\mu\nu\lambda;\sigma} - T_{51}^{\mu\nu\lambda;\sigma} \right) \, , \nn
	\tilde T_{26}^{\mu\nu\lambda;\sigma} &= (q_1^2-q_2^2)(t-u) \left( T_{26}^{\mu\nu\lambda;\sigma} + T_{27}^{\mu\nu\lambda;\sigma} \right) \, , \nn
	\tilde T_{27}^{\mu\nu\lambda;\sigma} &= (q_1^2+q_2^2-s)\left( T_{56}^{\mu\nu\lambda;\sigma} + T_{57}^{\mu\nu\lambda;\sigma} \right) - (q_1^2-q_2^2-s) T_{58}^{\mu\nu\lambda;\sigma} + (q_1^2-q_2^2) T_{72}^{\mu\nu\lambda;\sigma} \nn*
		&\quad + (t-u) \Bigl[ -T_{50}^{\mu\nu\lambda;\sigma} + T_{51}^{\mu\nu\lambda;\sigma} + (q_1^2-q_2^2-s) T_{29}^{\mu\nu\lambda;\sigma} + \frac{1}{2}(q_1^2-q_2^2+2s) T_{30}^{\mu\nu\lambda;\sigma} \nn*
		&\qquad\qquad\quad + (s-4M_\pi^2) \left( s T_1^{\mu\nu\lambda;\sigma} + T_{26}^{\mu\nu\lambda;\sigma} - T_{27}^{\mu\nu\lambda;\sigma} - T_{28}^{\mu\nu\lambda;\sigma} \right) \Bigr] \nn*
		&\quad + \frac{(t-u)^2}{2} \left( -T_{17}^{\mu\nu\lambda;\sigma} - T_{18}^{\mu\nu\lambda;\sigma} + T_{20}^{\mu\nu\lambda;\sigma} + T_{21}^{\mu\nu\lambda;\sigma} + T_{22}^{\mu\nu\lambda;\sigma} + T_{25}^{\mu\nu\lambda;\sigma} \right) \, , \nn*
	\tilde T_{28}^{\mu\nu\lambda;\sigma} &= (t-u) \left( (q_1^2-q_2^2-s) T_{50}^{\mu\nu\lambda;\sigma} + (q_1^2-q_2^2+s) T_{51}^{\mu\nu\lambda;\sigma} \right) \, .
\end{align}
The tensor structures $\tilde T_i^{\mu\nu\lambda;\sigma}$ can also be found in the supplementary material.

	\pagebreak

	\phantomsection
	\addcontentsline{toc}{section}{\numberline{}References}
	\bibliographystyle{utphysmod}
	\bibliography{Literature}
	
\end{document}